\documentclass[bibyear]{aa} 

\usepackage{natbib}

\usepackage{subfigure}

\usepackage{graphicx}

\usepackage[utf8]{inputenc}
\usepackage[english]{babel}

\usepackage{txfonts}
\newcommand{\crra}{Cram\'er-Rao}
\begin{document}

   \title{Analysis of the Bayesian Cram\'er-Rao lower bound in
     astrometry:}

   \subtitle{Studying the impact of prior information in the location
     of an object}

    \titlerunning{Bayesian Cram\'er-Rao bound in
     Astrometry}

     \authorrunning{Echeverria et al.}

   \author{Alex Echeverria\inst{1}, Jorge F. Silva,
          \inst{1},
          Rene A. Mendez\inst{2}
\and
Marcos Orchard\inst{1}}

\institute{Information and Decision Systems Group, Department of
  Electrical Engineering, Universidad de Chile, Av. Tupper 2007,
  Santiago, Chile\\
\email{aecheverria@ing.uchile.cl,
    josilva@ing.uchile.cl, morchard@ing.uchile.cl}
\and
Departamento de Astronom\'{\i}a, Facultad de Ciencias F\'{\i}sicas y
Matem\'aticas, Universidad de Chile, Casilla 36-D, Santiago, Chile\\
\email{rmendez@u.uchile.cl}
}

   \date{Received January, 2016}

\abstract
{ The best precision that can be
  achieved to estimate the location of a stellar-like object 
  is a topic of permanent interest in the astrometric community.
  }
{ We analyze bounds for the best position estimation of a stellar-like object on a CCD detector array in a
  Bayesian setting where the position is unknown,  but where we have access
  to a prior distribution. In contrast to a parametric setting where we 
  estimate a parameter from  observations,  the Bayesian approach 
  estimates a random object (i.e., the position is 
  a random variable) from observations that are statistically dependent on
  the position.}
{We characterize the  Bayesian Cram\'er-Rao (CR) that bounds the minimum mean square error
(MMSE) of the best estimator of the position of a point source on a
linear CCD-like detector, as a function of the properties of detector,
the source, and the background.
 }
 { 
  We quantify and analyze   the increase in
  astrometric performance from the use of a prior distribution of the
  object position, which is not available in the classical parametric setting.
  This gain is shown to be significant for various observational regimes, in
  particular in the case of faint objects or when the observations
  are taken under poor conditions. Furthermore, we present numerical evidence that
  the MMSE estimator of this problem 
  tightly achieves the Bayesian CR bound. This is a remarkable result, demonstrating that all
  the performance gains presented in our 
  analysis can be achieved with the MMSE estimator. 
  }
{The Bayesian CR bound 
  can be used as a benchmark indicator of the expected maximum positional precision
  of a set of astrometric measurements in which prior information can
  be incorporated. This bound can 
  be achieved through the conditional mean estimator, in contrast to the parametric
  case where  no unbiased estimator precisely reaches the CR bound.}


   \maketitle
%

\section{Introduction}
\label{sec_intro}

Astrometry, which relies on the precise determination of the relative
location of point sources, is the foundation of classical
astronomy and modern astrophysics, and it will remain a cornerstone of the
field for the 21st century. Historically, it was the first step in the
evolution of astronomy from phenomenology to a science that is rooted
in precise measurements and physical theory. Astrometry spans more
than two thousand years, from Hipparchus (ca. 130 BC) and Ptolemy (150
AD) to modern digital-based all sky surveys, from the ground and in space. The
dramatic improvement in accuracy reflects this historic time scale
(\citet{hog2011astrometry}, see,  e.g.,  his Figure~4). Nowadays, astronomers take
for granted resources such as the ESA Hipparcos mission, which yielded
a catalog of more than 100,000 stellar positions to an accuracy of
1~milliarcsecond, and look forward to the results of the ESA Gaia
astrometric satellite, which will deliver a catalog of over $10^9$
stars with accuracies smaller than 10-20 microarcseconds for objects
brighter than $V=15$ and a completeness limit of $V=20$.

The determination of the best precision that can be achieved to
determine the location of a stellar-like object has been a topic of
permanent interest in the astrometric community
\citep{van1975digital,lindegren1978photoelectric,auer1978digital,lee1983theoretical,winick1986cramer,jakobsen1992cramer,adorf1996limits,bastian2004maximum,lindegren2010}. 
One of the tools used to characterize this precision is the
\crra\ (CR) bound, which provides a lower bound for the
variance that can be achieved  to estimate (with an unbiased estimator)
the position of a point source (\citet{mendez2013analysis, mendez2014analysis};
\citet{lobos2015performance}), given the properties of the source and the
detector. In astrometry this CR bound offers meaningful closed-form
expressions that can be used to analyze the complexity of the
inference task in terms of key observational and design parameters, 
such as position of the object in the array, pixel resolution of the
instrument, and background. In particular, \citet{mendez2013analysis, mendez2014analysis} have developed closed-form expressions for this bound
and have studied its structure and dependency with respect to
important observational parameters. Furthermore, the analysis of the
CR bound allows us to address the problem of optimal pixel resolution of
the array for a given observational setting, and in general to
evaluate the complexity of the astrometric task with respect to the
signal-to-noise ratio (S/R) and different observational
regimes. Complementing these results, \citet{lobos2015performance} have studied the
conditions under which the CR bound can be achieved by a practical
estimator.  In that context, the least-squares estimator has been analyzed
to show the regimes where this scheme is optimal with respect to the
CR bound. On the other hand, \citet{Bouquillon2016} have applied the CR
bound to moving sources and compared the CR predictions to
observations of the Gaia satellite with the VLT Survey Telescope at
ESO, showing very good correspondence between the expectations and the
measured positional uncertainties.

In this work we extend the astrometric analysis mentioned above,
transiting from the classical parametric setting, where the position
is considered fixed but unknown, to the richer 
Bayesian setting, where the position is unknown but
the prior distribution of the object position is available. 
This changes in a fundamental way the nature of the
inference problem from a parametric context in which we are
estimating a constant parameter from a set of random observations
 to a random setting in which we estimate a random object (i.e., the
position is modeled as a random variable) from observations that are
statistically dependent with the position. 

The use of a Bayesian approach to astrometry is not new; we note,  for example, 
that recently \citep{michalik2015gaia} and \citep{michalik2015quasars} have proposed
analyzing part of the data from the Gaia astrometric satellite using a
suitable chosen set of priors in a Bayesian context. In
\citet{michalik2015gaia}, the feasibility of the Bayesian approach,
especially for the analysis of stars with poor observation histories,
is demonstrated through global astrometric solutions of simulated Gaia
observations.  In this case, the prior information is derived 
from reasonable assumptions regarding the distributions of
proper motions and parallaxes. On the other hand, in
\citep{michalik2015quasars} they show how the prior information regarding QSO
proper motions could provide an independent verification of the
parallax zero-point for the early reduction of Gaia data in the
context of the HTPM project, which uses Hipparcos stars as first epoch
(\citet{mignard2009hundred}; \citet{michalik2014joint}), or the TGAS project, which uses
the Tycho-2 stars as first epoch \citep{michalik2015tycho}.

An important new element of the Bayesian setting is the introduction of
a prior distribution of the object position. Therefore, the conceptual
question to address is to quantify and analyze the increase in
astrometric performance from the use of a prior distribution of the
object position, information that is not available (or used) in the classical parametric setting. To
the best of our knowledge, this direction has not been addressed
systematically by the community and remains an interesting open problem. In this work, we tackle this problem from a theoretical and numerical point of view, and  we provide some concrete practical implications.

On the theoretical side, we consider the counterpart of the CR bound
in the Bayesian context, which is known as the {\em Van Trees' inequality}, or
Bayesian CR bound \citep{van2004detection}. As was the case in the
parametric scenario, the Bayesian CR bound offers a tight performance
bound, more precisely a lower bound for the minimum mean square error (MSE)
of the best estimator for the position in the Bayes setting. The first
part of this work is devoted to formalize the Bayesian problem of
astrometry and to develop closed-form expressions for the Bayesian CR as
well as an expression to estimate the gain in performance. As in the
parametric case, different observational regimes are evaluated to
quantify the gain induced from the prior distribution of the object
location. On this, we introduce formal definitions for the prior
information and the information attributed to the observations, and
with these concepts the notion of when the information of the prior
distribution is relevant or irrelevant with respect to the information
provided by the observations. This last distinction determines when
there is a significant gain in astrometric precision from the prior
distribution. A powerful corollary of this analysis is that the
Bayes setting always offers a better performance than the parametric
setting, even in the worse-case
prior (i.e., that of a uniform distribution).

On the practical side, we consider some realistic experimental
conditions to evaluate numerically the gain of the Bayes setting with
respect to the parametric scenario. Remarkably, it is shown that the
gain in performance is significant for various observational regimes in astrometry,
which is particularly clear in the case of faint objects, or when the
observations are acquired under poor conditions (i.e., in the low
S/N regime). An alternative concrete way to illustrate this gain in
performance is elaborated in Sect.~\ref{sub_sec_equi_ob_brightness},
where we introduce the concept of the equivalent object brightness. On
the estimation side, we submit evidence that the minimum
mean square error (MMSE) estimator of this problem (the well-known
conditional mean estimator) tightly achieves the Bayesian CR lower
bound. This is a remarkable result, demonstrating that all the
performance gains presented in the theoretical analysis part of our
paper can indeed be achieved with the MMSE estimator, which in principle 
has  a practical implementation \citep{weinstein1988general}. We
end our paper with a simple example of what could be achieved
using the Bayesian approach in terms of the astrometric precision 
when new observations of varying quality are used and 
we incorporate as prior information data from the USNO-B all-sky
catalog.

\section{Preliminaries}
\label{sec_pre}

In this section we introduce the problem of astrometry as well as the
concepts and definitions that will be used throughout the paper. For
simplicity, we focus on the 1-D scenario of a linear array detector,
as it captures the key conceptual elements of the
problem\footnote{This analysis can be generalized to the 2-D case as
  shown in \citet{mendez2013analysis}.}.

\subsection{Astrometry} 
\label{sub_sec_astro_photo}

The problem we want to tackle is the inference of the position
of a point source with respect to the (known) relative
location of the picture elements of a detector array. This source is
parameterized by three scalar quantities, the   angular position in the sky (measured in seconds of arc [arcsec])  of the object
$x_c \in \mathbb{R}$ in the array,  
its intensity (or brightness, or flux) that we denote by $\tilde{F} \in
\mathbb{R}^+$, and a generic parameter that determines the width (or
spread) of the light distribution on the detector, denoted by $\sigma
\in \mathbb{R}^+$. These parameters induce a   probability over an observation space that we denote by $\mathbb{X}$. More precisely, given a point source represented by the triad
$(x_c,\tilde{F}, \sigma)$, it creates a nominal intensity profile in a
photon integrating device (PID), typically a CCD, which can be
generally written as:

\begin{equation}\label{eq_pre_1}
	\tilde{F}_{x_c, \tilde{F}}(x)=\tilde{F} \cdot
\phi(x-x_c,\sigma),
\end{equation}
where $\phi(x-x_c,\sigma)$ denotes the 1-D normalized
point spread function (PSF). In what follows, $\tilde{F}$ and $\sigma$
are assumed to be known and fixed, and consequently they are not
part of the inference problem addressed here.

In practice, the PSF described by Eq.~(\ref{eq_pre_1}) cannot be
observed with infinite precision, mainly because of three
disturbances (sources of uncertainty) that affect all measurements.
The first is an additive background noise, which captures the photon
emissions of the open (diffuse) sky and the noise of the instrument
itself (the read-out noise and dark-current,
\citet{howell2006handbook,tyson1986low}) modeled by $\tilde{B}_k$ in
Eq.~(\ref{eq_pre_2b}). The second is an intrinsic uncertainty between the
aggregated intensity (the nominal object brightness plus the
background) and the actual detection and measurement process by the
PID,  which is modeled by independent random variables that follow a Poisson
probability law. The third is the spatial quantization process associated
with the pixel-resolution of the PID as specified in
Eqs.~(\ref{eq_pre_2b}) and~(\ref{eq_pre_3}).  Including these three
effects, we have a countable collection of random variables (fluxes or
counts measurable by the PID) $\{I_k: k \in \mathbb{Z}\}$, where the
$I_k \sim Poisson(\lambda_k(x_c, \tilde{F}))$ are driven by the
expected intensity at each pixel element $k$. The underlying
(expected) pixel intensity is given by
\begin{equation}\label{eq_pre_2b}
	\lambda_k(x_c, \tilde{F}) \equiv\mathbb{E}\{I_k\}=
        \underbrace{\tilde{F} \cdot g_k(x_c)}_{\equiv
          \tilde{F}_k(x_c,\tilde{F})} + \tilde{B}_k,~\forall k\in
        \mathbb{Z}
\end{equation}
and 
\begin{equation}\label{eq_pre_3}
	g_k(x_c) \equiv \int^{x_k+\Delta x/2}_{x_k-\Delta x/2} \phi(x-
        x_c,\sigma) \, dx, \ \forall k \in \mathbb{Z},
\end{equation}
where $\mathbb{E}$ is the expectation value of the argument, while
$\left\{x_k: k \in \mathbb{Z}\right\}$ denotes the standard uniform
quantization of the real line-array with pixel resolution $x_{k+1}-x_k
= \Delta x>0$ for all $k \in \mathbb{Z}$. In practice, the PID has a
finite collection of $n$ measuring elements (or pixels), then a basic
assumption here is that we have a good coverage of the object of
interest, in the sense that for a given position of the source $x_c$,
it follows that
\begin{equation}\label{eq_pre_4}
	\sum_{k=1}^n g_k(x_c) \approx \sum_{k\in \mathbb{Z}} g_k(x_c)
        =\int_{-\infty}^{\infty} \phi(x-x_c,\sigma) \, dx = 1.
\end{equation}
 In Eq.~(\ref{eq_pre_3}), we have assumed the idealized
  situation that every pixel has the exact response function
  (equal to unity), or, equivalently, that the flat-field process has
  been achieved with minimal uncertainty.  This equation also assumes that the
  intra-pixel response is uniform. This is more important in the
  severely undersampled regime (see, e.g.,
  \citet[Fig.~1]{adorf1996limits}), which is not explored in this
  paper. However,  a relevant aspect of data calibration is achieving a
  proper flat-fielding, which can affect the correctness of our
  analysis and the form of the adopted likelihood function (more details below).

Finally, given the source parameters $(x_c, \tilde{F})$, the joint
{probability mass function} $\mathbb{P}$ (hereafter pmf)
of the observation vector $I^n=(i_1,...,i_n)$ (with values in $\mathbb{N}^n$) is given
by 
\begin{equation}\label{eq_pre_5}
\underbrace{\mathbb{P}(I^n=i^n=(i_1,...,i_n))}_{\equiv p_{x_c}(i^n)}
= p_{\lambda_1(x_c,\tilde{F})}(i_1) \cdot
p_{\lambda_2(x_c,\tilde{F})}(i_2) \cdots
p_{\lambda_n(x_c,\tilde{F})}(i_n),
\end{equation}
$\forall (i_1,...,i_n) \in \mathbb{N}^n,$ 
where $p_{\lambda}(x)=\frac{e^{-\lambda}\cdot \lambda^x}{x!}$ denotes
the pmf of the Poisson law \citep{gray2004introduction}\footnote{Throughout this
  paper, in general,  capital letters (e.g., $I^n$) denote
  a random variable (or, in this case, a random vector of $n$
  elements), and lower-case letters (e.g., $i^n$) denote a particular
  realization (or measured value) of the variable. This distinction
  will become particularly important in the Bayesian context described
  in the next section.}. The adoption of this probabilistic model
 is common in contemporary astrometry (e.g., in Gaia, see
\citet{Lindegren2008}).

  It is important to mention that Eq.~(\ref{eq_pre_5})
  assumes that the observations 
  are independent (although not identically
  distributed since they follow $\lambda_i$). This is only
  an approximation to the real situation since it implies
 that we are neglecting any electronic defects or
  features in the device,  such as  the cross-talk present in
  multi-port CCDs \citep{freyetal01}, or read-out correlations, such as
  the odd-even column effect in IR detectors \citep{mason07}, as well
  as calibration or data reduction deficiencies (e.g., due to
  inadequate flat-fielding; \citet{gawiet06}) that may alter this
  idealized detection process.
   In essence, we are considering an ideal detector that would satisfy the proposed
  likelihood function given by Eq.~(\ref{eq_pre_5}); in real
  detectors the likelihood function could be considerably more
  complex\footnote{In a real scenario, we may not even be
    able to write such a function owing to our imperfect
   characterization or limited knowledge of the detector device.}.
 Serious attempts have been made by manufacturers and observatories to minimize the impact of these
  defects, either by an appropriate electronic design or by adjusting
  the detector operational regimes (e.g., cross-talk can be reduced to
  less than 1 part in $10^4$ by adjusting the read-out speed and by a
  proper reduction process (see \citet{freyetal01}).

\subsection{Astrometric \crra\ lower bound} \label{subsec_cr_bounds}
\label{subsec:cr_bound}
The CR inequality offers a lower bound for the variance of the family of unbiased estimators.
More precisely, the CR theorem is as follows:

{\bf Theorem 1.}
(\citet{radhakrishna1945information,cramer1946contribution})
{
Let $\{I_k:k=1,...,n\}$ be a collection of independent observations that follow a
parametric pmf $p_{\theta^m}$ defined on $\mathbb{N}$. The parameters
to be estimated from $I^n=(I_1,...,I_n)$ will be denoted in general by
the vector $\theta^m=(\theta_1,\theta_2,...,\theta_m) \in \Theta =
\mathbb{R}^m$.  Let
 $$L(i^n;\theta^m) \equiv p_{\theta^m}(i_1)\cdot p_{\theta^m}(i_2)
\cdots p_{\theta^m}(i_n)$$
be the likelihood of the observation $i^n \in \mathbb{N}^n$ given
$\theta^m \in \Theta$.  
If the following condition is satisfied
\begin{equation} \label{cond2d}
\mathbb{E}_{I^n \sim p^n_{\theta^m}}\left\lbrace \frac{\partial
  \ln L(I^n; \theta^m) }{\partial \theta_j} \right\rbrace = 0,
\;\; \forall j \in \left\{1,\ldots, m \right\},
\end{equation}
then for any $\tau_n(\cdot):\mathbb{N}^n \rightarrow \Theta$ 
unbiased estimator of $\theta^m$ (i.e., $\mathbb{E}_{I^n \sim p^n_{\theta^m}} \left\lbrace
  \tau_n(I^n)\right\rbrace =\theta^m$) it follows that
\begin{equation}\label{varcr}
Var (\tau_n(I^n)_j) \geq [ \mathcal{I}_{\theta^m}(n)^{-1} ]_{j,j},
\end{equation}
where $\mathcal{I}_{\theta^m}(n)$ is the {\em Fisher information} matrix given by
\begin{equation}\label{fisher}
[ \mathcal{I}_{\theta^m}(n)]_{j,l} = \mathbb{E}_{I^n \sim
  p^n_{\theta^m}} \left\lbrace \frac{\partial \ln L(I^n;
  \theta^m)}{\partial \theta_j} \cdot \frac{\partial \ln L(I^n;
  \theta^m) }{\partial \theta_l} \right\rbrace,
\end{equation}
}
$\forall j,l \in \left\{1,\ldots, m \right\}$.

Returning to the observational problem in Sect.~\ref{sub_sec_astro_photo}, \citet{mendez2013analysis,mendez2014analysis} have characterized and analyzed the CR lower bound in  Eq.~(\ref{varcr}) for the isolated problem of astrometry and photometry, respectively, as well as the joint
problem of photometry and astrometry.  For completeness, we highlight their 
1-D astrometric result, which will be relevant for the discussion in subsequent sections of the paper:

{\bf Proposition 1.}
{
(\citet[pp. 800]{mendez2014analysis}) 
	{\label{pro_FI_photometry_astrometry} 
	Let us assume that $\tilde{F}$ and $\sigma \in \mathbb{R}^+$
        are fixed and known, and we want to estimate $x_c$ (fixed but
        unknown) from a realization of $I^n \sim p_{x_c} $ in
        Eq.~(\ref{eq_pre_5}), then the Fisher information in
         Eq.~(\ref{fisher}) is given by
	\begin{equation}\label{fi_astrometry}
		\mathcal{I}_{x_c}(n) = \sum_{k=1}^n \frac{ \left(
                  \tilde{F}\frac{d g_k(x_c)}{d x_c} \right)^2
                }{\tilde{F} g_k(x_c) + \tilde{B}_k},
	\end{equation}
	which from Eq.~(\ref{varcr}) induces a minimum variance
        bound for the {astrometric estimation problem}. More precisely
	\begin{equation}\label{fi_astrometryb}
          	\min_{\tau^n: \mathbb{N}^n \rightarrow \mathbb{R}} Var (\tau_n(I^n)) \geq \underbrace{\mathcal{I}_{x_c}(n)^{-1}}_{ \equiv \sigma_{CR}^2}.
        \end{equation}}
}
        The expression $\sigma_{CR}^2$ in Eq.~(\ref{fi_astrometryb}) is a shorthand for the {\em 1-D astrometric CR lower bound} in this parametric (classical) approach, as opposed to the Bayesian CR bound, which will be introduced in the next section.
      
\section{The Bayesian estimation approach in astrometry}
\label{sec_bayes}

We now consider a Bayesian setting \citep{moon2000mathematical} for the
problem of estimating the object location $x_c$ from a set of
observations $i^n \in \mathbb{N}^n$. The Bayes
scenario considers that the (hidden) position is a random variable
$X_c$ (i.e., a random parameter), as opposed to a fixed although
unknown parameter considered in the classical setting in Sect.~\ref{subsec:cr_bound}. 
The goal is to estimate $X_c$ from a realization of the observation random
vector $I^n$. In order for this inference
to be nontrivial, $X_c$ and $I^n$ should be statistically
dependent. In our case this dependency is modeled by the observation
equation in (\ref{eq_pre_5}).  More precisely, we have that the
conditional probability of $I^n$ given $X_c$ is given by
\begin{equation}\label{eq_sec_bayes_1}
	\mathbb{P}(I^n=i^n | X_c=x_c)= p_{x_c}(i^n)= \underbrace{\Pi^n_{k=1}
        p_{\lambda_k(x_c)}(i_k)}_{\equiv L(i^n;x_c)}
 \end{equation}
where $L(i^n;x_c)$ denotes the likelihood of observing $i^n$ given that
the object position is $x_c$ while $p_{\lambda_k(x_c)}(i_k)$ has been defined in 
Eq.~(\ref{eq_pre_5}).

The other fundamental element of the Bayesian approach is the ``{\em prior distribution}'' of $X_c$, given by a {\em
  probability density function} (pdf) $\psi(\cdot)$, i.e., for all
$B\subset \mathbb{R}$,
\begin{equation}\label{eq_sec_bayes_2}
	\mathbb{P}(X_c\in B)= \int_{B} \psi(x) \, dx.
 \end{equation}
Consequently, in the Bayes setting we know that, for all $A\subset
\mathbb{N}^n$ and $B\subset \mathbb{R}$, the joint distribution of the
random vector $(X_c,I^n)$ is given by
\begin{equation}\label{eq_sec_bayes_3}
	\mathbb{P}((X_c,I^n)\in B\times A)= \int_{B} \sum_{i^n\in A}
        p_{x}(i^n) \cdot \psi(x) \, dx.
\end{equation}
It is important to highlight the role of the prior distribution
of the object position $X_c$ because it is the key mathematical
object that allows us to pose the astrometry problem in the context of
Bayesian estimation.

For the estimation of the object location, we need to establish the
decision function $\tau^n(\cdot):\mathbb{N}^n \rightarrow \mathbb{R}$
that minimizes the MSE in inferring $X_c$ from a
realization of $I^n$.  More precisely, the optimal decision would be the solution of
the following problem:
\begin{equation}\label{eq_sec_bayes_4}
	\min_{\tau^n:\mathbb{N}^n \rightarrow \mathbb{R}}
        \mathbb{E}_{(X_c,I^n)} \left\{ \left( \tau^n(I^n) - X_c
        \right)^2 \right\}.
\end{equation}
The expectation value in Eq.~(\ref{eq_sec_bayes_4}) is taken with respect to
the joint distribution of both variables $(X_c, I^n)$
(see Eq.~(\ref{eq_sec_bayes_3})), and the minimum is taken over all
possible mappings (decision rules) from $\mathbb{N}^n$ to
$\mathbb{R}$.  On the right-hand-side (hereafter RHS) of
Eq.~(\ref{eq_sec_bayes_4}), $\tau^n(I^n)$ is the estimation of $X_c$
from $I^n$ throughout the decision rule $\tau^n$, also known as the
estimator \citep{lehmann1998theory}. The optimal MSE estimator, which is the
solution of Eq.~(\ref{eq_sec_bayes_4}), is known as the Bayes rule (or
estimator), which for the square error risk function has a known
theoretical solution function of the posterior density
$\mathbb{P}(X_c=x_c | I^n=i^n)$ \citep[chap. 8]{kay2010fundamentals}. More details are
presented in Sect.~\ref{sec_opt_estimator} (see in particular
Eq.~(\ref{eq_sec_opt_estimator_1})).

\subsection{Bayes \crra\ lower bound}\label{sub_sec:bayes_CR}
As was the case in the parametric setting presented in
Sect.~\ref{subsec_cr_bounds}, in the Bayes scenario it is also
possible and meaningful to establish bounds for the minimum MSE (MMSE
hereafter) in Eq.~(\ref{eq_sec_bayes_4}). This powerful result is
 known as the {\em van Trees inequality} or the Bayesian CR (BCR) lower
bound:
{\bf Theorem 2.}
{ 
 	\label{th_bayes_cr} \citep[Sec. ~2.4]{van2004detection}
	For any possible decision rule $\tau^n:\mathbb{N}^n
        \rightarrow \mathbb{R}$, it is true that
	\begin{equation}\label{eq_sec_bayes_5}
		\mathbb{E}_{(X_c,I^n)} \left\{ \left( \tau^n(I^n)-X_c
                \right)^2 \right\} \geq \left[ \mathbb{E}_{(I^n,X_c)} \left\{
                \left( \frac{d \ln \tilde{L}(X_c,I^n)}{d x} \right)^2
                \right\} \right]^{-1},
	\end{equation}
	where 
\begin{equation} \label{bayes_like}
\tilde{L}(x_c,i^n) \equiv p_{x_c}(i^n) \cdot \psi(x_c)=L(i^n;x_c)
\cdot \psi(x_c)
\end{equation}
is  shorthand for the joint density of $(X_c,I^n)$ (see
Eq.~(\ref{eq_sec_bayes_3})), and where $L(i^n;x_c)$ is the
likelihood of observing $i^n$ given that the object position is $x_c$,
for all $x_c\in \mathbb{R}$ and $i^n\in \mathbb{N}^n$ (see
Eq.~(\ref{eq_sec_bayes_1})).
}

This result turns out to be the natural extension of the parametric CR
lower bound to the Bayes setting (see Sect.~\ref{subsec:cr_bound}). 
We note in particular the similarities between
Eq.~(\ref{eq_sec_bayes_5}) and the expression in
Eq.~(\ref{varcr}). In the Bayes setting this result offers a lower
bound for the MSE of any estimator and, consequently, a lower bound
for the MMSE, i.e.,
\begin{equation}\label{eq_sec_bayes_5b}
	\min_{\tau^n: \mathbb{N}^n \rightarrow \mathbb{R}}
        \mathbb{E}_{(X_c,I^n)} \left\{ \left( \tau^n(I^n)-X_c
        \right)^2 \right\} \geq \left[ \mathbb{E}_{(X_c,I^n)} \left\{ \left(
        \frac{d \ln \tilde{L}(X_c,I^n)}{d x} \right)^2 \right\} \right]^{-1}.
\end{equation}

{\bf Proposition 2.}
{ 
If we analyze what we call {\em the Bayes-Fisher information} (BFI)
term (a function that depends on $F$ and $\psi$) on the
RHS of Eq.~(\ref{eq_sec_bayes_5b}), we can establish that
\begin{align}\label{eq_sec_bayes_6}
	\underbrace{\mathbb{E}_{(X_c,I^n)} \left\{ \left( \frac{d \ln
            \tilde{L}(X_c, I^n)}{d x} \right)^2 \right\}}_{\text{Bayes
            Fisher Information} = BFI(F,\psi)} &= \mathbb{E}_{X_c \sim \psi} \left\{ \mathcal{I}_{X_c}(n) \right\} \nonumber\\
           &+\underbrace{\mathbb{E}_{X_c \sim \psi} \left\{ \left( \frac{d \ln
               \psi(X_c)}{dx} \right)^2\right\}}_{\equiv \mathcal{I}(\psi)}.
\end{align}
}
(see Appendix \ref{app_derivation_bfi_identity} for the derivation of Eq.~(\ref{eq_sec_bayes_6}))

From Eq.~(\ref{eq_sec_bayes_6}), we conclude that the $BFI$ corresponds to the {\it average} Fisher information of the parametric setting (average with respect to $\psi$), 
plus a non-negative term associated with the prior distribution $\psi$ that we call the ``{\em prior information}'' and denote $\mathcal{I}(\psi)$. 
Finally, from Eq.~(\ref{eq_sec_bayes_5}) we have that
\begin{align}\label{eq_sec_bayes_7}
		\min_{\tau^n: \mathbb{N}^n \rightarrow \mathbb{R}}
                \mathbb{E}_{(X_c,I^n)} \left\{ \left( \tau^n(I^n)-X_c
                \right)^2 \right\} \geq \underbrace{\frac{1}{\mathbb{E}_{X_c\sim
                    \psi} \left\{ \mathcal{I}_{X_c}(n) \right\} +
                  \mathcal{I}(\psi)
		}}_{\equiv \sigma^2_{BCR}}, 
\end{align}
which will be called the Bayesian CR (BCR) lower bound (compare to Eq.~(\ref{fi_astrometryb})).

\subsection{Analysis and interpretation of the Bayes-Fisher information: $BFI(F,\psi)$}
\label{sub_sec_bayes_fisher}
At this point it is interesting to analyze the
$BFI$ in Eq.~(\ref{eq_sec_bayes_6}), which reduces to the
sum of two non-negative information components:
$$BFI(F,\psi) =\mathbb{E}_{X_c\sim \psi} \left\{ \mathcal{I}_{X_c}(n)
\right\} + \mathcal{I}(\psi).$$ 

The first term,  $\mathbb{E}_{X_c\sim \psi} \left\{ \mathcal{I}_{X_c}(n) \right\}$, can
be interpreted as the {\it average information} provided by the
observations $I^n$ to discriminate the random position $X_c$, which
is precisely the {\it average} Fisher information of the parametric
setting. We note, however, an important distinction: In the
  parametric setting, the CR lower bound depends directly on $x_c$
  (see Eqs.~(\ref{fi_astrometry}) and~(\ref{fi_astrometryb})), whereas
  in the Bayesian setting the BCR  does not depend
   {on a specific value $x_c$, but on a sort of prior average} (over the
  position) function of the spatial sharpness of $\psi(x)$.
 On the other hand, the second term on the RHS of 
Eq.~(\ref{eq_sec_bayes_6}), i.e., $\mathcal{I}(\psi)$, can be interpreted
as the information provided exclusively by the prior distribution
$\psi$ to decide $X_c$. This distinction is very important because
it allows us to evaluate regimes where the prior information of the
location for the source is relevant (or irrelevant), relative to the
information provided by the observations.
More precisely, we can say that

{\bf Definition 1.}
{The prior information, $\mathcal{I}(\psi)$, is said to be irrelevant relative to the
observations, if $\frac{\mathcal{I}(\psi)}{\mathbb{E}_{X_c\sim \psi}
  \left\{ \mathcal{I}_{X_c}(n) \right\} } \approx 0$.  Otherwise, it
is said to be relevant.}

{\bf Definition 2.}
{The information of the observations, $\mathbb{E}_{X_c\sim \psi} \left\{ \mathcal{I}_{X_c}(n) \right\}$, 
is said to be irrelevant relative to the prior distribution $\psi$, if $\frac{\mathbb{E}_{X_c\sim \psi}
  \left\{ \mathcal{I}_{X_c}(n) \right\} } {\mathcal{I}(\psi)} \approx
0$. Otherwise, it is said to be relevant.}

\section{Studying the gain in astrometric precision from the use of prior information}
\label{sec_com_cr}
In this section we evaluate how much gain in performance can be
obtained when we add prior information about the object position in
the Bayes setting, with respect to the baseline parametric case where 
it is not possible to account for that information. It is clear
that the knowledge of the distribution of $X_c$ should provide a gain
in the performance with respect to the parametric setting, where that
information is either not available or not used.  For this analysis,
the performance bounds given by the CR bound
in Eq.~(\ref{fi_astrometryb}) and the BCR bound
in Eq.~(\ref{eq_sec_bayes_7}) are compared.

We define the gain in performance, $Gain(\psi)$, attributed to the prior distribution 
$\psi$, as the improvement in astrometric precision of the best estimator 
of the Bayes setting with respect to the best estimator of the parametric setting. 
In formal terms, the gain is given by the reduction in MSE (from the parametric to the Bayes setting) 
in the process of estimating $X_c$ from $I^n$, i.e.,  
\begin{align}\label{eq_sec_com_cr_0}
Gain(\psi) \equiv \min_{\tau_{un}^n\mathbb{N}^n \rightarrow
  \mathbb{R} \text{ and $\tau_{un}^n$ is unbiased}}
\mathbb{E} \left\{ \left( \tau_{unbias}^n(I^n)-x
\right)^2 \right\} \nonumber\\
- \min_{\tau^n: \mathbb{N}^n \rightarrow
  \mathbb{R}} \mathbb{E} \left\{ \left( \tau^n(I^n)-X_c
\right)^2 \right\}, 
\end{align}  
where the first term on the RHS of Eq.~(\ref{eq_sec_com_cr_0}) represents the minimum MSE over the family of unbiased estimators of the position (i.e., estimators that do not have access to $\psi$), while the  second term on the RHS of Eq.~(\ref{eq_sec_com_cr_0}) is the  MMSE 
estimator of the Bayes setting, i.e. the best estimator over the family of mappings that have access to $\psi$.   The next result establishes a tight  lower bound for this gain, as a function of the information measures presented in Sect. \ref{sub_sec:bayes_CR}.

{\bf Proposition 3.}
{It follows that
\begin{align}\label{eq_pro_information_gain}
	Gain(\psi) &\geq  \underbrace{ \mathbb{E}_{X_c\sim \psi} \left\{
 {\mathcal{I}_{X_c}(n)}^{-1} \right\}}_{\equiv \sigma^2_{MCR}} -  \frac{1}{\mathbb{E}_{X_c\sim \psi} \left\{
                  \mathcal{I}_{X_c}(n) \right\} + \mathcal{I}(\psi)} \nonumber\\ 
                  &= \sigma^2_{MCR} - \sigma^2_{BCR}, 
\end{align}
where $\sigma^2_{MCR}$ on the RHS of Eq.~(\ref{eq_pro_information_gain}) denotes the average classical CR bound in Eq.~(\ref{fi_astrometryb}). (see Appendix \ref{app_pro_information_gain} for the derivation)}

As was expected, this gain 
is an explicit function of $\psi$, and in particular,  it is proportional to the prior information $\mathcal{I}(\psi)= \mathbb{E}_{X_c \sim \psi} \left\{ \left( \frac{d \ln \psi(X_c)}{dx} \right)^2\right\} \geq 0$.
One interesting scenario to analyze is the worse case prior that
happens when $X_c$ follows a uniform distribution over a bounded
(compact) set. In this case, it is simple to show that $
\mathcal{I}(\psi)=\mathbb{E}_{X_c \sim \psi} \left\{ \left( \frac{d
  \ln \psi(X_c)}{dx} \right)^2\right\}=0$ because a flat prior
distribution does not carry any information regarding the location of
the source and, consequently, Eq. (\ref{eq_pro_information_gain}) reduces
to
\begin{equation}\label{eq_sec_com_cr_4}
Gain(\psi) \geq \mathbb{E}_{X_c\sim \psi} \left\{
{\mathcal{I}_{X_c}(n)} ^{-1}\right\}- \mathbb{E}_{X_c\sim \psi}
\left\{ \mathcal{I}_{X_c}(n) \right\}^{-1} \geq 0.
\end{equation}
The non-negativity of $Gain(\psi)$,  explicitly indicated in the last inequality of
Eq.~(\ref{eq_sec_com_cr_4}),  follows from {\em Jensen's inequality}
\citep{perlman1974jensen} and from the fact that $1/x$ is a convex function
of $x$. This is a powerful result in favor of the Bayes approach,
indicating that even for the worse prior, where the random
parameter $X_c$ is uniformly distributed and $\mathcal{I}(\psi)=0$,
the optimal Bayes rule offers a better MSE than the best unbiased
parametric estimator, in other words, that $\sigma^2_{MCR} \ge
\sigma^2_{BCR}$ for any prior. We also note the important fact that if $\mathcal{I}(\psi)>0$
then $Gain(\psi)>0$, from Eq.~(\ref{eq_pro_information_gain}).

In the next section, we present a systematic analysis to evaluate and
compare the classical vs. the Bayesian CR bounds and the gain
$Gain(\psi)$, for various concrete scenarios of astrometric
estimation.

\section{Evaluation of the performance bounds}
\label{sub_sec_bounds_analysis}

In this section we quantify the gain in performance from the prior
information under some realistic astrometric settings. To do this, we
need to be more specific about the observation distribution in
Eq.~(\ref{eq_pre_5}), and the prior distribution in
Eq.~(\ref{eq_sec_bayes_2}).

\subsection{Astrometric observational setting}
\label{sub_sec_setting}
Concerning the observation distribution, we adopt some realistic
design variables and astronomical observing conditions to model the
problem, similar to those adopted in
\citet{mendez2013analysis,mendez2014analysis}. For the PSF (see Eq.~(\ref{eq_pre_1})),
various analytical and semi-empirical forms have been proposed,
for instance the ground-based model in \citet{King1971} and the
space-based model in \citet{bendinelli1987newton}. For this analysis, we
adopt a Gaussian PSF where $\phi(x,\sigma)= \frac{1}{\sqrt{2\pi} \,
  \sigma} e^{- \frac{x^2}{2 \sigma^2}}$ in Eq.~(\ref{eq_pre_3}), and
where $\sigma$ is the width of the PSF, assumed to be known. This PSF
has been found to be a good representation for typical
astrometric-quality ground-based data \citep{mendez2010proper}. In terms of
nomenclature, $FWHM \equiv 2 \sqrt{2 \ln 2} \,\, \sigma$ measured in
arcsec, denotes the {\em full width at half maximum} parameter, which
is an overall indicator of the image quality at the observing site
\citep{chromey2010measure}.

The background profile, represented by $ \left\{\tilde{B}_k:
k=1,...,n\right\}$ (see Eq.~(\ref{eq_pre_2b})), is a function of
several variables, like the wavelength of the observations, the phase of the moon
 (which contributes significantly to the diffuse sky background,
see Sect.~\ref{example}), the quality of the observing site, and the
specifications of the instrument itself.  We consider a uniform
background across the pixels underneath the PSF, i.e.,
$\tilde{B_k}=\tilde{B}$ for all $k$. To characterize the magnitude of
$\tilde{B}$, it is important to first mention that the detector does
not measure photon counts directly, but a discrete variable in {\em
  Analog to Digital Units} (ADUs) of the instrument, which is a
linear proportion of the photon counts \citep{howell2006handbook}.  This
linear proportion is characterized by a gain of the instrument $G$
in units of $e^-/$ADU. Here $G$ is just a scaling value, where we can
define $F \equiv \tilde{F}/G$ and $B \equiv \tilde{B}/G$ as the
intensity of the object and noise, respectively, in the specific ADUs
of the instrument.  Then, the background (in {ADUs}) depends on the
pixel size $\Delta x$~arcsec as \citep{mendez2013analysis}
\begin{equation}\label{eq_sub_sec_bounds_analysis_1}
B=f_s\Delta x+ \frac{D+RON^2}{G}~\mbox{ADU},
\end{equation}
where $f_s$ is the diffuse sky background in ADU/arcsec, while $D$
and $RON$, both measured in $e^{-}$, model the {\em dark-current} and
{\em read-out noise of the detector} on each pixel, respectively. We note
that the first component on the RHS of
Eq.~(\ref{eq_sub_sec_bounds_analysis_1}) is attributed to the site,
and its effect is proportional to the pixel size.  On the other hand,
the second component is attributed to errors of the PID, and it is
pixel-size independent. This distinction is important when analyzing the
performance as a function of the pixel resolution of the array (see
details in \citet[Sec. 4]{mendez2013analysis}). More important is the fact
that in typical ground-based astronomical observation, long exposure
times are considered, which implies that the background is dominated
by diffuse light coming from the sky (the first term on the RHS of
expression (\ref{eq_sub_sec_bounds_analysis_1})), and not from the
detector \citep[Sec. 4]{mendez2013analysis}.  In this context,
\citet{mendez2013analysis} have shown that
\begin{equation}\label{eq_sub_sec_bounds_analysis_2}
\mathcal{I}_{x_c}(n) = \frac{\tilde{F}^2 }{2 \pi \sigma^2 \tilde{B}}
\cdot \sum_{k=1}^{n} \frac{\left( e^{-\gamma(x^-_k-x_c)} -
  e^{-\gamma(x^+_k-x_c)} \right) ^2}{\left( 1 + \frac{1}{\sqrt{2 \pi}
    \, \sigma}\frac{\tilde{F}}{\tilde{B}} \cdot
  \int_{x^{\_}_k}^{x^+_k} e^{-\gamma(x-x_c)} \, dx \right)},
\end{equation}
where $\gamma(x) \equiv \frac{1}{2}(\frac{x}{\sigma})^2$, with
$x^{\_}_k = x_k - \frac{\Delta x}{2}$ and $x^+_k= x_k + \frac{\Delta
  x}{2}$.  Furthermore, in the so-called high-resolution regime, i.e.,
when $\Delta x/\sigma \ll 1$, the following limiting (faint and bright
source) closed-form expression for $\mathcal{I}_{x_c}(n)$ can be
derived (see details in \citet[Sec. 4.1.]{mendez2013analysis}):
\begin{equation} \label{eq_sub_sec_bounds_analysis_2b}
\mathcal{I}_{x_c}(n)^{-1} \approx \left\{
			\begin{array}{cc}
			\frac{\sqrt{\pi}}{2 \, (2 \, \ln 2)^{3/2}}
                        \cdot \frac{\tilde{B}}{\tilde{F}^2} \cdot
                        \frac{FWHM^3}{\Delta x} & \mbox{if $\tilde{F}
                          \ll \tilde{B}$} \\
			\frac{1}{8 \, \ln 2} \cdot \frac{1}{\tilde{F}}
                        \cdot FWHM^2 & \mbox{if $\tilde{F} \gg
                          \tilde{B}$}.
			\end{array} \right.
\end{equation}

Concerning the prior distribution, we consider a Gaussian distribution
with mean $\mu$ and variance $\sigma_{priori}>0$, therefore
$\psi(x)=\frac{1}{\sqrt{2\pi} \, \sigma_{priori}}
  e^{- \frac{(x-\mu)^2}{\; \; \; \; \; 2\sigma^2_{priori}}}$, for which,
\begin{equation}\label{eq_sub_sec_bounds_analysis_3}
\mathcal{I}(\psi) = \mathbb{E}_{X_c \sim \psi} \left\{ \left( \frac{d
  \ln \psi(X_c)}{dx} \right)^2\right\}=\frac{1}{\sigma^2_{priori}}.
\end{equation}
Then, under these assumptions, the expression for the BCR lower bound
for the astrometric problem in Eq.~(\ref{eq_sec_bayes_7}) becomes
\begin{align}\label{eq_sub_sec_bounds_analysis_4}
&(\sigma^2_{BCR})^{-1}= \mathbb{E}_{(X_c,I^n)} \left\{ \left( \frac{d \ln
          \tilde{L}(X_c, I^n)}{d x} \right)^2 \right\} \nonumber\\ 
	&=  \frac{\tilde{F}^2 }{2 \pi \sigma^2 \tilde{B}}
        \mathbb{E}_{X_c \sim \mathcal{N}(\mu,\sigma_{priori})} \left\{
        \sum_{k=1}^{n} \frac{\left( e^{-\gamma(x^-_k-x_c)} -
          e^{-\gamma(x^+_k-x_c)} \right) ^2}{\left( 1 +
          \frac{1}{\sqrt{2 \pi} \, \sigma}\frac{\tilde{F}}{\tilde{B}}
          \cdot \int_{x^{\_}_k}^{x^+_k} e^{-\gamma(x-x_c)} \, dx
          \right)} \right\} \nonumber\\
          &+ \frac{1}{\sigma^2_{priori}}.
\end{align}
This expression can be evaluated numerically if we know all the
parameters of the problem, this is done in
Sect.~\ref{sub_sec_numerical}.

\subsection{BCR lower bound at the high and low signal-to-noise regimes}
\label{sub_sec_extreme_regimes}
Before evaluating the expression in expression
(\ref{eq_sub_sec_bounds_analysis_4}), we can derive specialized
closed-form expression for two relevant limiting scenarios in
astrometry.  First the scenario of a {\em source dominated regime},
when $\tilde{F} \gg \tilde{B}$ (i.e., the estimation on a very bright
object).  For that we can fix the prior information
$\sigma_{priori}>0$, $\tilde{B}$, $\Delta x$, and $FWHM$, and take the
limit when $\tilde{F} \rightarrow \infty$\footnote{Similarly, we can
  fix $\sigma_{priori}$, $\tilde{F}$, $\Delta x$, and $FWHM$, and take
  the limit when $\tilde{B} \rightarrow 0$.}.  If we do so, it is
simple to verify that the prior information becomes irrelevant
(Definition 3)  relative to the information of
the observations $I^n$, and, consequently, the BCR bound in
Eq.~(\ref{eq_sub_sec_bounds_analysis_4}) reduces to
\begin{align}\label{eq_sub_sec_bounds_analysis_5}
&\sigma^2_{BCR}=\left[ \mathbb{E}_{X_c \sim \mathcal{N}(\mu,\sigma_{priori})} \left\{
\mathcal{I}_{X_c}(n) \right\} \right]^{-1} \nonumber\\
 &= \left[ \frac{\tilde{F}^2 }{2 \pi \sigma^2
  \tilde{B}} \cdot \mathbb{E}_{X_c \sim
  \mathcal{N}(\mu,\sigma_{priori})} \left\{ \sum_{k=1}^{n}
\frac{\left( e^{-\gamma(x^-_k-x_c)} - e^{-\gamma(x^+_k-x_c)} \right)
  ^2}{\left( 1 + \frac{1}{\sqrt{2 \pi} \,
    \sigma}\frac{\tilde{F}}{\tilde{B}} \cdot \int_{x^{\_}_k}^{x^+_k}
  e^{-\gamma(x-x_c)} \, dx \right)} \right\} \right]^{-1}.
\end{align}
This is the case when the prior statistics of $X_c$ does not have any
impact on the performance of the estimation of the object location,
because of the high fidelity (informatory content) of the
observations.

The second scenario is {\em the background dominated regime}, when
$\tilde{F} \ll \tilde{B}$ (i.e., the estimation on a very faint
object). For this analysis we can fix $\sigma_{priori}>0$,
$\tilde{B}$, $\Delta x$, and the $FWHM$, and take the limit when
$\tilde{F} \rightarrow 0$\footnote{Alternatively, we can fix
  $\sigma_{priori}$, $\tilde{F}$, $\Delta x$, and $FWHM$, and take the
  limit when $\tilde{B} \rightarrow \infty$.}. In this case the
information of the observation is irrelevant relative to the prior
information $\mathcal{I}(\psi)$ 
(Definition 4) and, consequently, the BCR bound in
Eq.~(\ref{eq_sub_sec_bounds_analysis_4}) reduces to
\begin{equation}\label{eq_sub_sec_bounds_analysis_6}
\sigma^2_{BCR} = \mathcal{I}(\psi)^{-1} = \sigma^2_{priori}.
\end{equation}
Therefore,  when the observations are irrelevant
(non-informative), the MMSE of the estimation of $X_c$
(Eq.~(\ref{eq_sec_bayes_7})) reduces to the prior variance of
$X_c$. Hence, it is direct to verify that the best MSE estimator (MMSE) in this context is the
prior mean, i.e., 
$$\mathbb{E}_{X_c\sim \mathcal{N}(\mu,\sigma_{priori})} \left\{ X_c
\right\} =\mu,$$ which, as expected, does not depend on the
observations $I^n$. 

We note that, in general, $\mathcal{I}(\psi)^{-1}$ is
an upper bound for the expression in Eq.~(\ref{eq_sec_bayes_7}), which
represents the worse case scenario from the point of view of the
quality of the observations.

\subsection{Numerical evaluation and analysis}
\label{sub_sec_numerical}
For the site conditions, we consider the scenario of a ground-based
station located at a good site with clear atmospheric conditions and
the specifications of current science-grade PIDs, where
$f_s=2\,000$~ADU/arcsec, $D=0$, $RON=5$~e$^-$, $G=2$~e$^-$/ADU and
$FWHM=0.5$~arcsec or $FWHM=1$~arcsec (with these values $B=313$~ADU
for $\Delta x=0.2$~arcsec using
Eq.~(\ref{eq_sub_sec_bounds_analysis_1})). In terms of scenarios of
analysis, we explore different pixel resolutions for the PID array
$\Delta x \in [0.1, 2.0]$ measured in arcsec, and different signal
strengths $F \in \left\{268, 540, 1\,612 \right\}$~ADU\footnote{These
  are the same values explored in \citet[Table 3]{mendez2013analysis}.}.
We note that increasing $F$ implies increasing the signal-to-noise (S/N) of the problem, which can be approximately measured by the
ratio $F/B$. On a given detector plus telescope setting, these
different $S/R$ scenarios can be obtained by changing appropriately
the exposure time (open shutter) that generates the image (for further
details, see Eq.~(\ref{snr})).

\begin{figure*}[h!]
\centering
\subfigure[$FWHM=1$~arcsec]
{\includegraphics[width=0.45\textwidth]{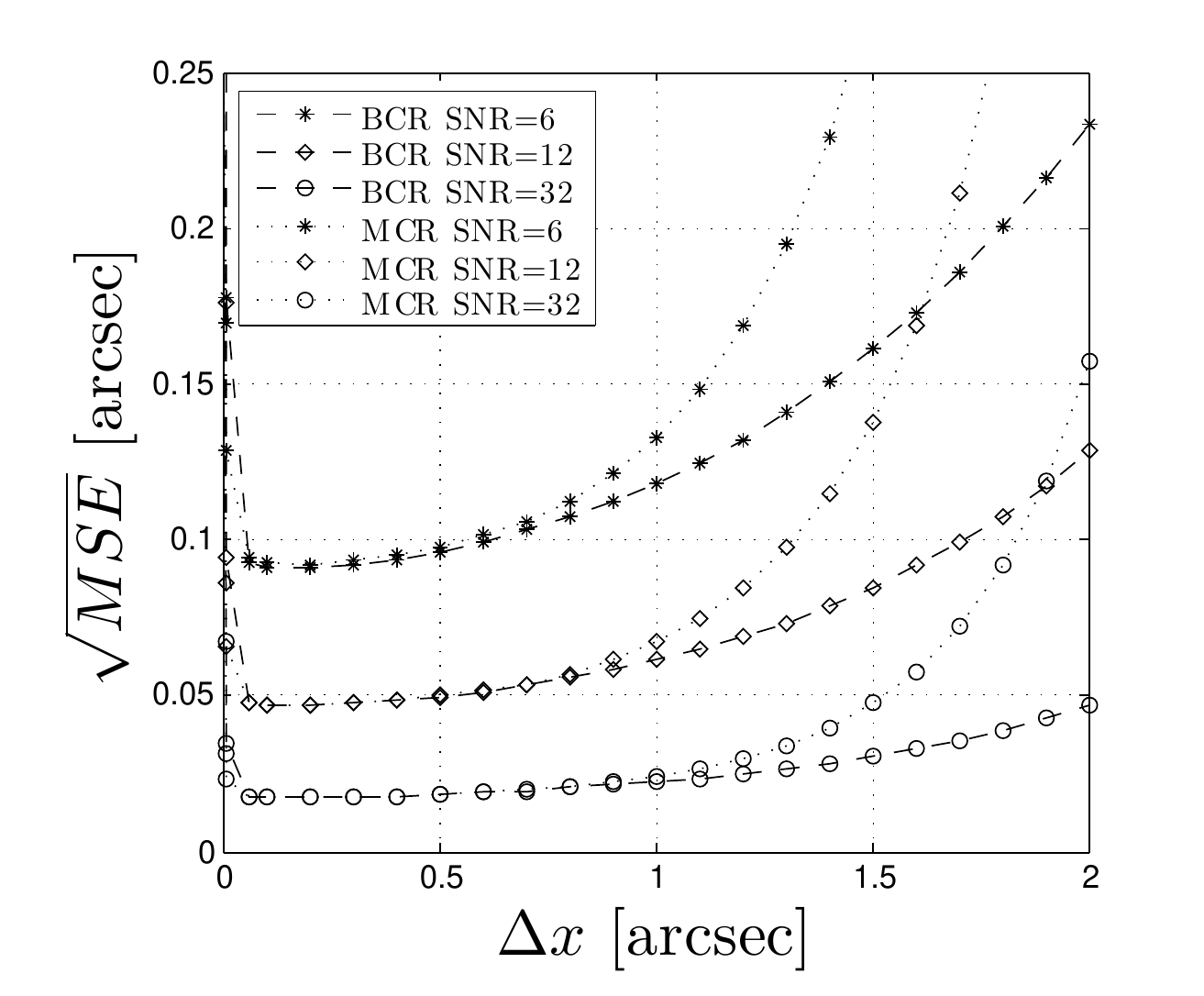}}\hspace{1 cm}
\subfigure[$FWHM=0.5$~arcsec]
{\includegraphics[width=0.45\textwidth]{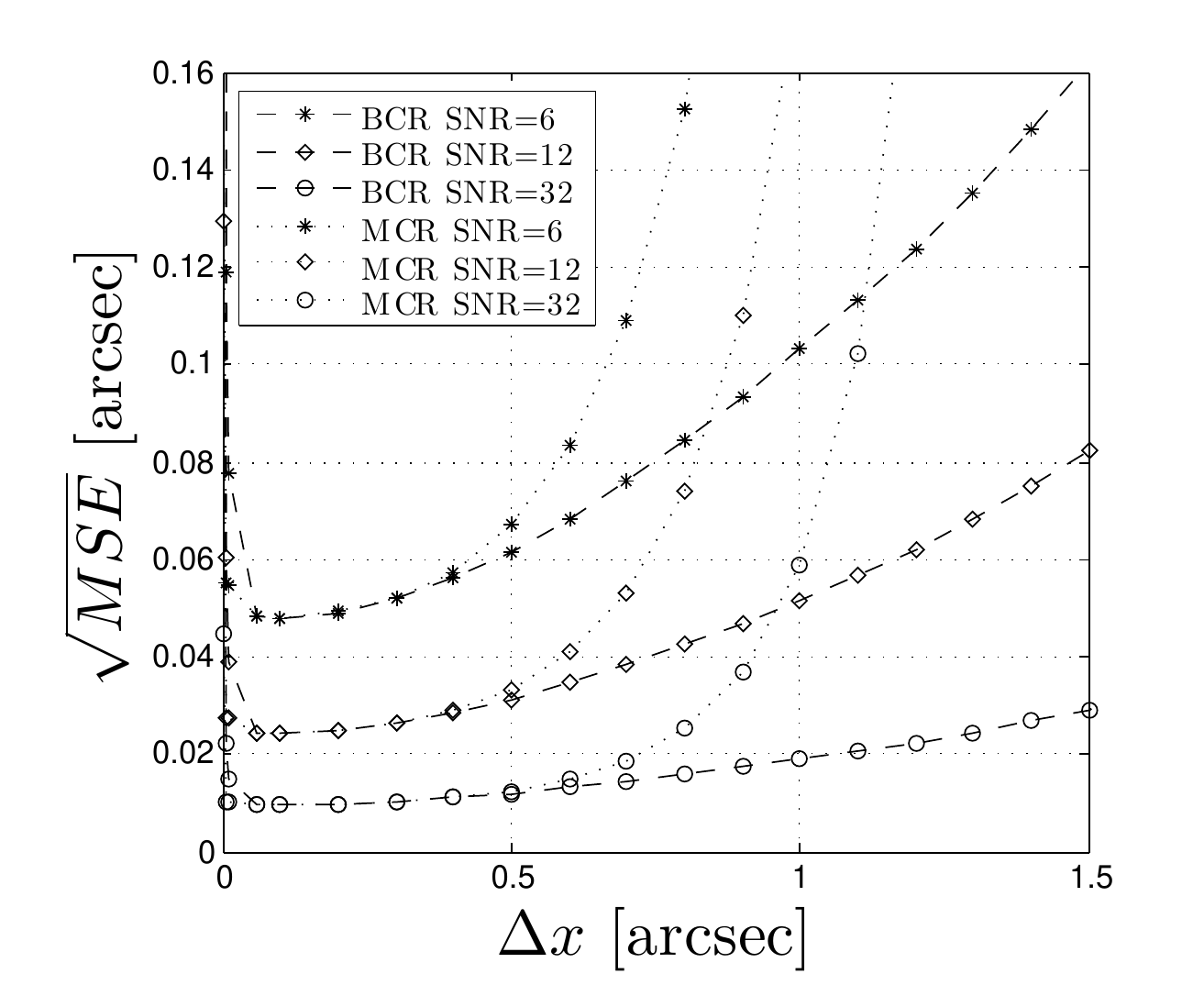}}
\caption{Relationship between the classical MCR (from
  Eqs.~(\ref{eq_sec_com_cr_2})
  and~(\ref{eq_sub_sec_bounds_analysis_2})), and the BCR (from
  Eq.~(\ref{eq_sub_sec_bounds_analysis_4})) lower bounds as a function
  of pixel size for three different $S/N$ regimes and two $FWHM$ for the
  case of $\sigma_{priori}=0.5$~arcsec. As can be seen $\sigma_{BCR}
  \le \sigma_{MCR}$ in all regimes.}
\label{BCR y MCR, sigma 0.1}
\end{figure*}

Figure~\ref{BCR y MCR, sigma 0.1} shows the parametric and Bayes CR
bounds for three $S/R$ regimes as a function of pixel size of the PID,
and for two $FWHM$ scenarios (0.5 and 1.0~arcsec).  We first note, as
the theory predicts, that the BCR bound is below the classical CR
bound in all cases, and that the gap (the performance gain
$Gain(\psi)$ in Eq.~(\ref{eq_pro_information_gain})) increases as a function of increasing the pixel size in
all cases.  In fact the difference between the bounds becomes relevant
for a pixel size larger than $\sim$0.8~arcsec in Fig.~\ref{BCR y
  MCR, sigma 0.1}a and bigger than $\sim$0.6~arcsec in Fig.~\ref{BCR
  y MCR, sigma 0.1}b. These results can be interpreted  as follows:  As
$\Delta x$ increases, the astrometric quality of the observation
deteriorates and the prior information $\mathcal{I}(\psi)$ becomes more and more
relevant in the Bayes context, information that is not available in
the parametric scenario.  This explains the non-decreasing monotonic
behavior of $Gain(\psi)$ as a function of $\Delta x$ for all the
scenarios. If we look at one of the figures, and analyze the gain
$Gain(\psi)$ as a function of the $S/N$ regime, we notice that the pixel
size $\Delta x$ at which the prior information $\mathcal{I}(\psi)$ becomes
relevant, in the sense that $Gain(\psi) > \tau$ for some fixed
threshold $\tau$, increases with the $S/N$.  In other words, for faint
objects the prior information is relevant for a wider range of pixel
sizes, than in the case of bright objects.

\begin{figure*}[h!]
\centering
\subfigure[$FWHM=1$~arcsec]
{\includegraphics[width=0.45\textwidth]{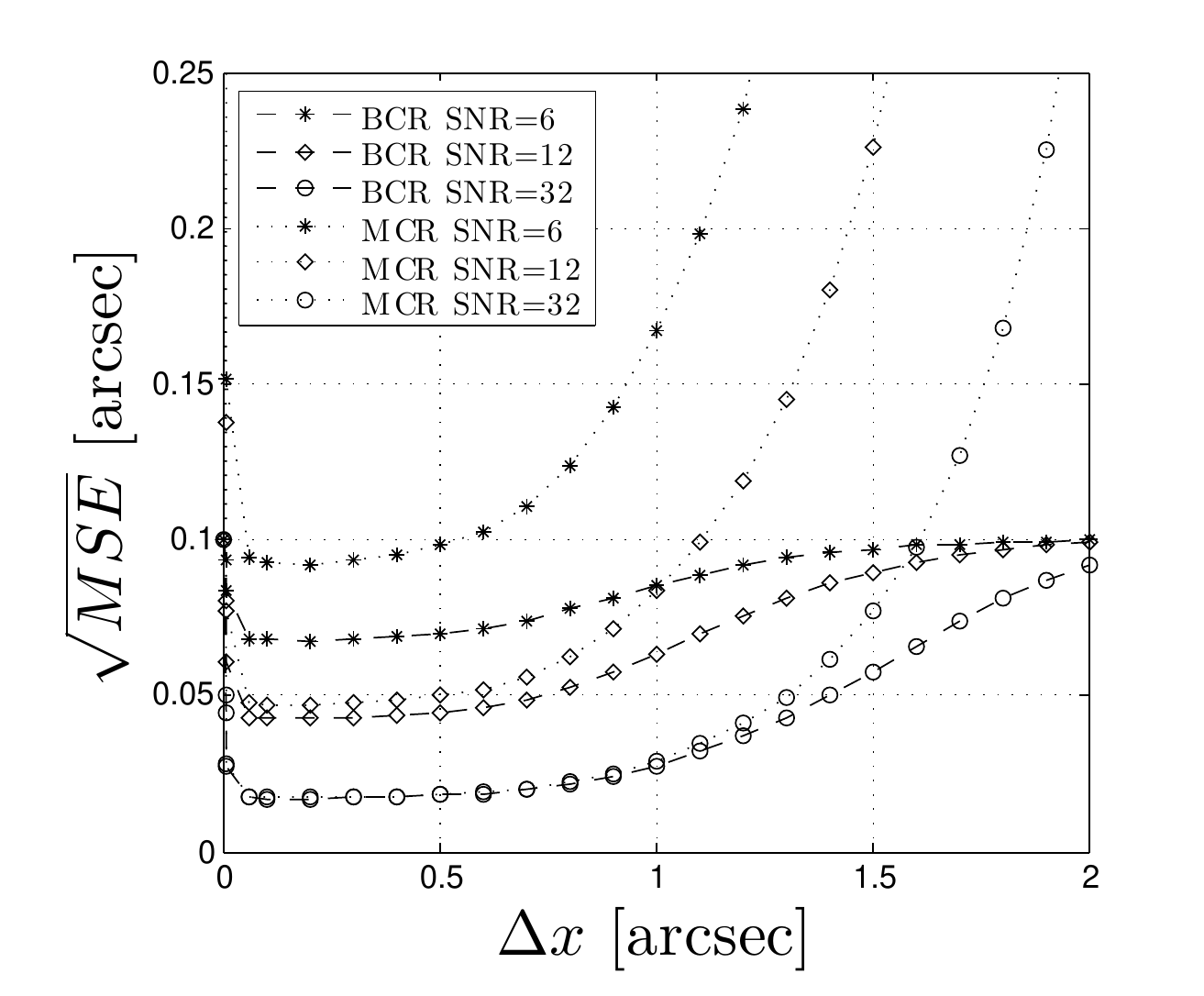}}\hspace{1 cm}
\subfigure[$FWHM=0.5$~arcsec]
{\includegraphics[width=0.45\textwidth]{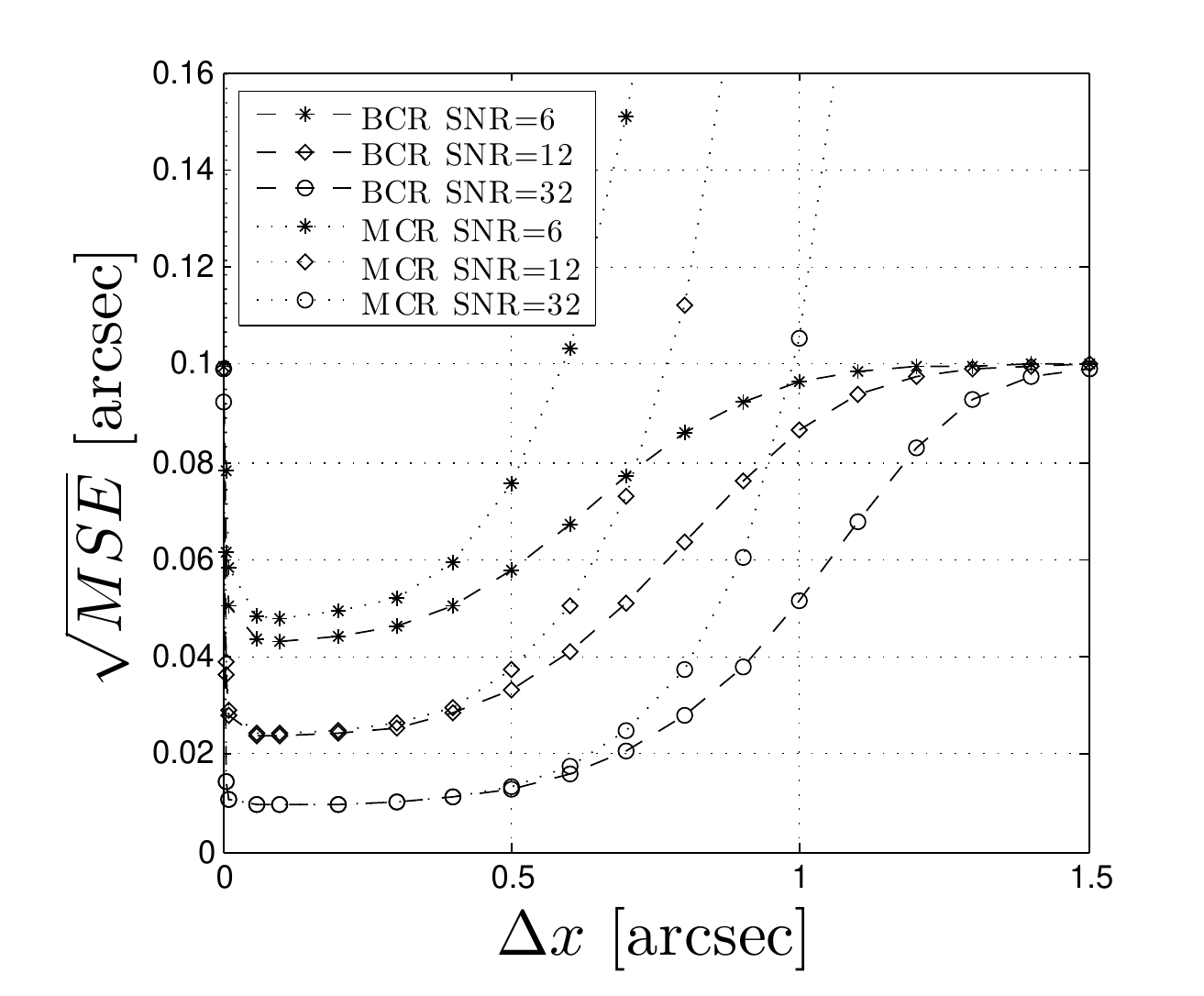}}
\caption{Same as Fig.~\ref{BCR y MCR, sigma 0.1} but with
  $\sigma_{priori}=0.1$~arcsec.}
\label{BCR y MCR, sigma 0.5}
\end{figure*}

\begin{figure*}[h!]
\centering
\subfigure[$FWHM=1$~arcsec]
{\includegraphics[width=0.45\textwidth]{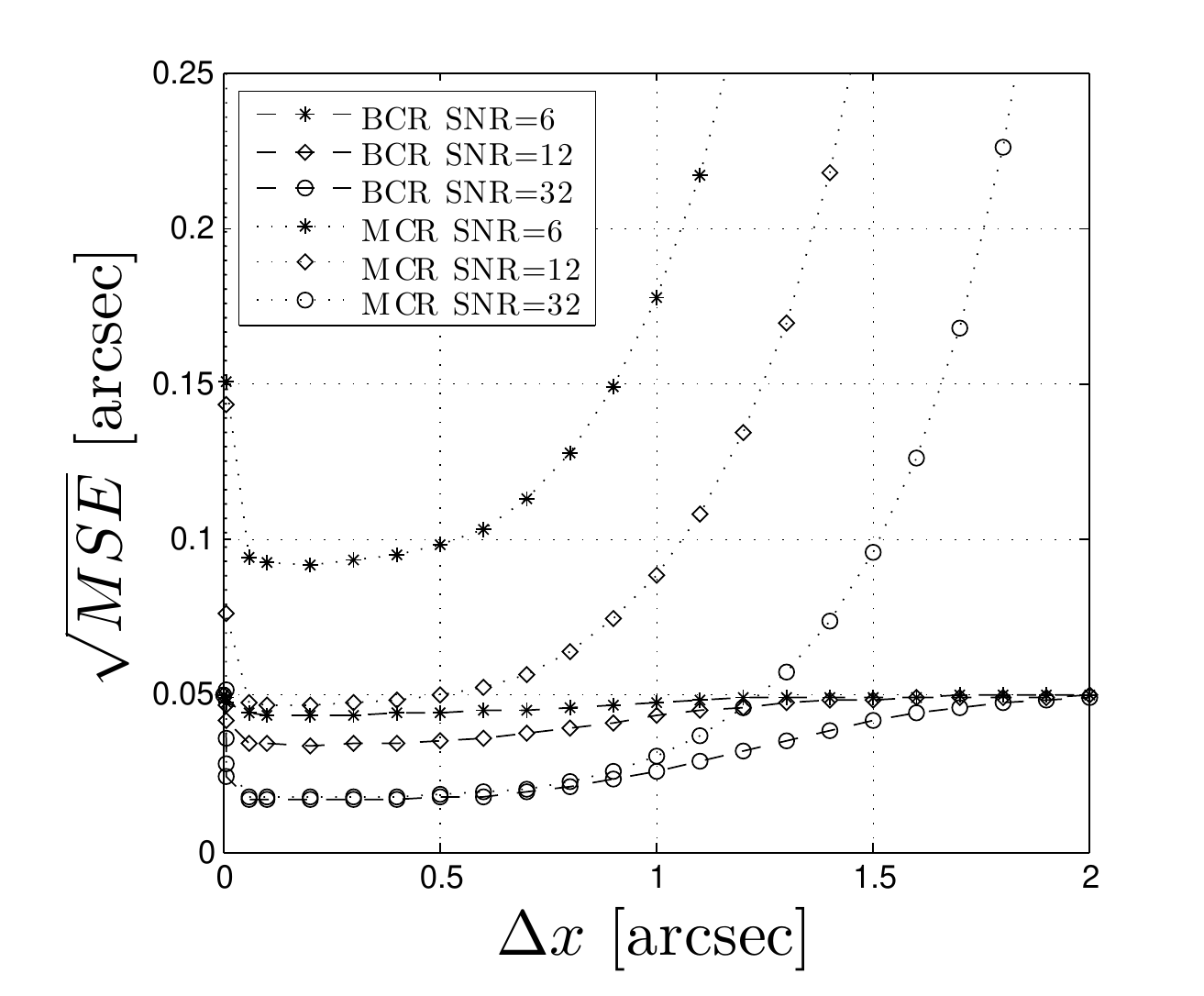}}\hspace{1 cm}
\subfigure[$FWHM=0.5$~arcsec]
{\includegraphics[width=0.45\textwidth]{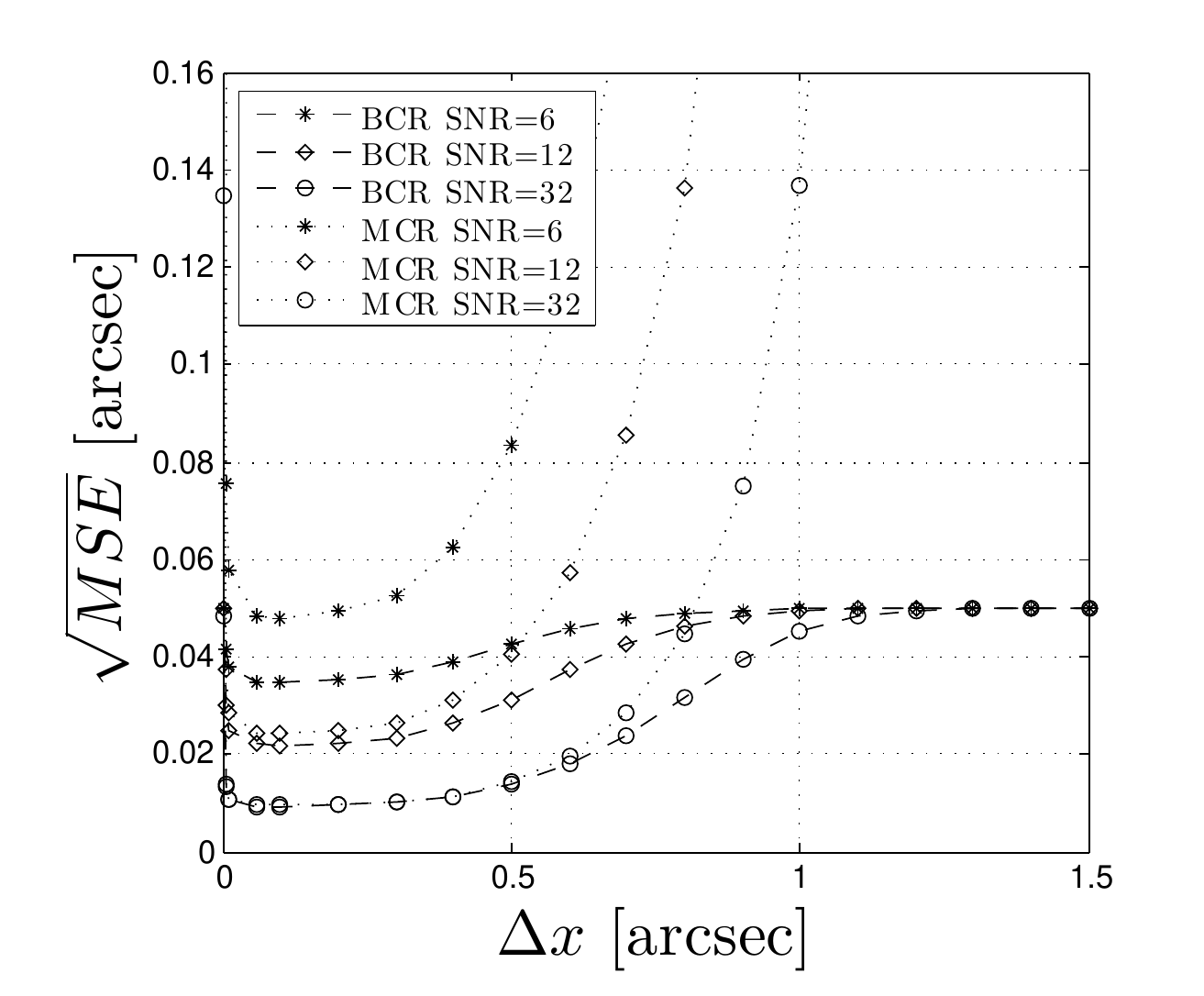}}
\caption{Same as Fig.~\ref{BCR y MCR, sigma 0.1} but with
  $\sigma_{priori}=0.05$~arcsec. In this figure and in
  Fig.~\ref{BCR y MCR, sigma 0.5} we see that while $\sigma_{MCR}$
  increases without bound as the quality of the observations
  deteriorate, $\sigma_{BCR}$ is bounded by the prior information in
  accordance with Eq.~(\ref{eq_sub_sec_bounds_analysis_6}) (see also
  comments in the text after that equation).}
\label{BCR y MCR, sigma 0.05}
\end{figure*}

Figs.~\ref{BCR y MCR, sigma 0.5} and \ref{BCR y MCR, sigma 0.05}
exhibit the trends for the bounds for the same experimental conditions
as in Fig.~\ref{BCR y MCR, sigma 0.1}, but with a significantly
smaller prior variance ($\sigma_{priori} \in \left\{ 0.1, 0.05
\right\}$~arcsec).  Therefore, we increase the prior information
$\mathcal{I}(\psi)$ to see what happens in the performance gain. In this
scenario, the gain $Gain(\psi)$ is significantly non-zero for all
pixel resolutions, meaning that even for very small pixel size (in the
range of 0.1-0.2~arcsec) the Bayes setting offers a boost in the
performance, which is very significant for faint objects, see for
instance the case of $FWHM=1$~arcsec and $S/N = 6$ (Fig.~\ref{BCR y MCR,
  sigma 0.5}a). From Figs.~\ref{BCR y MCR, sigma 0.5} and \ref{BCR y
  MCR, sigma 0.05}, another interesting observation is that the BCR
bound converges to its upper bound limit, provided by the prior
information $\mathcal{I}(\psi)^{-1} = \sigma^2_{priori}$, as $\Delta x$
increases, which is very clear in Fig.~\ref{BCR y MCR, sigma 0.05}a
for $\Delta x>1$ in all the $S/N$ regimes.  We note that this upper bound
limit is absent in the trend of the classical CR limit, as the
performance of the best estimator in the parametric setting
deteriorates with the loss of quality of the observations without a
bound.

To conclude, we can say that having a source of prior
information on the object position offers a gain in the performance of
astrometry especially for faint objects, which can be substantial even
for an excellent instrument (with a small noise and good pixel
resolution) and optimum observational conditions (small background and
small $FWHM$). On the other hand, when the observational conditions
deteriorate, the Bayes setting becomes significantly better than the
parametric approach for a wide range of object brightness. These
curves provide a justification in favor of the use of the Bayes
approach, in particular for the estimation of the position of faint
objects. Therefore, if we have access to a source of prior information
(e.g., a previous catalog), this can complement the information of
the observations (fluxes) and introduce a gain in the performance of
astrometry. The next section goes a bit deeper into this analysis,
while Sect.~\ref{example} presents a simple comparison with a real
catalog.

\section{Equivalent object brightness}
\label{sub_sec_equi_ob_brightness}

The information provided by the prior described in the previous
sections can also be presented in the form of an equivalent
(ficticious) increase in the flux of the source.  If we have a
non-zero prior information $\mathcal{I}(\psi)>0$, and if we fix all the parameters
associated with the observational conditions, i.e., $\Delta x$, $FWHM$,
$B$, and $G$, what value should be given to the equivalent intensity $F_{par}$ of a
fictitious object in the parametric (classical) estimation context (which has a true flux $F$) 
such that
\begin{equation}\label{eq_sub_sec_equi_ob_brightness_1}
	\sigma_{BCR}^2(F,\psi) = BFI(F, \psi)^{-1} =
        \sigma^2_{MCR}(F_{par}, \psi).
\end{equation}
The BCR bound (derived from the Bayes-Fisher information, $BFI$),
which are a function of $F$ and $\psi$, have been defined in
Eq.~(\ref{eq_sec_bayes_7}) and (\ref{eq_sec_bayes_6}) respectively,
while $\sigma_{MCR}^2(F_{par},\psi)$ denotes the average classical CR
bound, function of the ficticious object with "{\em equivalent}"
intensity $F_{par}$ and $\psi$ (see the expression on the RHS of
Eq.~(\ref{eq_sec_com_cr_2}))\footnote{The other dependencies on the
  bounds are considered implicit as the focus of this analysis is on
  the brightness of the object.}. In other words, for an object with
true intensity $F$, we want to find the intensity of a brighter object
$F_{par}$ in the parametric case (where the prior
distribution $\psi$ is not available) such that the classical CR and
the BCR bound are the same. It is clear that $F_{par} \geq F$, and the
difference is proportional to $\mathcal{I}(\psi)>0$\footnote{This comes from the
  fact that $Gain(\psi) = \sigma^2_{MCR}(F, \psi) - \sigma^2_{BCR}(F,
  \psi) \geq 0$ in Eq.~(\ref{eq_sec_com_cr_3b}).}.

\begin{figure}[h!]
\centering
\subfigure[Small flux]
{\includegraphics[width=0.45\textwidth]{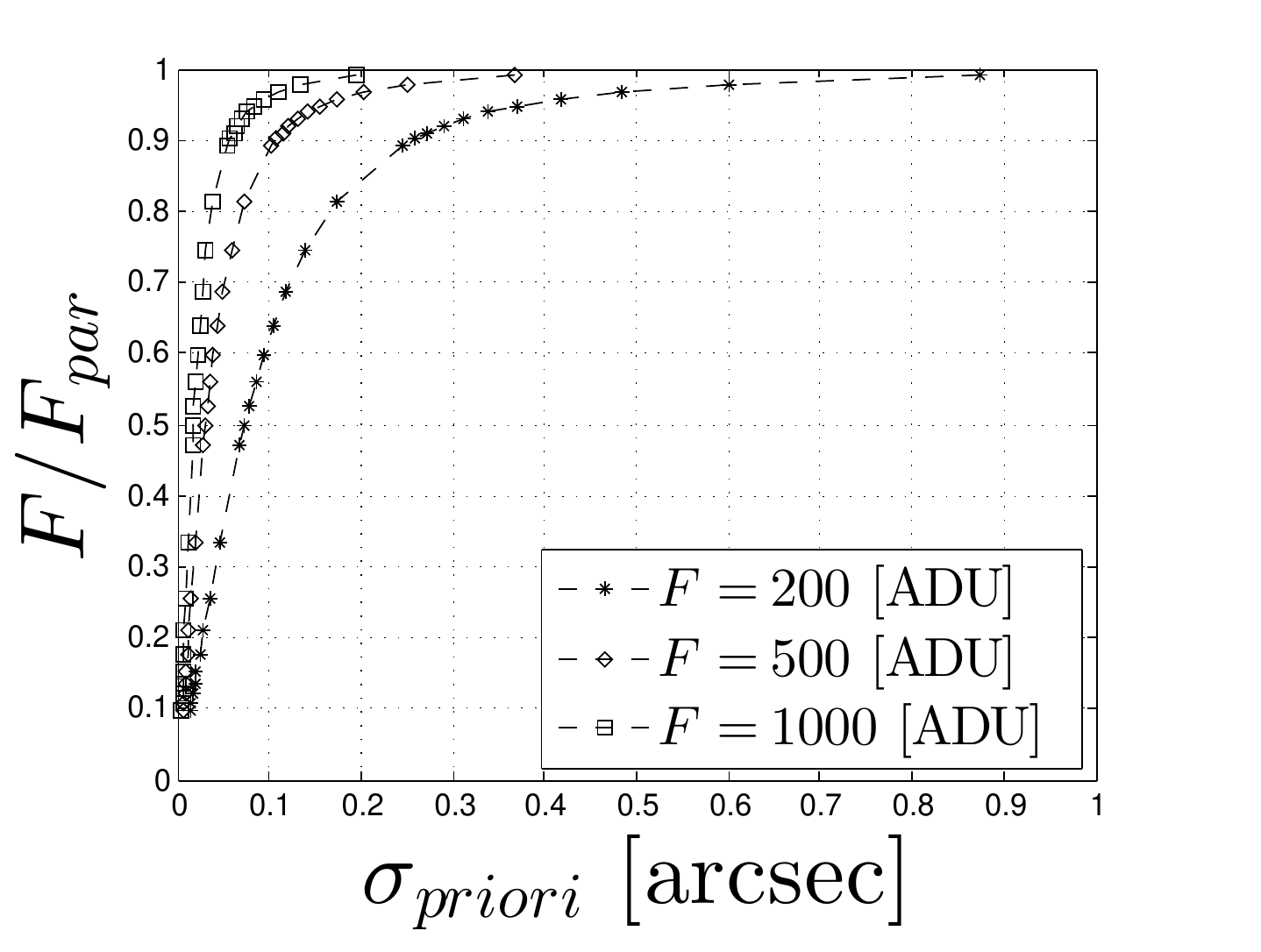}}\hspace{1 cm}
\subfigure[Large flux]
{\includegraphics[width=0.45\textwidth]{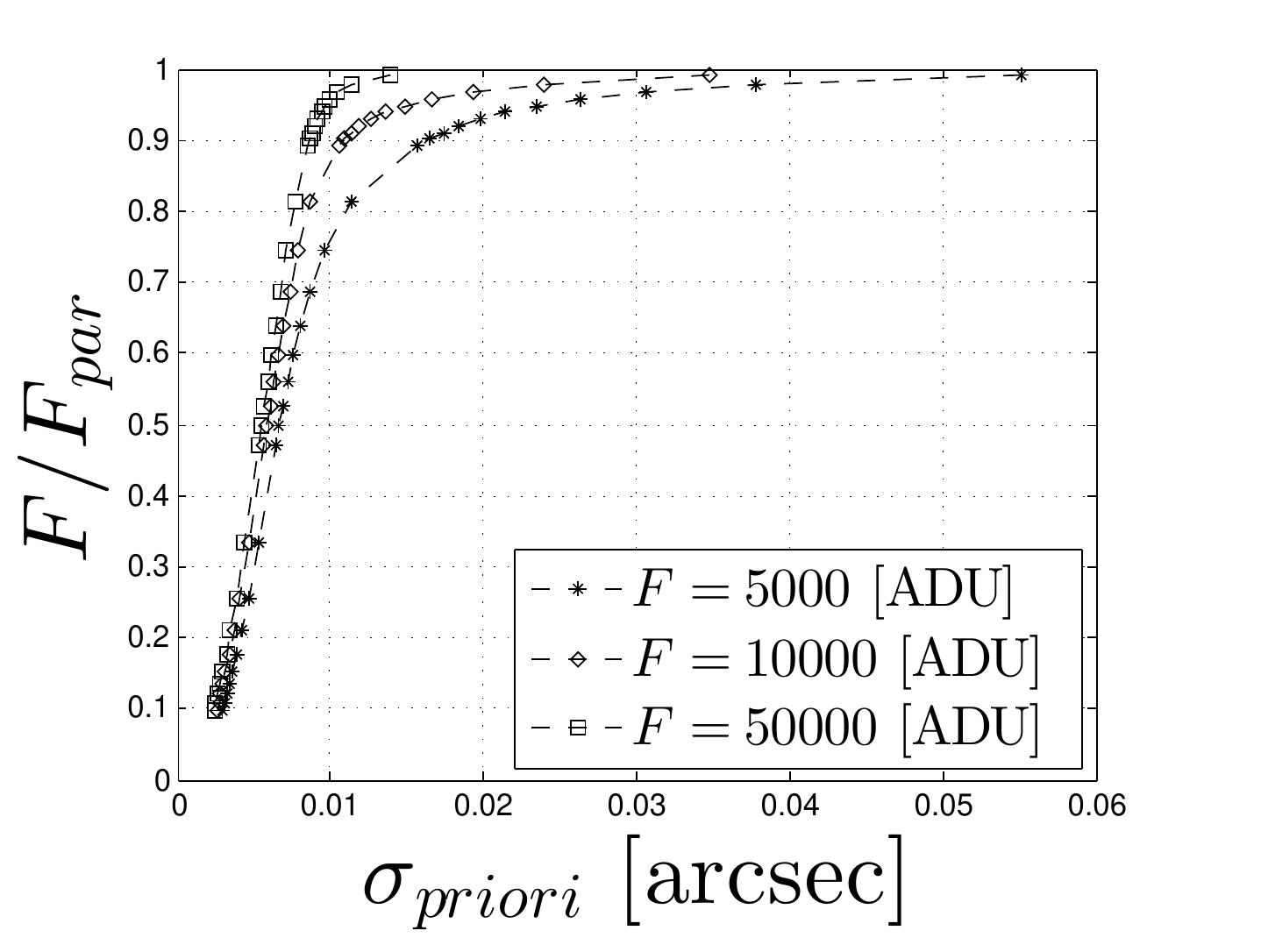}}
\caption{Ratio of the flux for an object with true flux $F$ and its
  equivalent flux $F_{par}$ obtained by the condition indicated in
  Eq.~(\ref{eq_sub_sec_equi_ob_brightness_1}), as a function of the
  prior information given by $\sigma_{priori}$.}
\label{Gain}
\end{figure}
 
Fig.~\ref{Gain}a illustrates the ratio $\frac{F}{F_{par}} \in (0,1)$
as a function of the prior information $\mathcal{I}(\psi)=1/\sigma_{priori}^2$.
For this we consider three scenarios of object intensity $F\in
\left\{200, 500, 1\,000 \right\}$~ADU, and we plot the ratio as a
function of $\sigma_{priori}$ in the range $[0,1]$~arcsec.  As
expected, when $\sigma_{priori} \rightarrow \infty $ then we have that
$\frac{F}{F_{par}} \rightarrow 1$; on the other hand, when
$\sigma_{priori} \rightarrow 0$ then $\frac{F}{F_{par}} \rightarrow
0$.  This last regime is more interesting to analyze, as it is the
case when $F$ is only a fraction of $F_{par}$. In particular, at low
$S/N$ (panel (a) in Fig.~\ref{Gain}), when the prior variance
represented by $\sigma_{priori}$ is in the range of $(0.05,
0.1)$~arcsec, then $F$ needs only to be between 20\% and up to 50\%
smaller than $F_{par}$, depending on the value of $F$, to yield the
same astrometric CR bound. For this particular example another 
way of presenting this result is that a level of prior information
$\sigma_{priori} \leq 0.1$~arcsec makes the object in a parametric
context the equivalent of an object 1.25 to 2 times brighter (or,
equivalently, a gain between 0.25 to 0.75 magnitudes) than its real
intensity when using a Bayesian approach. This is a remarkable
observation, and provides a concrete way to measure the impact of
prior information in astrometry. Fig.~\ref{Gain}b shows the case of
brighter point sources, where $\sigma_{priori}$ needs to be
significantly lower to observe the gains presented in
Fig.~\ref{Gain}a, as expected.

\section{Comparing the BCR lower bound with the performance of the optimal Bayes estimator}
\label{sec_opt_estimator}

One of the very important advantages of the Bayes estimation, in
comparison with the parametric scenario, is that the solution of the
MMSE estimator in Eq.~(\ref{eq_sec_bayes_4}), i.e., the Bayes rule,
has an analytical expression  function of the joint distribution of $(X_c,I^n)$
 that is available for this problem. In contrast, in the
  parametric scenario, there is no prescription on how to build an
  unbiased estimator that reaches the CR lower bound in
  Eq.~(\ref{fi_astrometryb}), unless certain very restrictive
  conditions are met (see \citet{mendez2013analysis}, especially their Eqs.~(5)
  and~(46)). Unfortunately these conditions are not satisfied in the
  astrometric case using PID detectors (see \citet{lobos2015performance}, especially
  their Sect.~(3.1) and Appendix~A), so in the parametric case there
  is no unbiased estimator that can precisely reach the CR
  bound. 
  
 Returning to our problem, the Bayes rule is the well-known posterior mean of $X_c$ given a
realization of the observations. More formally, for all $i^n\in
\mathbb{N}^n$ the MMSE estimator is \citep{weinstein1988general}
\begin{align}\label{eq_sec_opt_estimator_1}
\tau^n_{Bayes}(i^n) &\equiv \mathbb{E}_{X_c|I^n=i^n} \left\{X_c
\right\} = \frac{\int_{x\in \mathbb{R}} x \cdot \psi(x) p_{x}(i^n) dx
}{\int_{x\in \mathbb{R}} \psi(\bar{x}) p_{\bar{x}}(i^n) d\bar{x}}\\
\label{eq_sec_opt_estimator_1b} &=\int_{x\in \mathbb{R}} x
\cdot p_{X_c|I^n}(x|i^n) dx.
\end{align}
It should be noted that $\psi(x) \cdot p_{x}(i^n)$ is the joint density of the
vector $(X_c, I^n)$, the denominator of the RHS of
Eq.~(\ref{eq_sec_opt_estimator_1}) is the marginal distribution of
$I^n$, which we denote by $p_{I^n}(i^n)\equiv \int_{x\in \mathbb{R}}
\psi(\bar{x}) p_{\bar{x}}(i^n) d\bar{x}$ and, consequently,
$p_{X_c|I^n}(x|i^n)= \frac{\psi(x) \cdot p_{x}(i^n)}{p_{I^n}(i^n)}$ in
Eq.~(\ref{eq_sec_opt_estimator_1b}) denotes the posterior density
of $X_c$, evaluated at $x_c$, conditioned to $I^n=i^n$.

Furthermore, the performance of the MMSE estimator
$\tau^n_{Bayes}(\cdot)$ has the following analytical expression
\begin{align}\label{eq_sec_opt_estimator_2}
&\underbrace{\mathbb{E}_{(X_c,I^n)} \left\{ \left( \tau^n_{Bayes}(I^n)
  - X_c \right)^2 \right\}}_{MMSE} = \mathbb{E}_{I^n} \left\{
\mathbb{E}_{X_c | I^n} \left\{ \left( \tau^n_{Bayes}(I^n) - X_c
\right)^2 \right\} \right\} \nonumber\\
&=\mathbb{E}_{I^n} \left\{ \underbrace{\mathbb{E}_{X_c | I^n}
  \left\{\left( \mathbb{E}_{X_c|I^n} \left\{X_c \right\} - X_c
  \right)^2 \right\}}_{Var(X_c|I^n)} \right\}\\
\label{eq_sec_opt_estimator_2b} &=\sum_{i^n\in \mathbb{N}^n} p_{I^n}(i^n) \cdot \underbrace{\int_{x \in
  \mathbb{R}} (\tau^n_{Bayes}(i^n)-x)^2 \cdot p_{X_c|I^n}(x|i^n) \, dx}_{Var(X_c|I^n=i^n)},
\end{align}
which can be interpreted as the average variance of $X_c$ given
realizations of $I^n$.

Therefore, revisiting the inequality in expression
(\ref{eq_sec_bayes_7}), it is essential to analyze how tight the
BCR bound is or, equivalently, how large  the difference between the
MMSE in Eq.~(\ref{eq_sec_opt_estimator_2b}) and the BCR bound is.  To
answer this important question, in the next subsection we conduct
some numerical experiments to evaluate how close  the BCR bound is to
the performance of the optimal estimator given by
Eq.~(\ref{eq_sec_opt_estimator_2b}) under some relevant observational
regimes, and in various scenarios of prior information.

\subsection{Numerical results}
\label{subsec_experiments_mmse_bcr}

\begin{figure}[h!]
\centering
\subfigure[$\sigma_{priori}$=0.1~arcsec] 
{\includegraphics[width=0.45\textwidth]{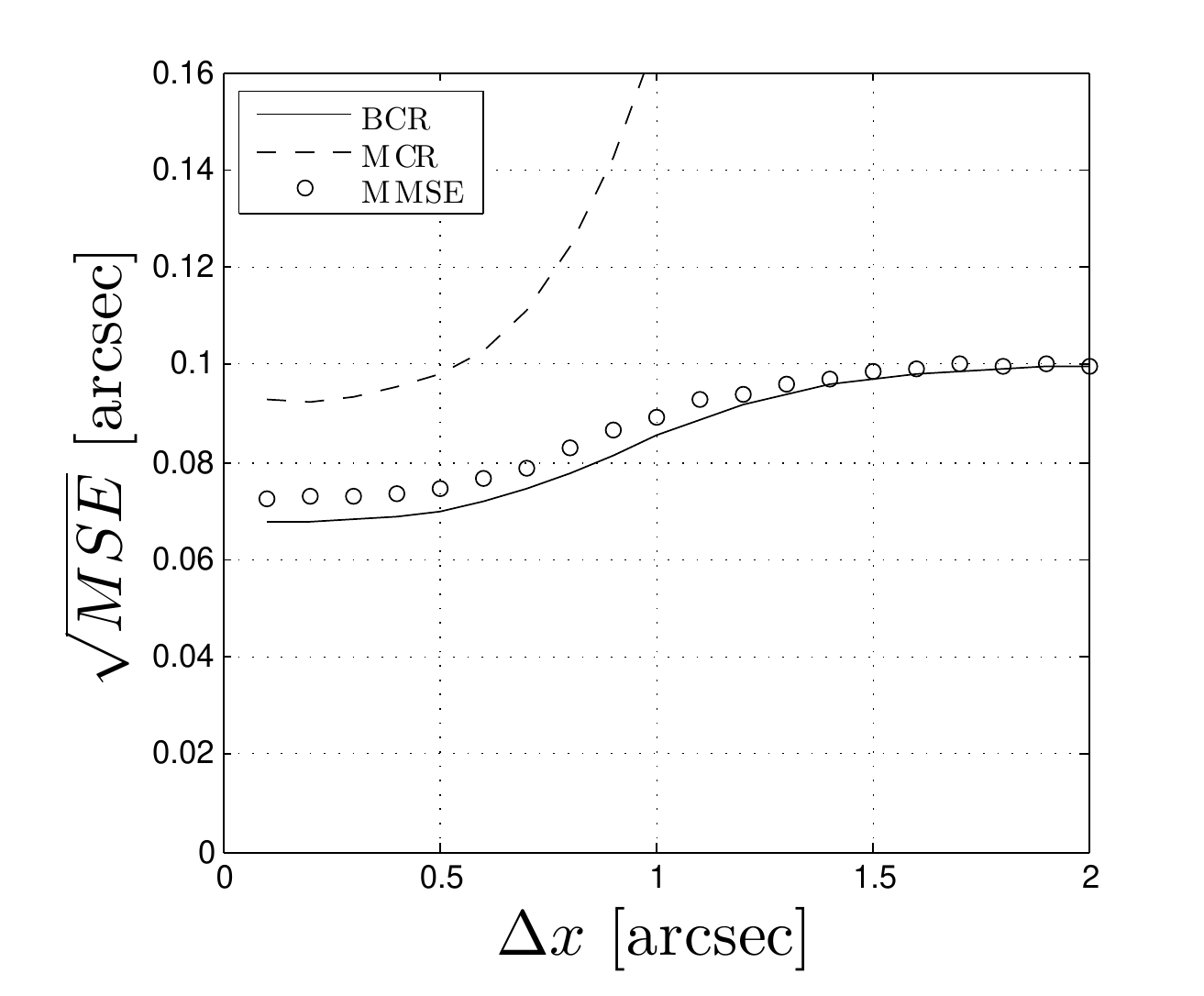}}\hspace{1 cm}
\subfigure[$\sigma_{priori}$=0.05~arcsec] 
{\includegraphics[width=0.45\textwidth]{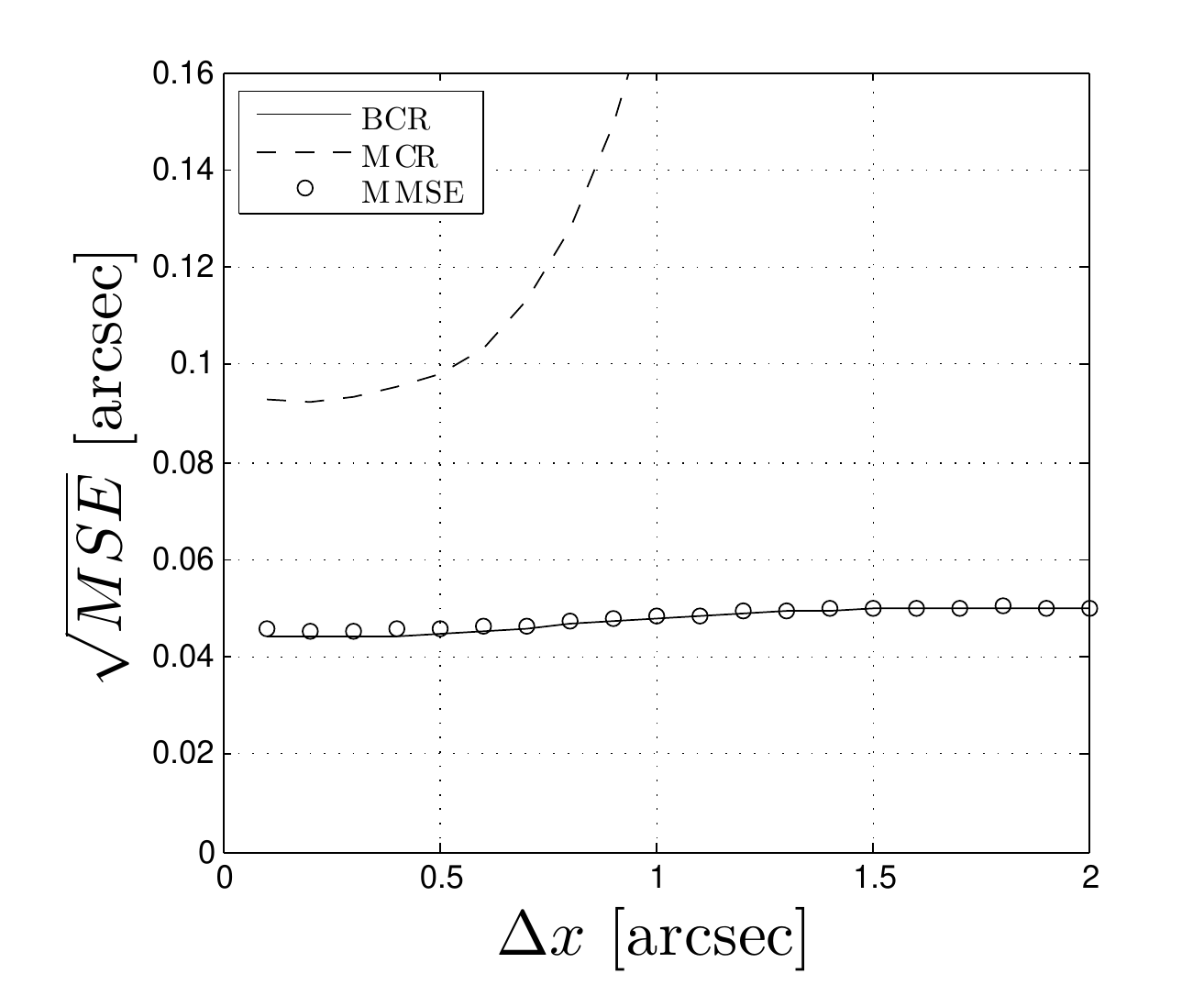}}
\caption{Comparison between the MMSE from
  Eq.~(\ref{eq_sec_opt_estimator_2b}) and the BCR bound from Eq.~(\ref{eq_sec_bayes_7}) for two
  different $\sigma_{priori}$ scenarios, considering a $S/N$ = 6. In
  addition, the MCR bound from Eq.~(\ref{eq_sec_com_cr_2}) is plotted
  to highlight the information gain.}
\label{SNR=6Db}
\end{figure}

\begin{figure}[h!]
\centering
\subfigure[$\sigma_{priori}$=0.1~arcsec]
{\includegraphics[width=0.45\textwidth]{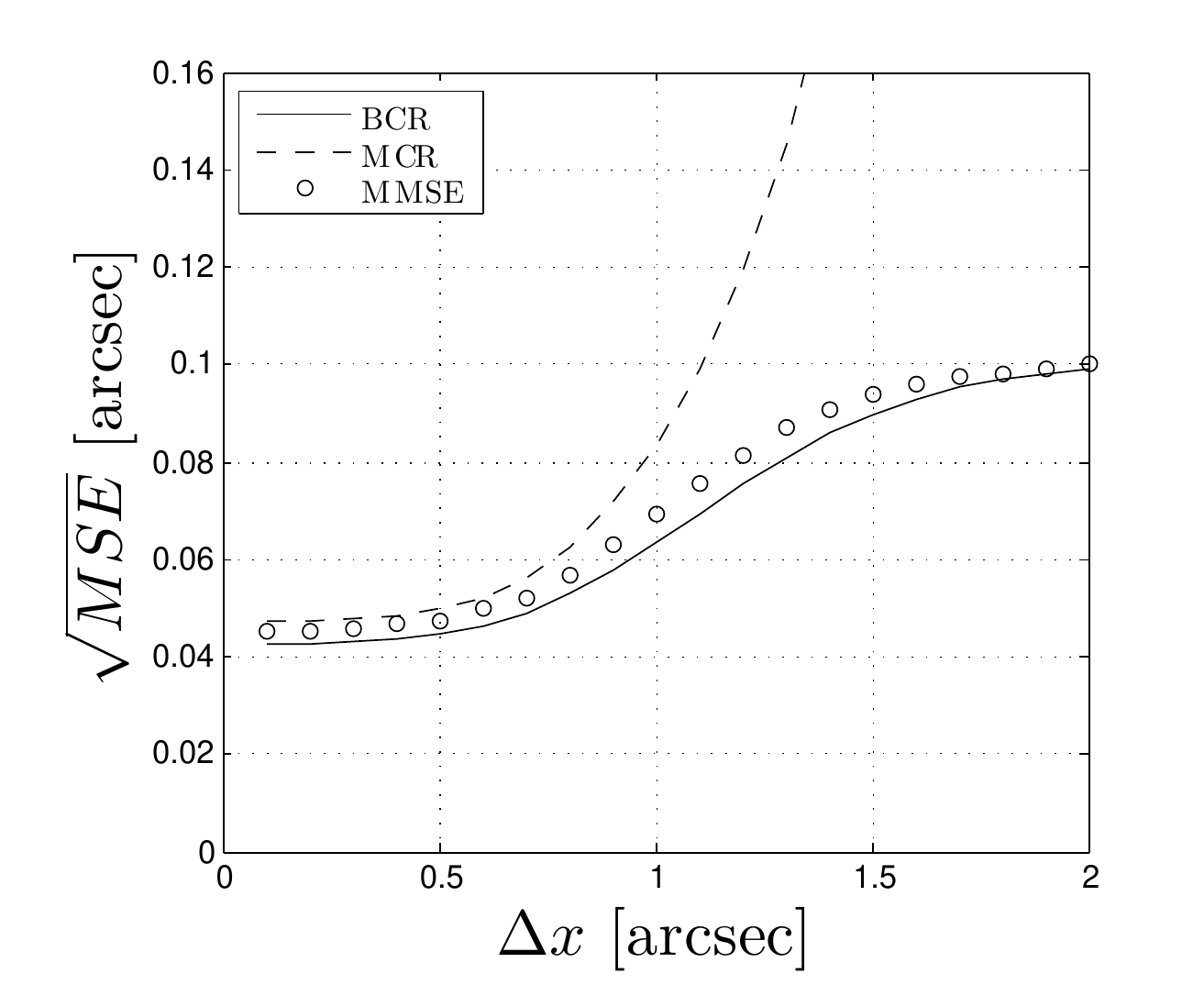}}\hspace{1 cm}
\subfigure[$\sigma_{priori}$=0.05~arcsec]
{\includegraphics[width=0.45\textwidth]{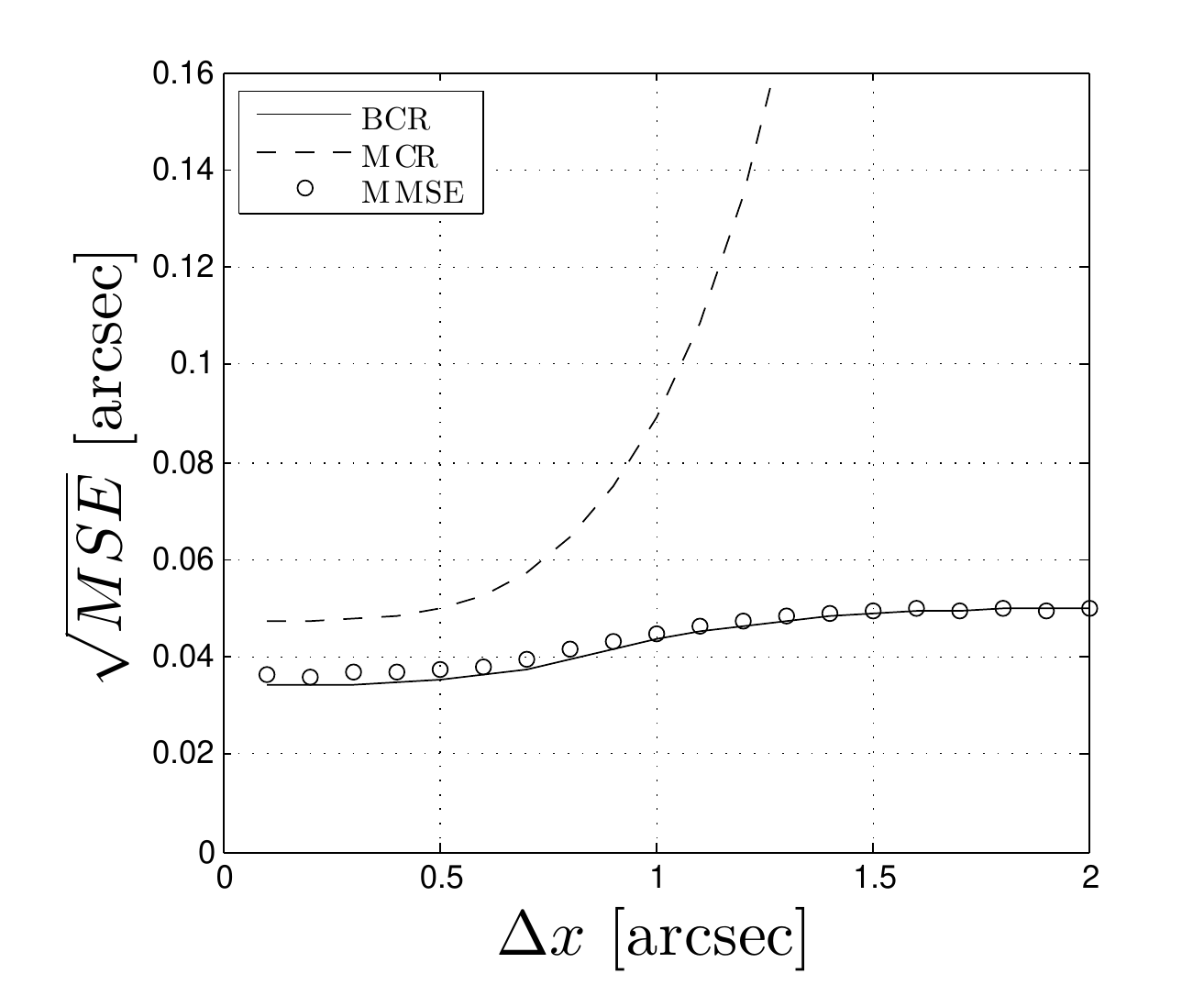}}
\caption{Same as Fig.~\ref{SNR=6Db} but considering  $S/N=12$.}
\label{SNR=12Db}
\end{figure}

\begin{figure}[h!]
\centering
\subfigure[$\sigma_{priori}$=0.1~arcsec]
{\includegraphics[width=0.45\textwidth]{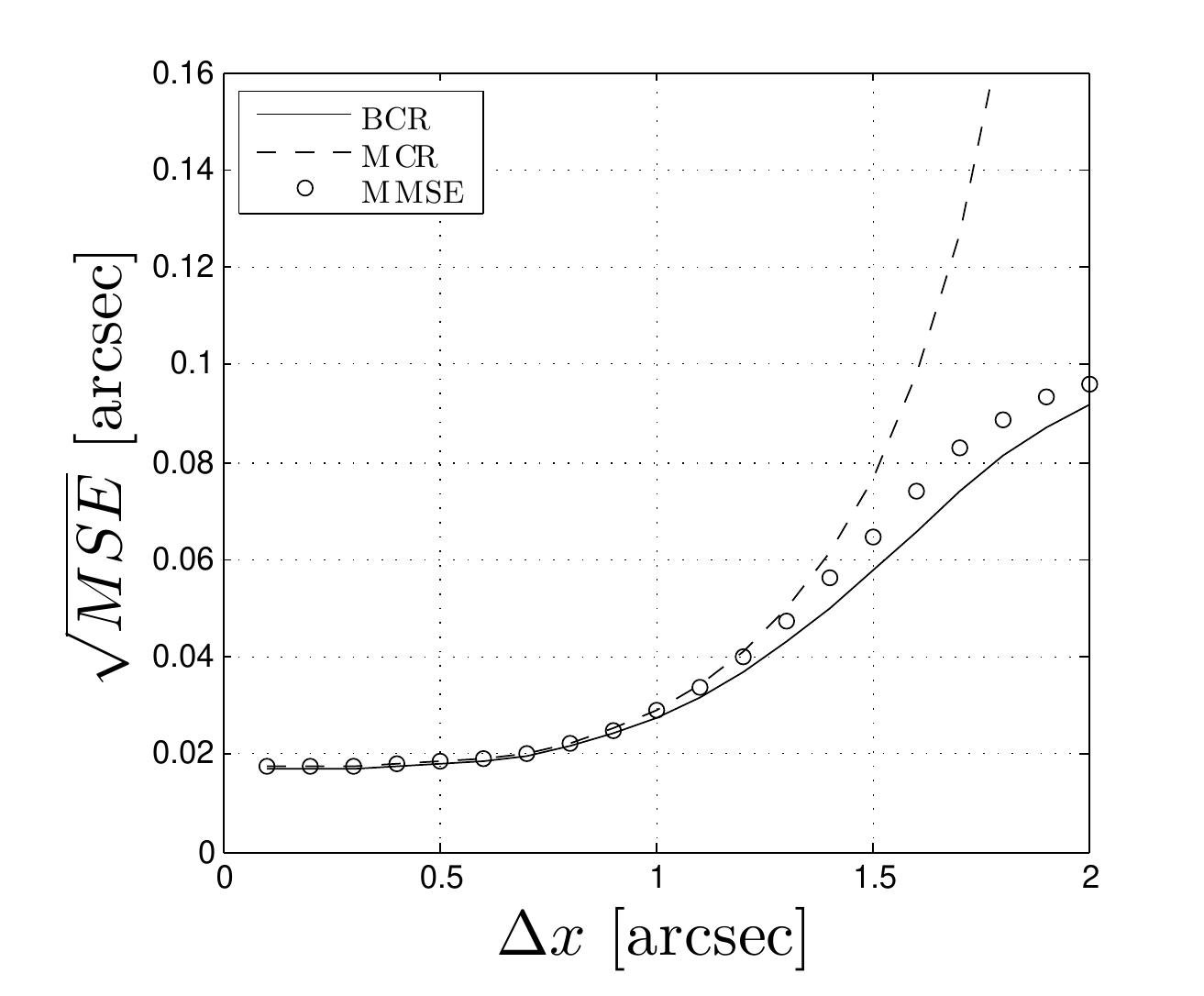}}\hspace{1 cm}
\subfigure[$\sigma_{priori}$=0.05~arcsec]
{\includegraphics[width=0.45\textwidth]{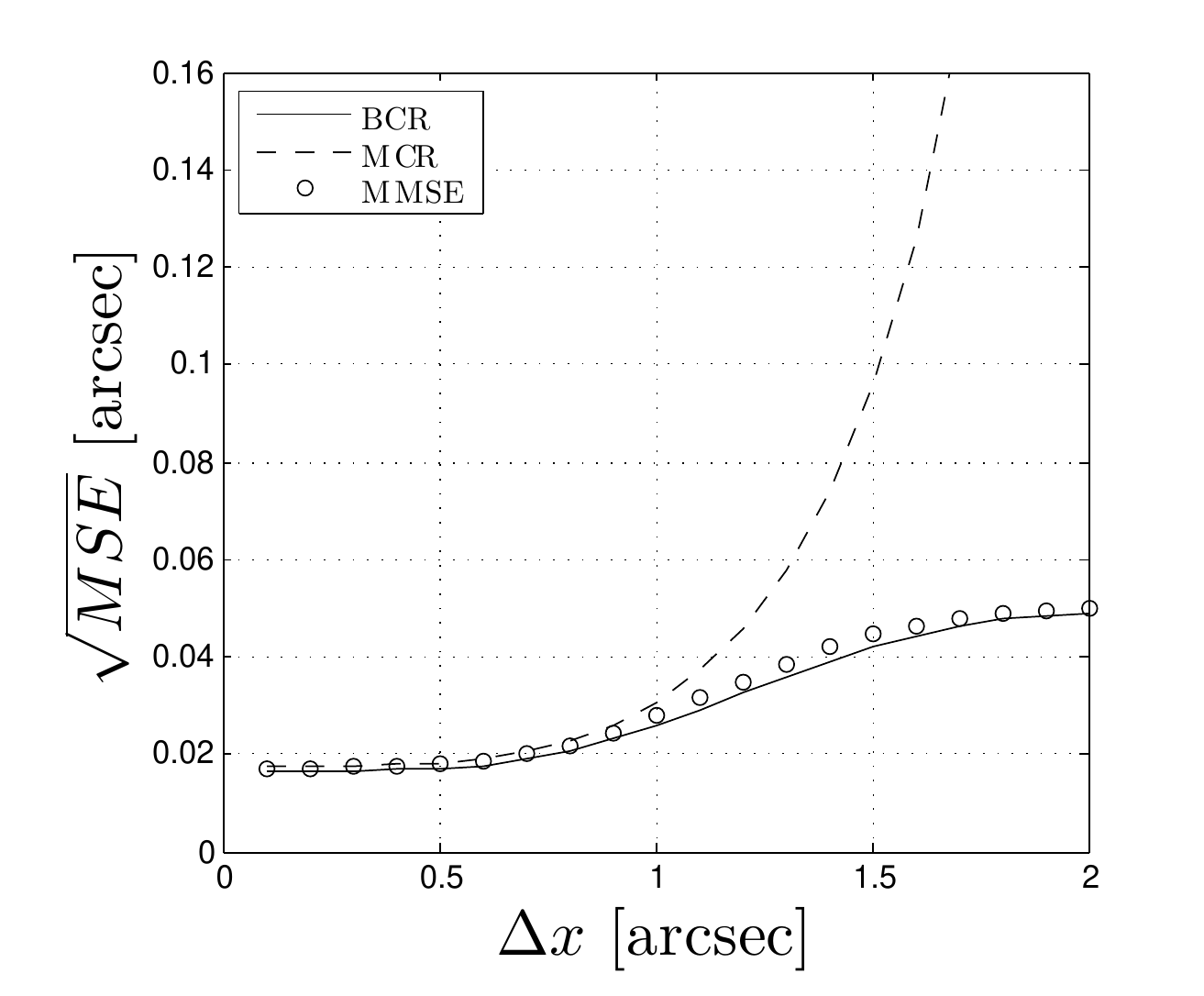}}
\caption{Same as Fig.~\ref{SNR=6Db} but considering  $S/N=32$.}
\label{SNR=32Db}
\end{figure}

Figs.~\ref{SNR=6Db}, \ref{SNR=12Db}, and \ref{SNR=32Db} present the
MMSE from Eq.~(\ref{eq_sec_opt_estimator_2b}) side by side with the
BCR bound in different observational regimes. From these results we
can say that, for all practical purposes, the optimal Bayes rule in
Eq.~\eqref{eq_sec_opt_estimator_1b} (used to determine the location of
a point source)  achieves the BCR lower bound. Consequently, we
conclude that for astrometry, the Bayes rule offers a concrete and
implementable way to achieve the theoretical gain analyzed and studied
in Sects.~\ref{sec_com_cr}, \ref{sub_sec_numerical}, and
\ref{sub_sec_equi_ob_brightness} of this work. 

\section{An example of the BCR bound applied on real data}
\label{example}

In this section we provide a simple example, based on real
data, of how a priori information could be used to improve
the quality of astrometric estimation.

\subsection{Overall description of purpose}
\label{purpose}

In Bayesian statistics, any piece of information that could be
rendered in a mathematical form is in principle a possible prior.
In the example we develop here, we use the positional
information provided by a catalog that reports a (dependable)
uncertainty on those positions. We assume that the catalog
positions are distributed following a Gaussian distribution given by
the reported uncertainty, centered on the catalog value (see
Eq.~(\ref{eq_sub_sec_bounds_analysis_3})). Since the focus of this
paper is on the uncertainty bounds and not on the estimation of the
actual values (the coordinates) themselves, the catalog values are
actually irrelevant. We will further assume that the prior values are
not biased in any way, i.e., that there is no mismodeling (see also
the paragraph above Eq.~(\ref{eq_sec_com_cr_1})) otherwise, combining
the old and new values may render biased results (this issue,
which is beyond the scope of the current paper, will be explored in a
forthcoming report). Finally, we assume that we carry out new
observations with pre-defined equipment (telescope+imaging camera)
with known properties and under controlled conditions ($FWHM$, $S/N$ at
a certain flux), so that we will be able to evaluate the BCR and MCR bounds in Eq.~(\ref{eq_pro_information_gain}) for each object
in the catalog.
If we fix the observational conditions, the only
remaining free parameter is the distribution in flux of the observed
objects (i.e, the apparent luminosity function). Since the goal of
this example is to illustrate how the catalog data can
have an impact on the expected quality of the positions derived from
the new observations, we can compare the number of objects that
satisfy a certain maximum positional uncertainty threshold as a
function of brightness in the classical vs. the Bayesian
approaches to illustrate the gains of the latter. A description of how
do this, and the main results, are presented in the following sections.

\subsection{Prior information and incremental observations}
\label{details}

For the purposes outlined in the previous section, we will make use
of the USNO-B1 catalog (\citet{monet2003usno}) which provides astrometric
positions and their uncertainties, and photographic photometry on
various optical bands for a complete set of stars brighter than $V
\sim 21$ with an overall 0.2~arcsec astrometric accuracy at J2000 and
$\sim$0.3~mag photometric accuracy. In addition, for each entry the
catalog provides a star/galaxy index, which is believed to be 85\%
accurate at distinguishing stars from non-stellar sources. In order to
have a clear star/galaxy separation and the best possible photometry,
both of which become less accurate in dense stellar regions, we have
extracted a representative area of $20 \times 20$~arc-minutes$^2$
around the SGP from the online version of the catalog available
through the VizierR web page within the CDS service
(http://cds.u-strasbg.fr/).

\begin{figure}[h!]
\centering
\subfigure{
{\includegraphics[width=0.45\textwidth]{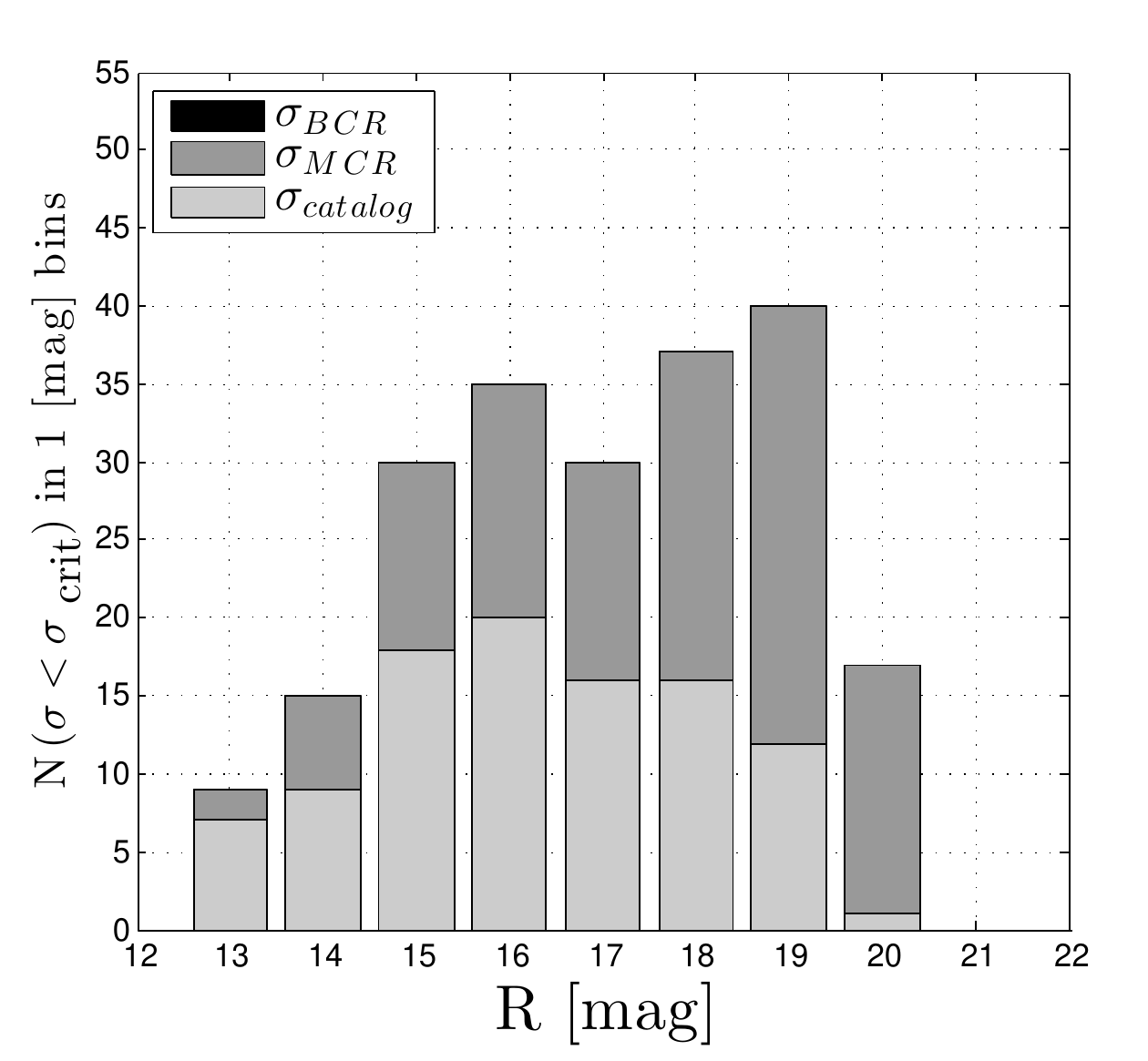}}}
\subfigure{
{\includegraphics[width=0.45\textwidth]{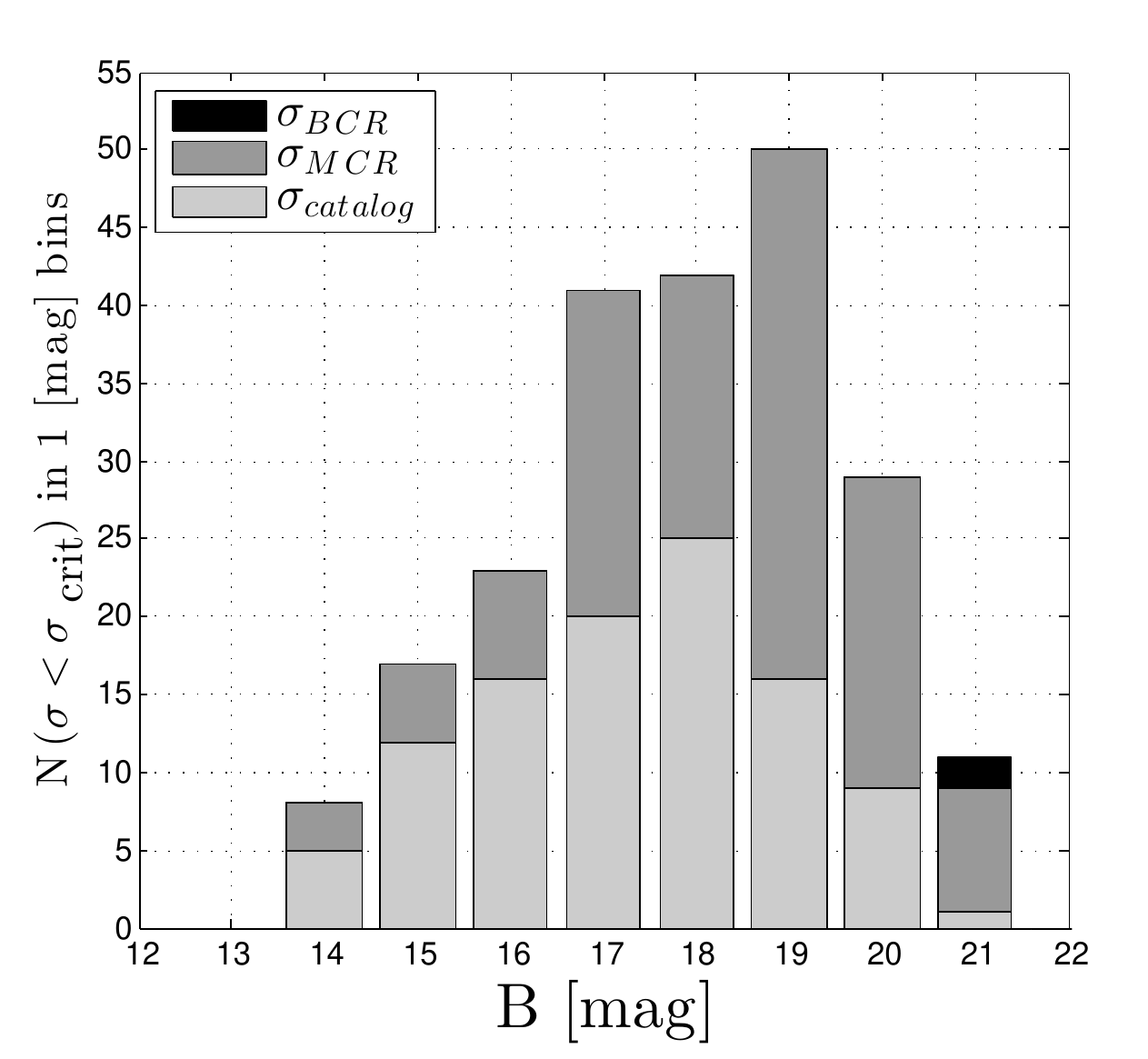}}}
\caption{Histogram of the number of stars (in bins of 1~mag) in an
  area of $20 \times 20$~arc-minutes$^2$ towards the SGP from the
  USNO-B1 catalog that satisfy the criteria that their astrometric
  position has a positional uncertainty less that a threshold value of
  $\sigma_{\mbox{crit}}$=0.1~arcsec. Light gray is for
  those objects that satisfy the astrometric threshold in the
  catalog directly (106 objects). Light+dark gray is the number of
  objects that would satisfy the criteria when one observation with
  EFOSC2@NTT is carried out on the same sample, according to the
  expectations from the classical MCR, for a $FWHM$ of
  0.7~arcsec. Finally, Light+dark gray+black is the number of objects
  when the BCR is considered (in this case $\sigma_{\mbox{priori}}$ is
  the catalog value for the positional uncertainty of each individual object). The left panel is for
  the R band, right panel for the B band.}
\label{Histo1}
\end{figure}

Our test catalog contains $2\,700$ entries of which 226 objects
satisfy the joint criteria that their star/galaxy index is $\ge 5$
(and are thus considered to be stars to a high level of certainty) and
that have valid values for their photometry (i.e., B or R~mag $\ne
99.99$\footnote{To further improve the photometry we took the average
  of the two values of B and R provided in the catalog for objects
  with valid photometry.}) and their astrometry ($\sigma_{RA} \ne
999$, $\sigma_{DEC} \ne 999$). Since the purpose of this exercise is
to illustrate the impact of prior information (provided in this case
by the USNO-B1 catalog), we further trimmed the sample to those
objects with a slightly smaller (but still realistic) uncertainty
threshold in their astrometric positions equal to 0.1~arcsec (because
our analysis has been done for a linear array, we are considering only
one coordinate, with an uncertainty which is the mean of the values
declared in the catalog for RA$\, \cos$(DEC) and DEC
coordinates). This sample contains 106 objects, and their histogram as
a function of brightness is depicted in Fig.~\ref{Histo1}. We have
chosen this astrometric uncertainty threshold as a value useful for fiber-fed or multi-slit
target positioning in typical spectrographs. We will now evaluate how
these histograms change through new observations when we incorporate
the prior information.

We assume that we obtain new astrometric observations with a typical
optical imaging PID. As a representative case we have taken the
throughput and other specifications of the EFOSC2 instrument (with CCD\#40)
 currently installed at the NTT 3.58m telescope of the
ESO-La Silla observatory\footnote{These can be found in the
  Observatory web pages:
  http://www.eso.org/sci/facilities/lasilla/instruments/efosc.html}. In
particular, we adopt a RON of 9.2~e$^-$ (normal read-out mode), a
(negligible) dark current $D=7~e^-$/pix/hr, a gain $G=1.33~e^-$/ADU,
and an effective pixel size (in 2x2 binned mode) of $\Delta
x=0.24$~arcsec. The throughput of EFOSC2 at the NTT is constantly
monitored\footnote{See
  https://www.eso.org/sci/facilities/lasilla/instruments/efosc/inst/zp/.html},
and these ``zero points'' (on different pass bands) are used to
compute the flux (in e$^-$/s) for objects of a given apparent
magnitude as observed by EFOSC2, but also to estimate the flux that
would be observed by telescopes of other aperture (of similar optical
characteristics as the NTT), or through detectors of different pixel
size (of similar characteristics to CCD\#40), as explained below.

To fully incorporate the background $f_s$ (see
Eq.~(\ref{eq_sub_sec_bounds_analysis_1})) in our CR estimations, we
need to consider its main contributor, i.e., the changing sky
brightness according to moon phase. For this we use the table of sky
brightness in mag/arcsec$^2$ at different pass bands and different
moon ages provided by \citet{walker1987noao}; although it was
measured for the sky above CTIO, it can be taken as representative for
a good astronomical site. Our calculations turned out to be rather
insensitive to moon phase, mostly because of the relatively bright
magnitude limit of the USNO-B1 catalog, and we therefore adopted in
what follows a moon age of 7 days (gray time) corresponding to a sky
brightness of 21.6~mag~arcsec$^{-2}$ in the B band and
20.6~mag~arcsec$^{-2}$ in the R band.

We  also assume that our new observations are carried out in such
a way that we are able to secure a minimum $S/N$ of 3.0 for the faintest
objects available in our SGP sample from the USNO-B1 catalog. This
defines an exposure time $t$ that is the solution of the equation
(see also \citet{mendez2013analysis}, their Eq.~(28))
\begin{equation} \label{snr}
S/N = \frac{F\cdot t}{\sqrt{F\cdot t + N_{pix} \cdot (f_s\cdot t\cdot\Delta x
    +RON^2)}}
\end{equation}
where $F$ is the flux in e$^-$/s of the faintest USNO-B1 sources as
seen by EFOSC2@NTT, $N_{pix}$ is the number of pixels under the
(chosen) aperture for computing the flux under the PSF (we adopt
$N_{pix}$ such that the $S/N$ includes 99.5\% of the source flux),
while the other terms have all been previously defined. With the
derived exposure time, we discard any catalog object for which the
peak count is predicted to be larger than the detector saturation
level of 65,535~ADUs. This eliminates a few objects in our
histograms with B, R$\le$ 11.

\subsection{Analysis of MCR and BCR predictions}
\label{anal}

If we secure one observation with the exposure time calculated from
Eq.~(\ref{snr}) at the NTT, and if we assume that the uncertainty of
their astrometric positions tightly approach the theoretical limits
(which is justified by our results in
Sect.~\ref{subsec_experiments_mmse_bcr}), then the number of targets
(effectively available on the sky, since they are on the USNO-B1
catalog) whose positional uncertainty is
smaller than the adopted threshold of 0.1~arcsec would increase, as
shown by the light+dark gray areas in Fig.~\ref{Histo1}. To compute
the CR bounds we have assumed a $FWHM=0.7$~arcsec. Obviously, the NTT
data are predicted to be of much higher astrometric quality than the data
derived from the plates, which is clearly seen as the significant
increase in the number of sources that satisfy the astrometric
threshold $\sigma_{\mbox{crit}}$. Finally, in black, we show the
number of sources that would be added to the light+dark gray
histogram, as a function of magnitude, if we compute the BCR, i.e., incorporating the prior information from the
catalog. In this case, the prior information does not
play a significant role as the increase in the number of sources is
minimal. This shows again that since the NTT data alone is of much
higher quality, the information contained in the catalog is less
relevant.

\begin{figure}[h!]
\centering
\subfigure{
{\includegraphics[width=0.45\textwidth]{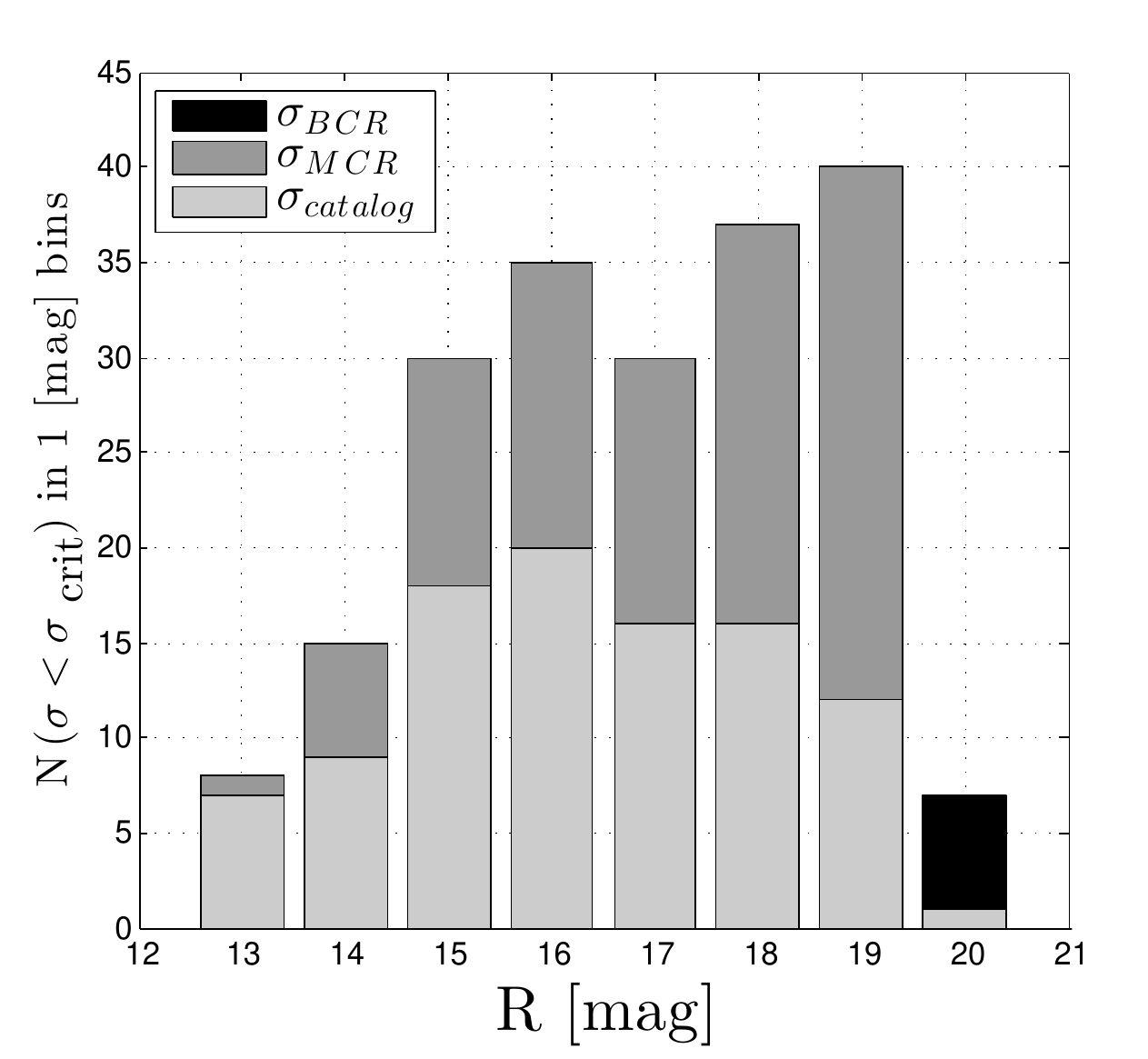}}}
\subfigure{
{\includegraphics[width=0.45\textwidth]{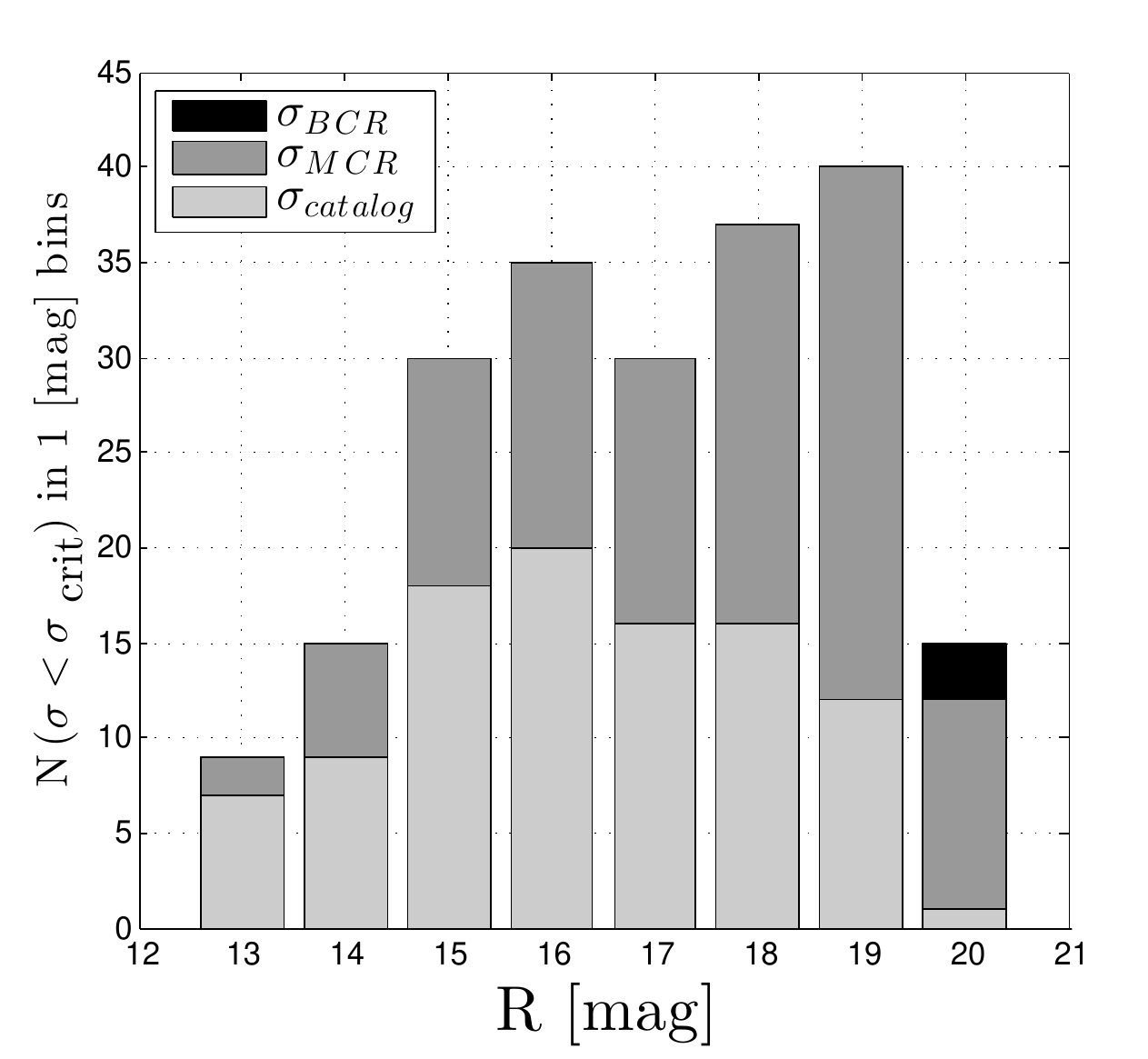}}}
\caption{Similar to Fig.~\ref{Histo1}, except that both panels are in
  the R band. The left panel is for EFOSC2@NTT with a degraded seeing
  of 1.2~arcsec $FWHM$, while the right panel is for optimistic sky
  conditions of 0.7~arcsec but on a 1~m class telescope equipped with
  a camera equivalent to EFOSC2. These histograms show the evolving
  role of the a priori information in improving the astrometry
  depending on sky conditions and the equipment used. The benefits of
  using the Bayesian approach is noticeable at the faintest bins.}
\label{Histo2}
\end{figure}

A scenario where we consider an observation at the NTT under less
favorable sky conditions ($FWHM = 1.2$~arcsec) is shown in the left
panel of Fig.~\ref{Histo2}. Compared to the left panel of
Fig.~\ref{Histo1} we see that the two faintest bins do benefit
from the prior position, as expected. If we insist on sharp images,
but now using a smaller aperture telescope of 1~m (in comparison with
the 3.58~m of the NTT), the predicted behavior will be that of
Fig.~\ref{Histo2} (right panel). This last case is interesting:
While in most apparent magnitude bins the contribution from the
current observations almost doubles the number of targets per bin in
comparison with the objects from USNO-B1 that satisfy the positional
threshold, the impact of the Bayesian CR in the faintest bin is most
significant. This result is in line with the gains for faint targets
described in Sects.~\ref{sub_sec_numerical}
and~\ref{sub_sec_equi_ob_brightness}, and shown in Figs.~\ref{BCR y
  MCR, sigma 0.1} to~\ref{Gain}.

\begin{figure}[h!]
\centering
\subfigure{
{\includegraphics[width=0.45\textwidth]{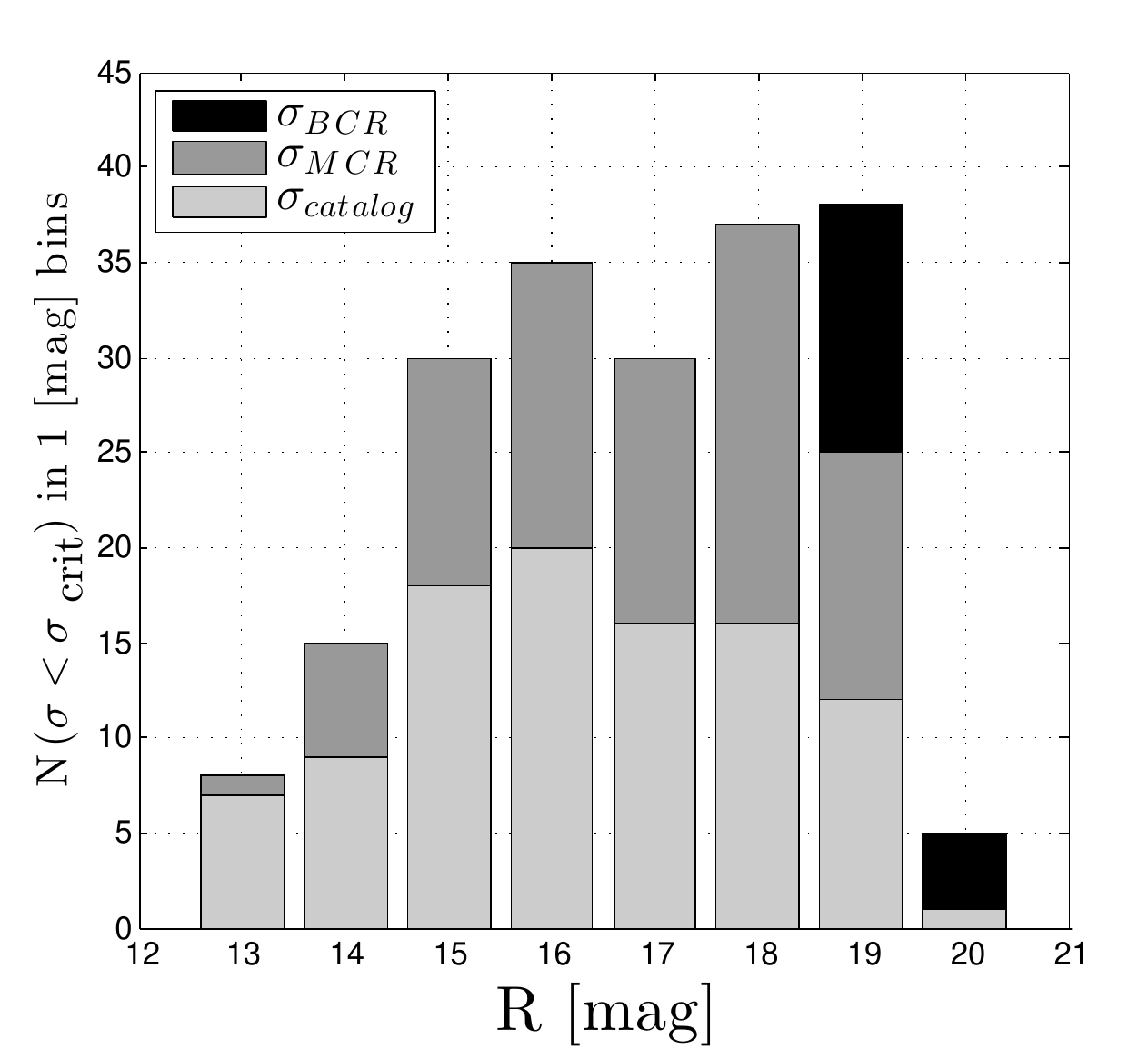}}}
\subfigure{
{\includegraphics[width=0.45\textwidth]{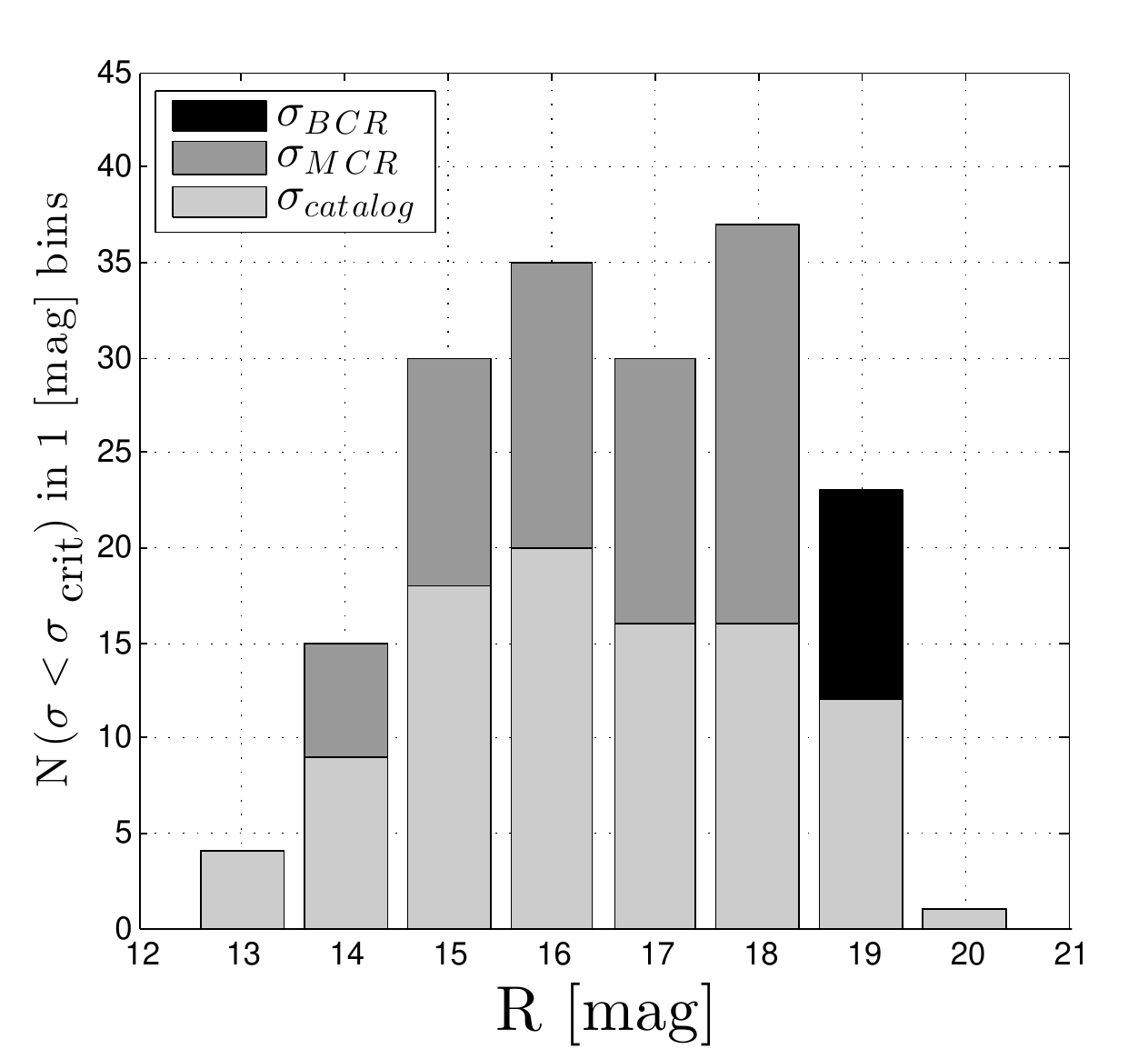}}}
\caption{Similar to Fig.~\ref{Histo2}, except that both panels are the
  predictions for observations with a 1~m class telescope. The left
  panel is for a $FWHM$ of 1.2~arcsec, while the right panel is for a
  $FWHM$ of 2.0~arcsec. As the quality of the new observations
  degrades, their contribution  to the sample, as well as that
  using prior information, is displaced to brighter magnitudes. The
  gain from using the information in the catalog is relevant in both
  cases at the faintest bins, as in Fig.~\ref{Histo2}: At 19th
  magnitude we have 38 objects in the Bayesian scenario, and 25
  objects in the parametric approach, i.e., an increase of more than
  50\% for the left panel. In the right panel the increase is even
  more important: 23 objects in the Bayesian approach and 12 from the
  catalog (equal to the number from MCR), an increase of $\sim
  90$\%.}

\label{Histo3}
\end{figure}

Finally, two cases (which are perhaps more realistic)  are presented in
Fig.~\ref{Histo3} but utilizing a 1~m telescope for a moderate seeing
(left panel) and a less-than-ideal seeing (right panel). As the image quality 
of the new observations deteriorates, we can see that 
the contribution of these observations to the number of objects with
good astrometric positions decreases, while the use of prior information
reinforces even further the faintest bins. For example, in
Fig.~\ref{Histo3}a, the overall number of objects with good astrometry
increases by about 9\% when using the BCR (or 12\% for
Fig.~\ref{Histo3}b), although the increase is quite dramatic at the
lowest $S/N$ bins (see Fig.~\ref{Histo3} caption).

We note that in all the computations above, we have taken the catalog as
is, but we know that
in the faintest bins, the real number of targets will be larger than
shown in Figs.~\ref{Histo1} to~\ref{Histo3} due to
incompleteness. However, since these objects are not in the catalog, and
hence there is no prior information on them, they do not
affect the impact that prior information has on their positions, which
was the purpose of this exercise (the minimum astrometry uncertainty
for these extra objects will be solely determined by their classical
MCR).

\subsection{Improvements in the mean errors of individual positions}
{In the previous section, we give a overview of the impact of
using a priori information on starcounts as a function of magnitude
when we impose a constraint on the minimum acceptable astrometric
accuracy. This is relevant for the evaluation of the bulk performance
of surveys that convey astrometric information, but for a practitioner
astrometrist, it might be more relevant to be able to evaluate the
actual improvement in astrometric precision of the catalogued objects
using the BCR in comparison with the classical MCR. In what follows we
present this information for the same observational scenarios
described in the previous section, giving more details on the
expected improvement in the mean error of individual positions.}

\begin{figure*}[h!]
\centering
\centerline{
\subfigure[FWHM=0.7 arcsec, aperture=3.5 m]{\includegraphics[width=0.33\textwidth]{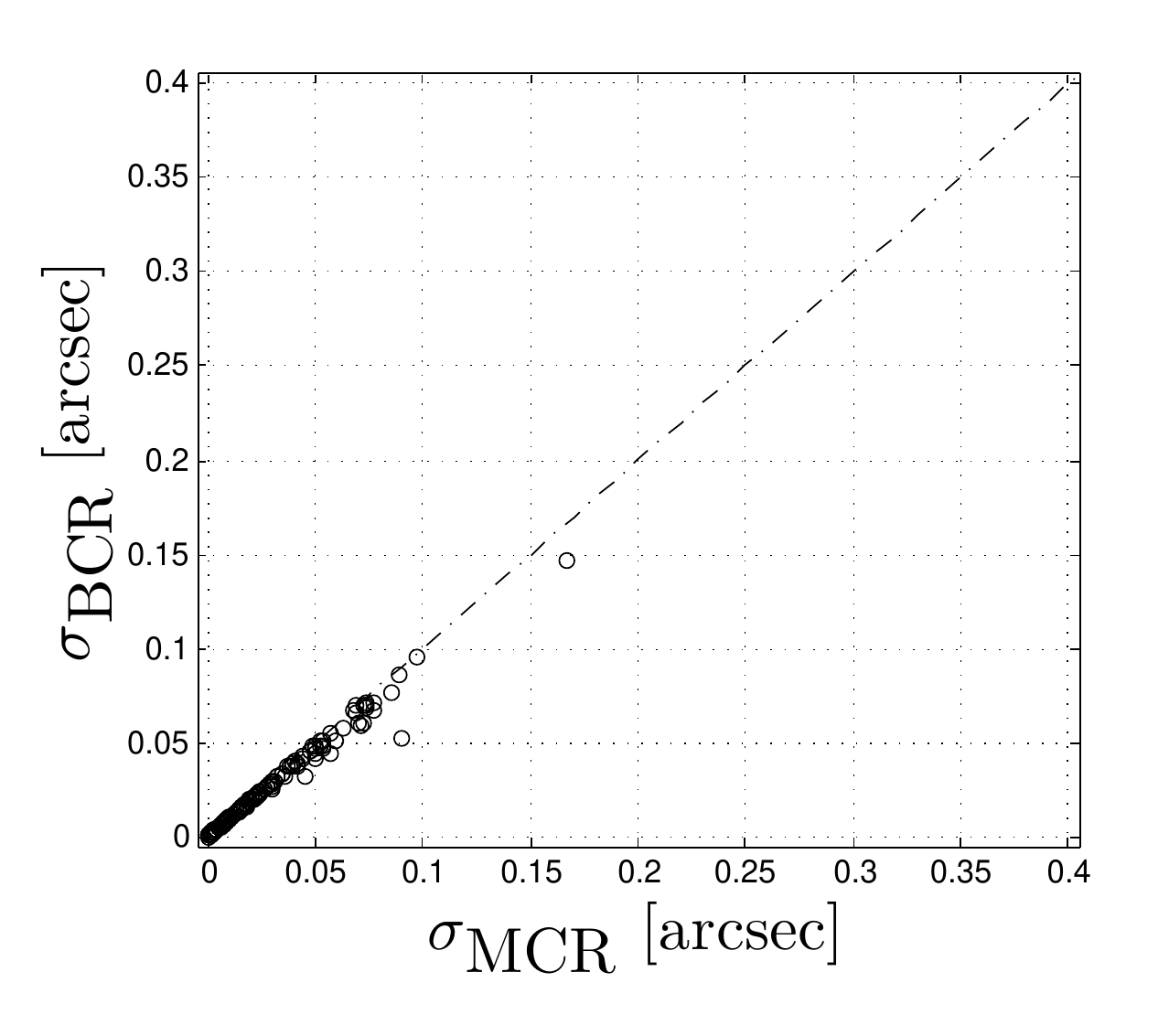}}
\subfigure[FWHM=0.7 arcsec, aperture=1 m]{\includegraphics[width=0.33\textwidth]{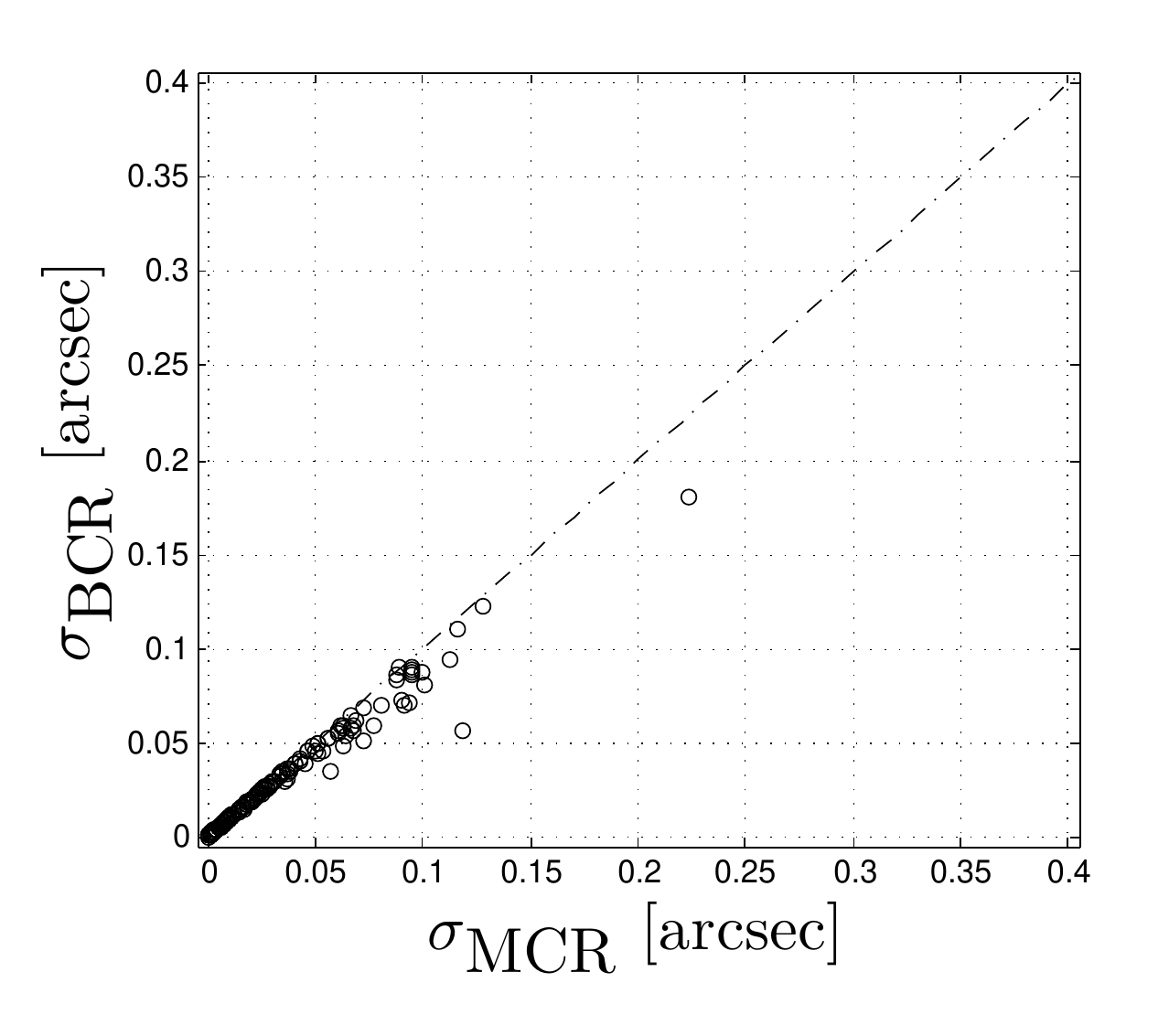}}
\subfigure[FWHM=1.2 arcsec, aperture=3.5 m]{\includegraphics[width=0.33\textwidth]{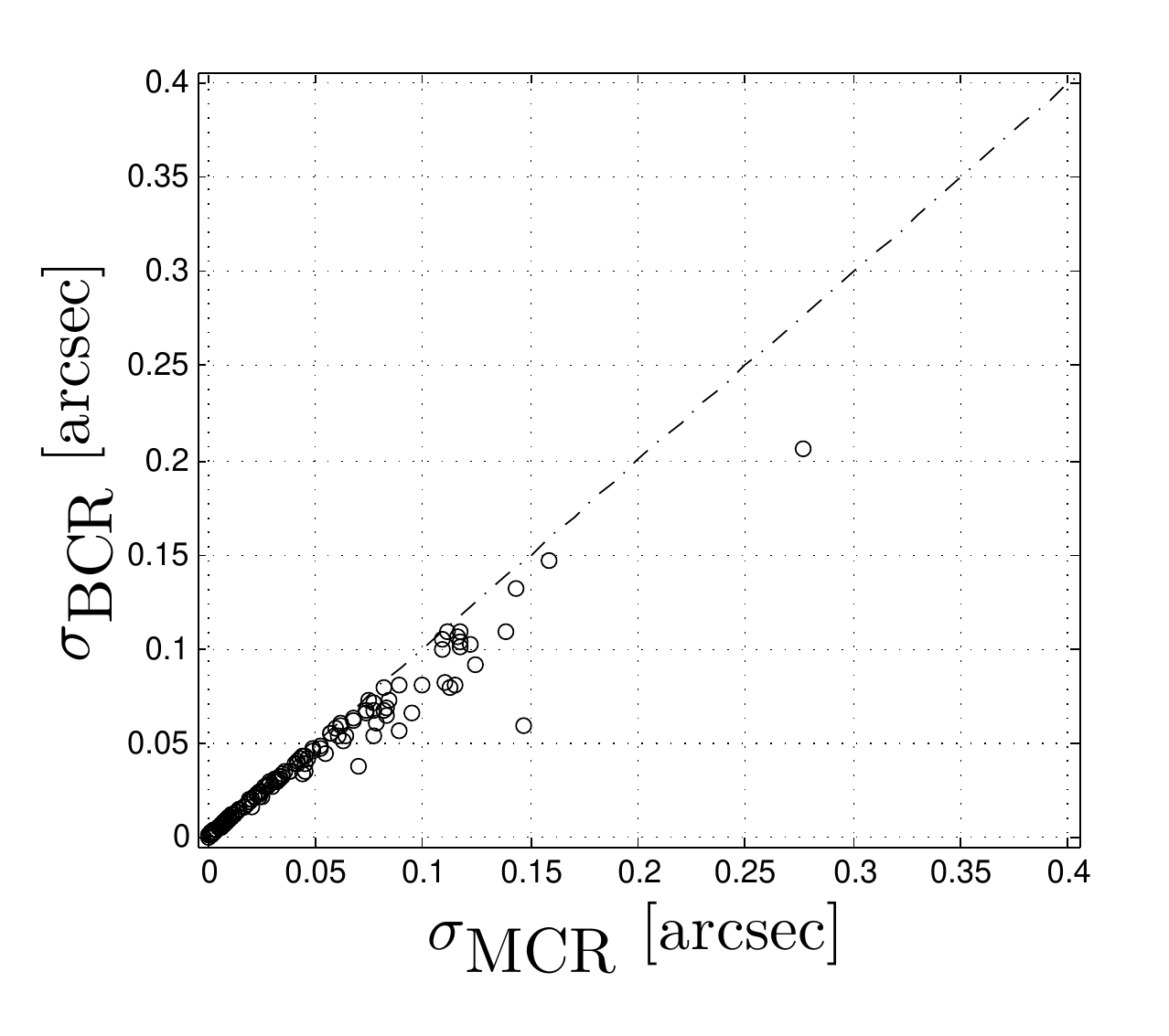}}}
\centerline{
\subfigure[FWHM=1.2 arcsec, aperture=1 m]{\includegraphics[width=0.33\textwidth]{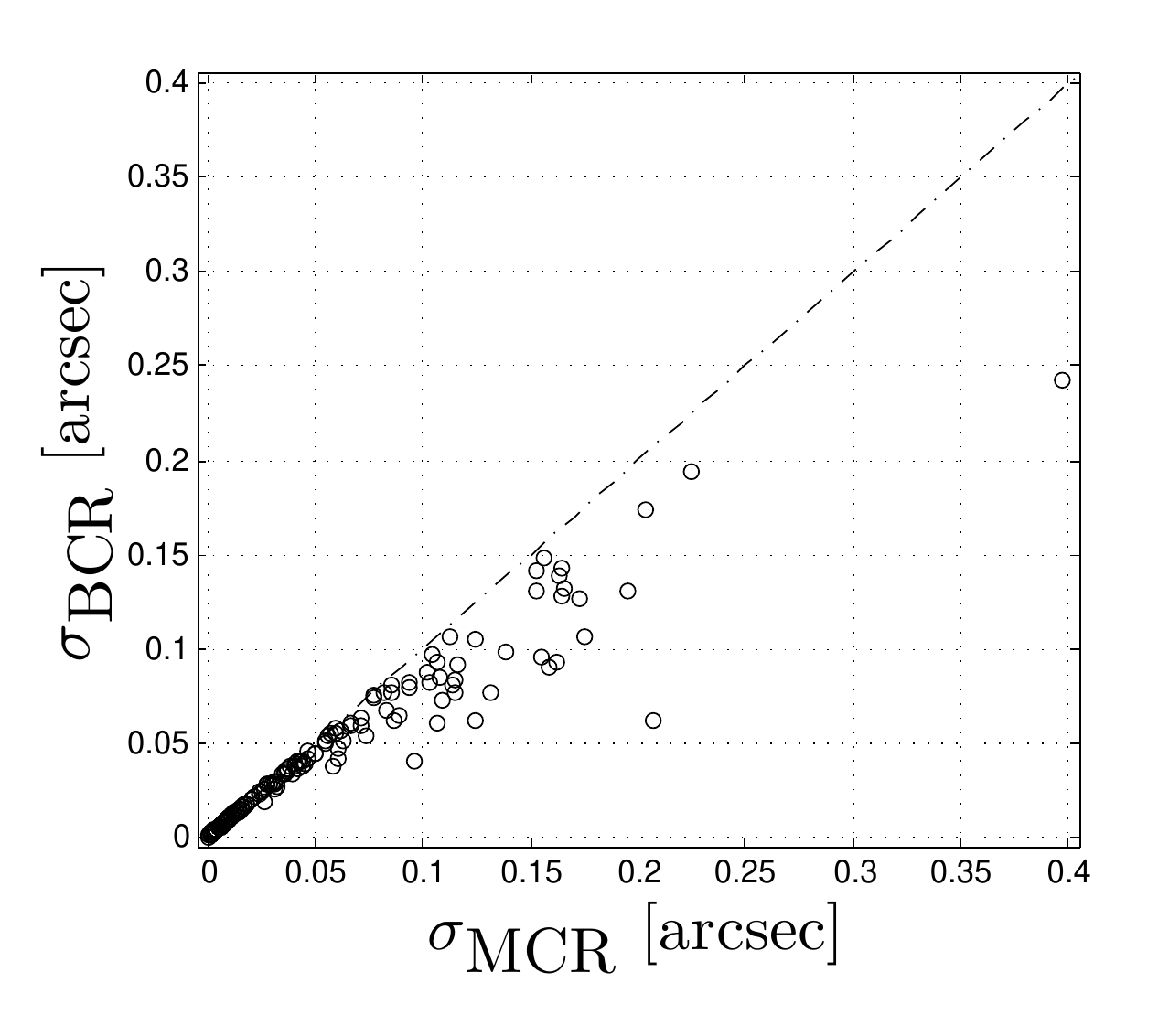}}
\subfigure[FWHM=2 arcsec, aperture=1 m]{\includegraphics[width=0.33\textwidth]{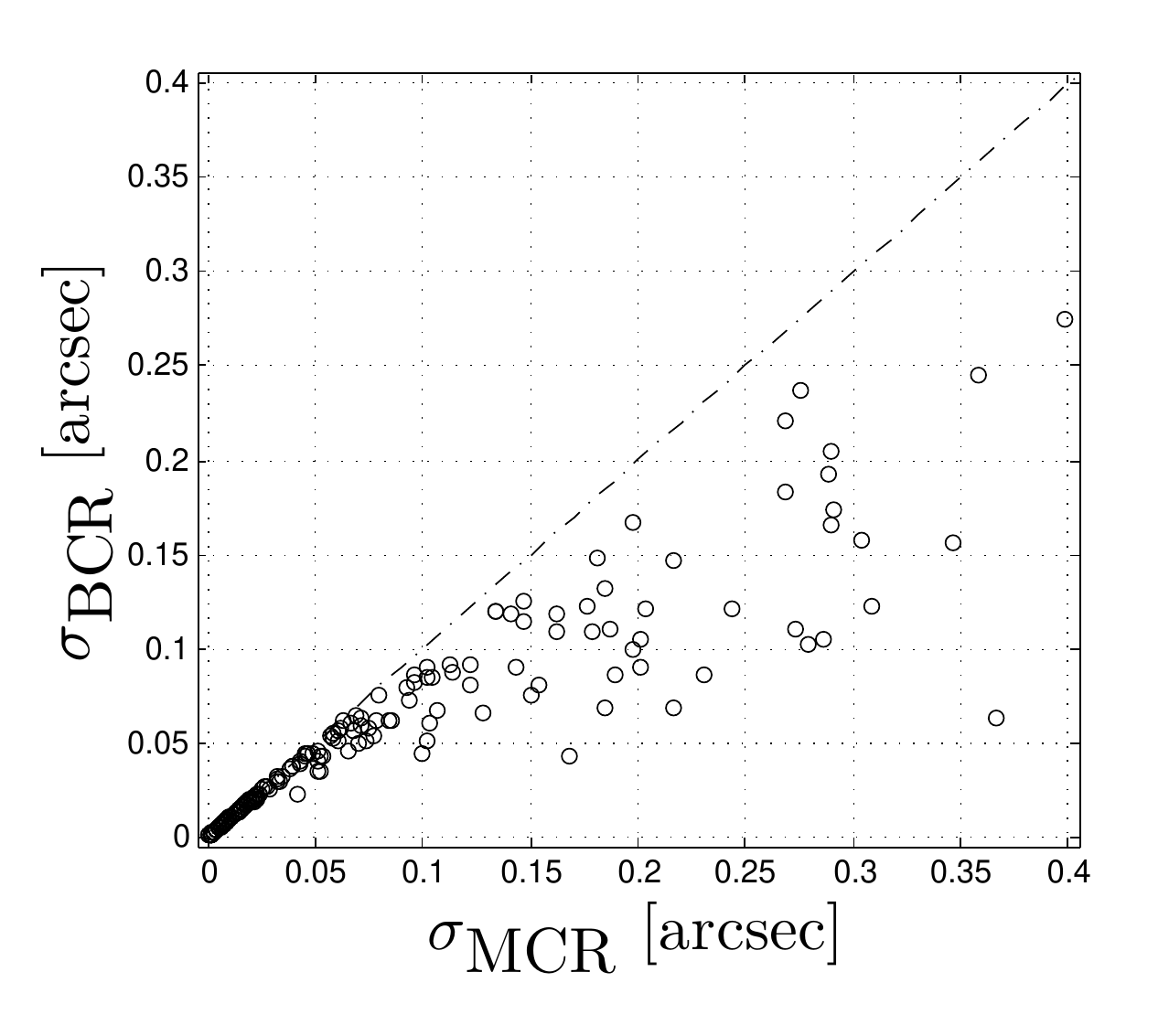}}}
\caption{BCR vs. MCR bounds for the whole SGP sample analyzed in this paper. The dashed line indicates the one-to-one relationship, i.e., no gain of the BCR with respect to the parametric case given by the MCR bound.}
\label{s_BCRvss_MCR}
\end{figure*}

In Fig.~\ref{s_BCRvss_MCR} we show the BCR {\it vs.} the MCR bounds
(both in arcsec) for all of the 226 objects in the USNO-B1 stellar-like
SGP catalog described above. The diagonal dashed line indicates the
locus of objects for which the use of a Bayesian approach does not
lead to any improvement over the classical parametric case. As
explained in Proposition~3, Eq.~(\ref{eq_sec_com_cr_4}), and in
  Sect.~\ref{sub_sec_extreme_regimes}, we predict that all the
  objects will be located below that line or, at most, on the line,
  which is obviously the case in all the observational scenarios. As
  was already hinted in the histograms presented in the previous
  section, when the (new) observations become of poorer and poorer
  quality or, equivalently, when the a priori information becomes more
  and more relevant (in the precise sense defined in
  Sect.~\ref{sub_sec_bayes_fisher}), a larger number of objects
  start to populate the bottom of these $\sigma_{\mbox{BCR}}$ {\it
    vs.}  $\sigma_{\mbox{MCR}}$ diagrams. The two objects that appear
  below the line even in the case when the observations are of good
  quality (see in particular panels (a), (b), and (c) in
  Fig.~\ref{s_BCRvss_MCR}) have very high quality positions reported
  on the catalog and therefore, for them, the a priori information
  is quite relevant, always. Another important point to highlight from
  these plots is that, of course, as the quality of the observations
  deteriorate, both the $\sigma_{\mbox{MCR}}$ and
  $\sigma_{\mbox{BCR}}$ increase, which pushes the catalog objects
  up and to the right in the diagrams. However,  the use of the Bayesian
  approach ensures that the deterioration of the latter is hampered by
  the use of a priori information, which the MCR does not use, hence this explains
  the increased scattering of points to the right and below the
  diagonal line from panels (a)-(e) in Fig.~\ref{s_BCRvss_MCR}.

\begin{figure*}[h!]
\centering
\centerline{
\subfigure[FWHM=0.7 arcsec, aperture=3.5 m]{\includegraphics[width=0.33\textwidth]{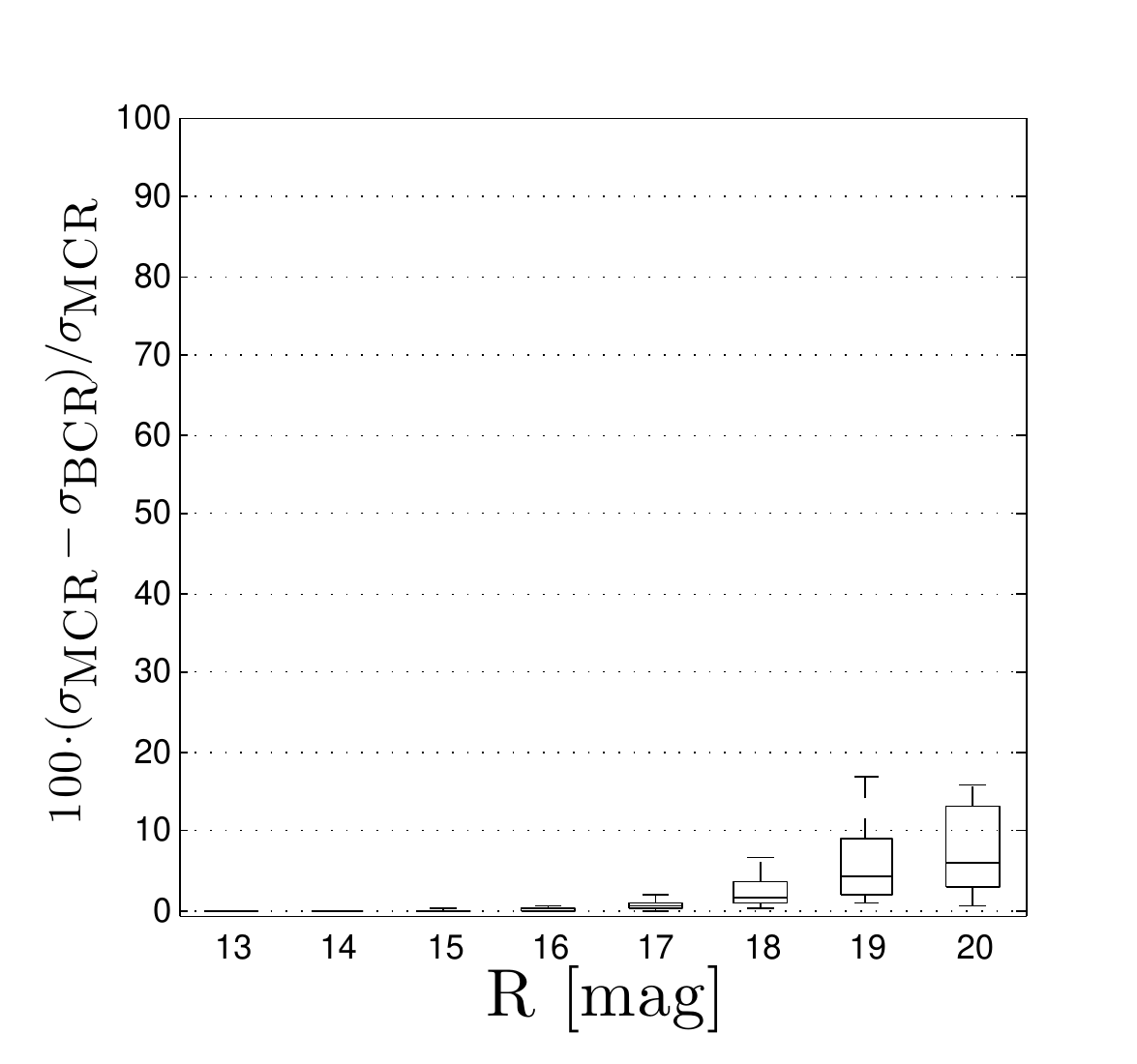}}
\subfigure[FWHM=0.7 arcsec, aperture=1 m]{\includegraphics[width=0.33\textwidth]{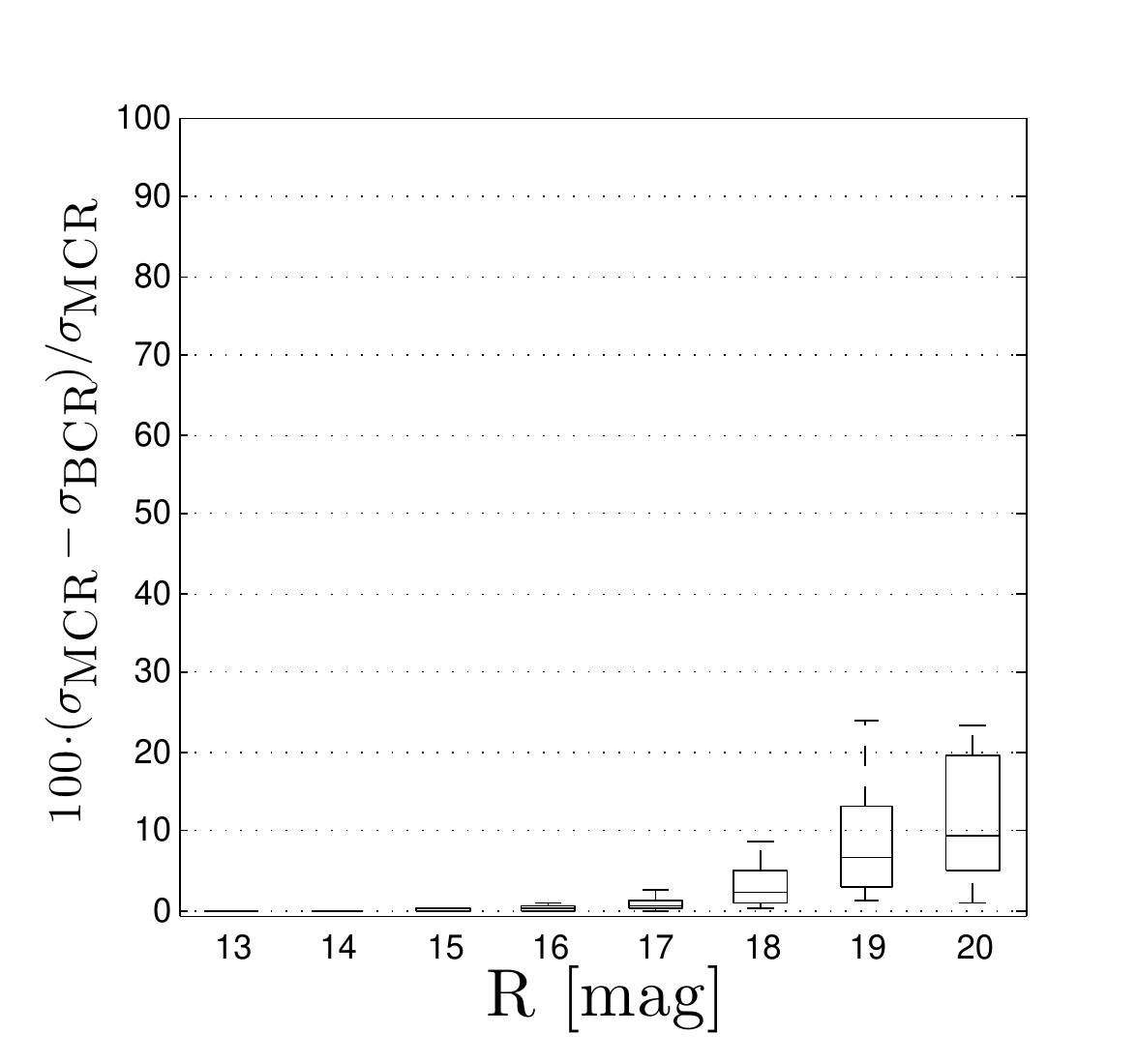}}
\subfigure[FWHM=1.2 arcsec, aperture=3.5 m]{\includegraphics[width=0.33\textwidth]{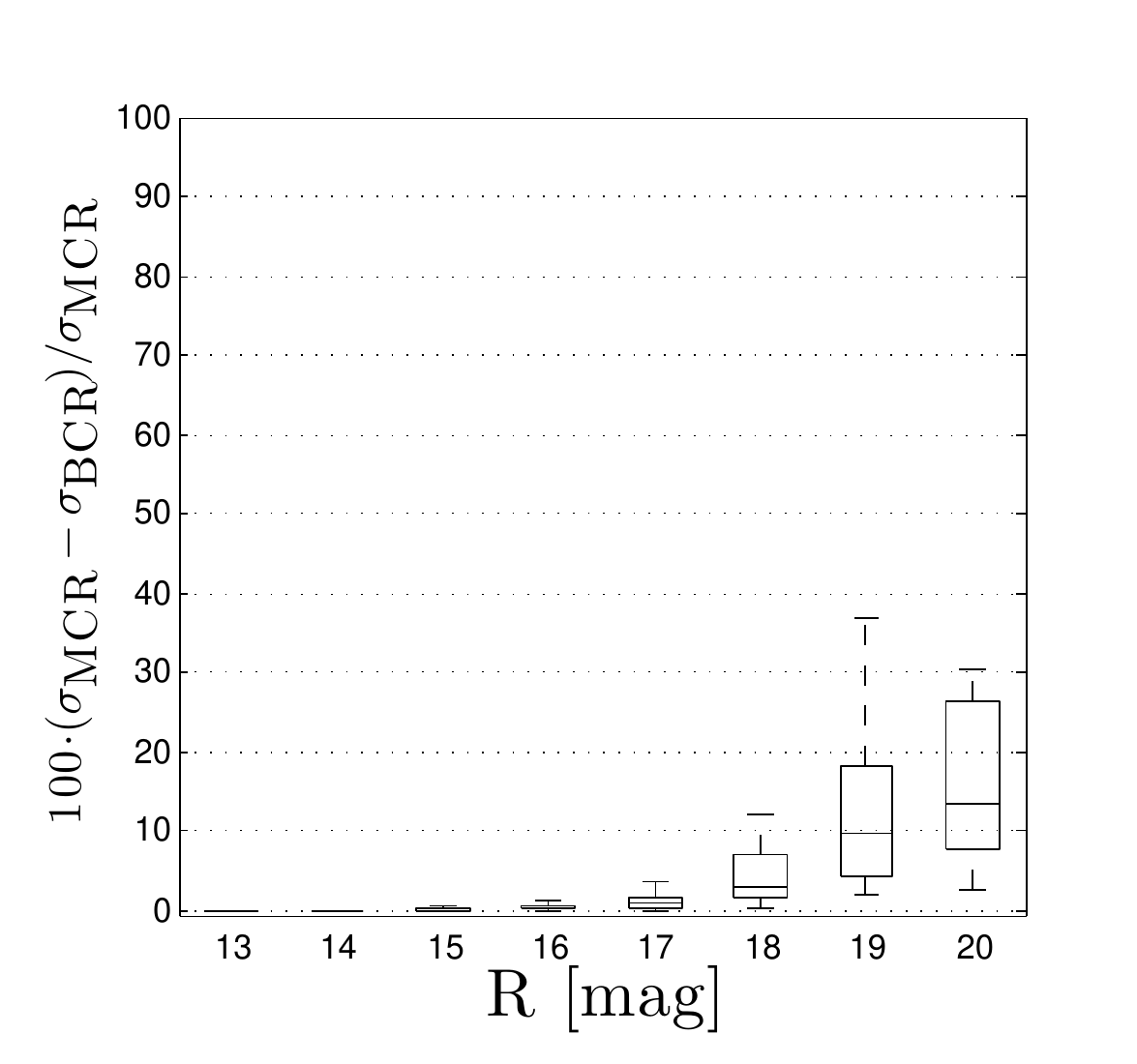}}}
\centerline{
\subfigure[FWHM=1.2 arcsec, aperture=1 m]{\includegraphics[width=0.357\textwidth]{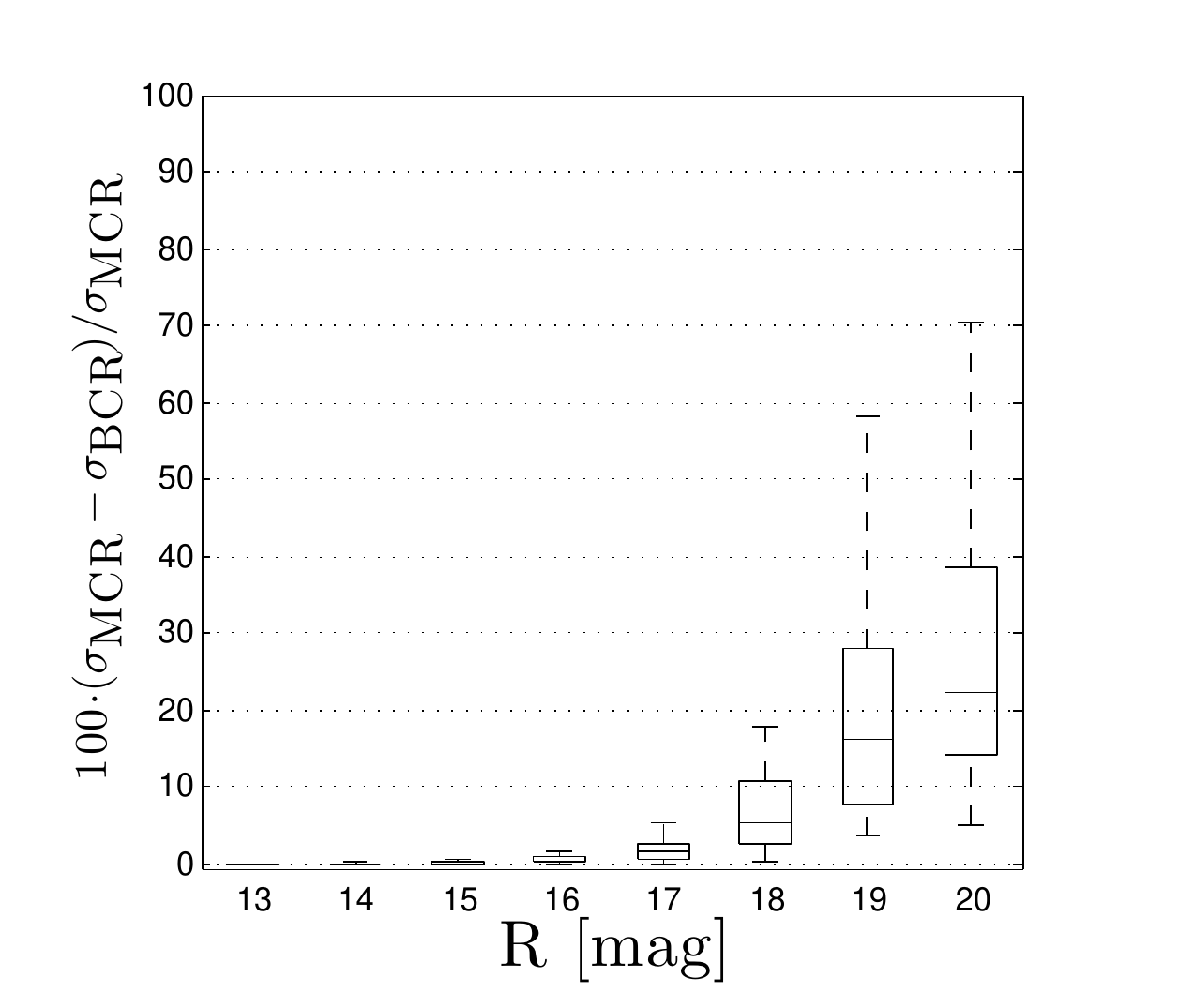}}
\subfigure[FWHM=2 arcsec, aperture=1 m]{\includegraphics[width=0.33\textwidth]{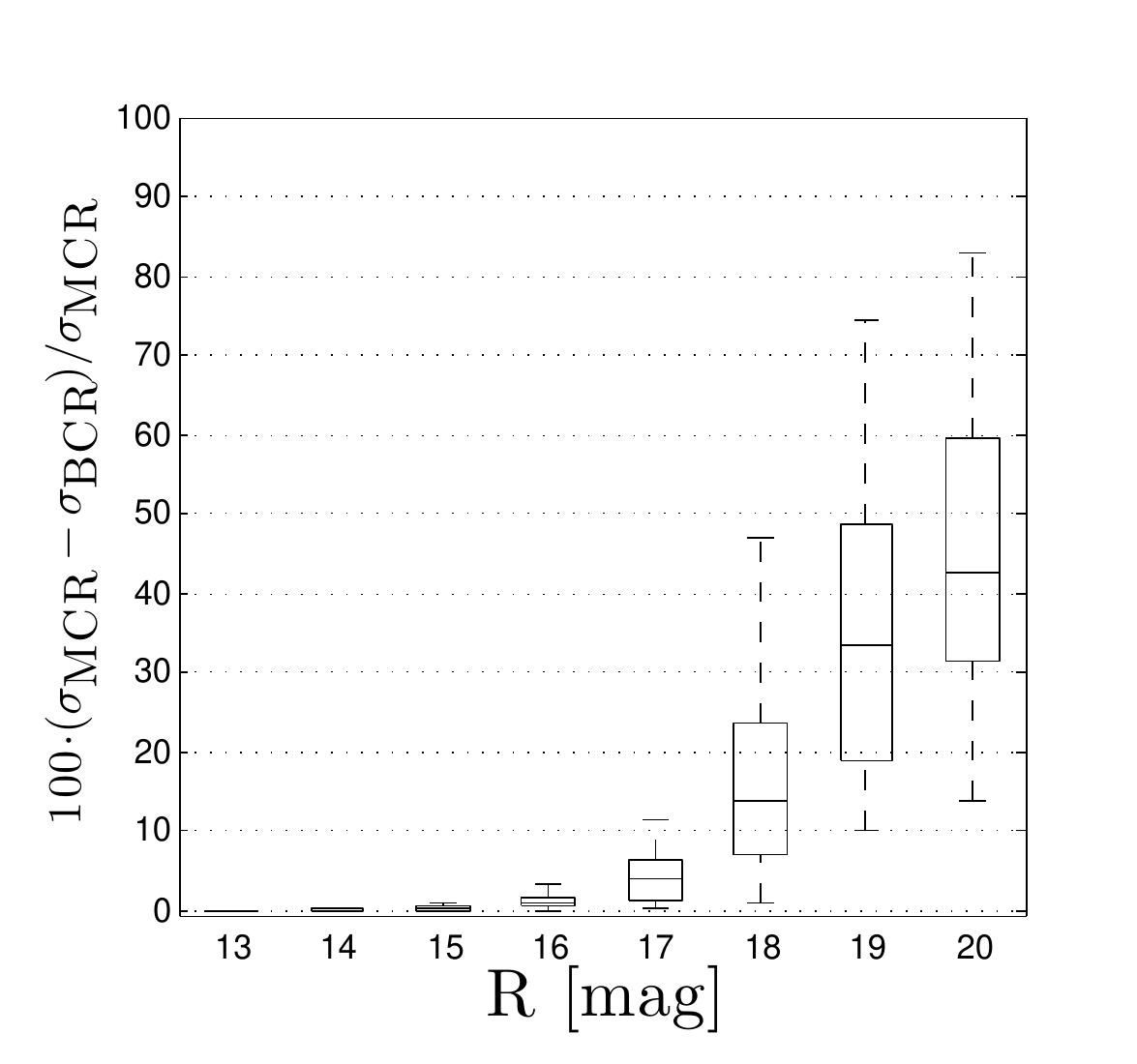}}}
\caption{Fractional improvement of astrometric precision as a function
  brightness in the R band for the whole SGP USNO-B1 sample, in
  centered 1 magnitude bins, for the different observational scenarios
  presented in the previous section. The box boundary indicates the
  1st and 3rd quartiles, the horizontal line in the center of the box
  is the median, and the upper and lower ticks show respectively the largest and
  smallest value in that magnitude bin.}
\label{fract_improv}
\end{figure*}

As clearly shown by Figs.~\ref{SNR=6Db}, \ref{SNR=12Db}, and
\ref{SNR=32Db} in  Sect.~\ref{subsec_experiments_mmse_bcr}, an
underlying parameter determining the gains of the Bayesian approach is
the source flux, which is not well rendered by
Fig.~\ref{s_BCRvss_MCR}. To address more clearly this particular
aspect, Fig.~\ref{fract_improv} shows the relative improvement on
astrometric precision as a function of magnitude for the whole USNO-B1
SGP sample.  This improvement is measured in terms of a relative gain 
in astrometric performance, i.e., for each sample of the database
we compute
 $100 \left( \frac{\sigma_{MCR}- \sigma_{BCR}}{\sigma_{MCR}} \right)$.
This array of figures ratifies the usefulness of a Bayesian
strategy to mitigate the inevitable deterioration of astrometric
precision as a function of flux (see Eq.~(30) in
\citet{mendez2014analysis}). Remarkably, this becomes particularly dramatic when
the observations are performed under adverse conditions (panels (d)
and (e) in Fig.~\ref{fract_improv}), where one can expect
improvements of between 10\% and 50\% in the astrometric
precision for the faintest objects when using the Bayesian approach as
described here.

\section{Conclusions and outlook}
\label{final}

In this work we provide a systematic analysis of the best precision that can be achieved to
determine the location of a stellar-like object on a CCD-like detector
array in a Bayesian setting. 
This setting changes in a radical way the nature of the
inference task in hand: from a parametric context --- in which we are
estimating a constant (or parameter) from a set of observations
--- to a setting in which we estimate a random object (i.e., the
position is modeled as a random variable) from observations that are
statistically dependent with the position. A key new element of the
Bayesian setting is the introduction of a prior distribution of the
object position: We systematically quantify and analyze 
the gain in astrometric performance from the use of a prior
distribution of the object position, information which is not available (or used) in the classical
parametric setting. We tackle this problem from a
theoretical and experimental point of view.

We derive new closed-form expressions for the Bayesian CR and expressions
to estimate the gain in astrometric precision. Different observational regimes
are evaluated to quantify the gain induced from the prior distribution
of the object position. An insightful corollary of this analysis is that
the Bayes setting always offers a better performance than the
parametric setting, even in the worse-case prior (i.e., that of a
uniform distribution).

We evaluate numerically the benefits of the Bayes setting with respect to
the parametric scenario under realistic experimental conditions: We
find that the gain in performance is significant for various
observational regimes, which is particularly clear in the case of
faint objects, or when the observations are taken in poor conditions
(i.e., in the low signal-to-noise regime). In this context we
introduce the new concept of the equivalent object brightness. We also
submit evidence that the minimum mean square error estimator of this
problem (the well-known conditional mean) tightly achieves
the Bayesian CR lower bound, a remarkable result, demonstrating that
all the performance gains presented in the theoretical analysis part
of our paper can indeed be achieved with the minimum mean square error
estimator, which has in principle a practical implementation. We
finalize our paper with a simple example of what could be achieved
using the Bayesian approach in terms of the astrometric precision of
positional measurements with new observations of varying quality, when
we incorporate data from existing catalog as prior information.

In a forthcoming paper we expect to extend the present analysis of the
BCR to the case of photometric estimations and to implement the
MMSE estimator in Eq.~(\ref{eq_sec_opt_estimator_1}). With the help of
simulations we will be able to quantify the gains (and risks) of using
a Bayesian approach with priors for the joint estimation of photometry
and astrometry of point sources. For example, \citet{michalik2015gaia}
have pointed out that a Bayesian approach would lead to biased
astrometric results (particularly for proper motions and parallaxes),
and should be avoided if a reasonable solution can be found using a
parametric approach. The conditions and extent to which this statement
is true could be precisely verified through detailed simulations, thus
providing a range where a Bayesian scheme might be safely applied.
  
\section{Acknowledgments}
This material is based on work supported by grant of CONICYT-Chile,
Fondecyt 1140840. In addition, the work of J. F. Silva and M. Orchard
is supported by the Advanced Center for Electrical and Electronic
Engineering, Basal Project FB0008. R.A.M.  and J.F.S. acknowledge 
support from a CONICYT-Fondecyt grant \# 1151213, and R.A.M.  from the Project IC120009
 Millennium Institute of Astrophysics (MAS) of the Iniciativa
Cient\'{\i}fica Milenio del Ministerio de Econom\'{\i}a, Fomento y
Turismo de Chile. R.A.M also acknowledges ESO/Chile for hosting him
during his sabbatical leave during 2014, the period in which part of this
work was started.

This research has made use of the VizieR catalog access tool, CDS,
Strasbourg, France. The original description of the VizieR service was
published in \citet{ochsenbein2000vizier}.

\appendix

\section{Derivation of the Bayes-Fisher information identity in Eq.~(\ref{eq_sec_bayes_6})}
\label{app_derivation_bfi_identity}
Using the definition of $\tilde{L}(X_c, I^n)$ in Eq.~(\ref{bayes_like}), it follows that
\begin{align}\label{eq:app_derivation_bfi_identity_1}
	&\mathbb{E}_{(X_c,I^n)} \left\{ \left( \frac{d \ln
            \tilde{L}(X_c, I^n)}{d x} \right)^2 \right\} = \mathbb{E}_{(X_c,I^n)} \left\{
        \left( \frac{d \ln L(I^n; X_c)}{d x} \right)^2 \right\} +\nonumber\\
        &2\cdot \mathbb{E}_{(X_c,I^n)} \left\{ \frac{d \ln L(I^n;
          X_c)}{d x} \cdot \frac{d \ln \psi(X_c)}{dx} \right\}
     + \mathbb{E}_{X_c} \left\{ \left( \frac{d \ln
          \psi(X_c)}{dx} \right)^2\right\}\\
	   \label{eq:app_derivation_bfi_identity_1b}
	   &= \mathbb{E}_{X_c} \underbrace{\mathbb{E}_{I^n|X_c}
             \left\{ \left( \frac{d \ln L(I^n; X_c)}{d x} \right)^2
             \right\}}_{\mathcal{I}_{X_c}(n)} + \mathbb{E}_{X_c}
           \left\{ \left( \frac{d \ln \psi(X_c)}{dx}
           \right)^2\right\}\\
	   \label{eq:app_derivation_bfi_identity_1c}
	 &= \mathbb{E}_{X_c} \left\{ \mathcal{I}_{X_c}(n) \right\} +
		\mathbb{E}_{X_c} \left\{ \left( \frac{d \ln \psi(X_c)}{dx} \right)^2\right\}.
\end{align}
The first line, Eq.~(\ref{eq:app_derivation_bfi_identity_1}), comes from the definition of $\tilde{L}(x,i^n)$
in Eq.~(\ref{bayes_like}). The second line,
Eq.~(\ref{eq:app_derivation_bfi_identity_1b}), follows directly from the fact that for
every position $x_c\in \mathbb{R}$, $\mathbb{E}_{I^n|X_c=x_c} \left\{
\frac{d \ln L(I^n; x_c)}{d x} \right\}=0$ (see
Eq.~(\ref{cond2d})).  The last line, Eq.~(\ref{eq:app_derivation_bfi_identity_1c}), 
results from the definition of the Fisher information for the scalar
case in Eq.~(\ref{fisher}). 

\section{Proof of the lower bound of $Gain(\psi)$ in Eq.~(\ref{eq_pro_information_gain})}
\label{app_pro_information_gain}
\textit{Proof:}
Let us consider a fixed prior distribution $\psi$. To evaluate the
performance of the baseline parametric case, let us consider an
arbitrary unbiased estimator $\tau^n(\cdot)$ of the position.  Then,
conditioned on a position $X_c=x$, we have that $Var(\tau^n(I^n)) =
\mathbb{E}_{I^n|X_c=x} \left\{ \left( \tau^n(I^n)-x \right)^2
\right\}$, this from the fact that the estimator is
unbiased. Therefore, from the CR lower bound in
Sect.~\ref{subsec_cr_bounds}, it follows that
\begin{equation}\label{eq_sec_com_cr_1}
 \mathbb{E}_{I^n|X_c=x} \left\{ \left( \tau^n(I^n)-x \right)^2
 \right\}= Var(\tau^n(I^n)) \geq \frac{1}{\mathcal{I}_x(n)}, \ \ \forall x\in
 \mathbb{R}.
\end{equation}
From the expression above, the average performance of $\tau^n(\cdot)$, with
respect to the statistics of $X_c\sim \psi$, is not smaller than
$\mathbb{E}_{X_c} \left\{ \frac{1}{\mathcal{I}_{X_c}(n)} \right\}$,
which is the average CR lower bound (with respect to $\psi$). In other
words, for any unbiased estimator $\tau_{unbias}^n(\cdot) \in \mathcal{T}^n$
of the object position ($\mathcal{T}^n$ denotes the collection of unbiased estimator), we have that
\begin{equation}\label{eq_sec_com_cr_2}
 \mathbb{E}_{(X_c, I^n)} \left\{ \left( \tau_{unbias}^n(I^n)-X_c
 \right)^2 \right\} \geq \underbrace{ \mathbb{E}_{X_c\sim \psi} \left\{
 \frac{1}{\mathcal{I}_{X_c}(n)} \right\}}_{\equiv \sigma^2_{MCR}}.
\end{equation}
We note that $\sigma^2_{MCR}$ is the average classical CR bound,
which, as we see below (Eq.~(\ref{eq_sec_com_cr_3b})), is
the right figure of merit to compare with the Bayesian CR lower bound. 
In other words, Eq.~(\ref{eq_sec_com_cr_2}) offers a lower bound for the MSE of any
unbiased estimator with no access to the prior probability law of
$X_c$. 

On the other hand, we have that using the prior distribution of
$X_c$, the BCR on Eq.~(\ref{eq_sec_bayes_7}) offers a lower bound for
the MSE of any estimator.  If we assume for a moment that the optimal
MSE estimator solution of Eq.~(\ref{eq_sec_bayes_4}), which we denote
by $\hat{\tau}^n_{Bayes}$, achieves the BCR lower bound\footnote{In
  Section \ref{subsec_experiments_mmse_bcr} we show that this
  hypothesis is indeed true for many realistic astrometric scenarios.} in
Eq.~(\ref{eq_sec_bayes_7}), then we can compute the
performance gain as follows:
\begin{align}\label{eq_sec_com_cr_3}
&Gain(\psi) = \min_{\tau_{unbias}^n \in \mathcal{T}^n}
\mathbb{E}_{(X_c, I^n)} \left\{ \left( \tau_{unbias}^n(I^n)-x
\right)^2 \right\} - \nonumber\\ 
&\min_{\tau^n: \mathbb{N}^n \rightarrow
  \mathbb{R}} \mathbb{E}_{(X_c,I^n)} \left\{ \left( \tau^n(I^n)-X_c
\right)^2 \right\}\nonumber\\
		&= \min_{\tau_{unbias}^n \in \mathcal{T}^n}
\mathbb{E}_{(X_c, I^n)} \left\{ \left( \tau_{unbias}^n(I^n)-x
\right)^2 \right\} \nonumber\\
&- \frac{1}{\mathbb{E}_{X_c\sim \psi} \left\{
  \mathcal{I}_{X_c}(n) \right\} + \mathbb{E}_{X_c \sim \psi} \left\{
  \left( \frac{d \ln \psi(X_c)}{dx} \right)^2\right\}} \\
		\label{eq_sec_com_cr_3b}
		&\geq \mathbb{E}_{X_c\sim \psi} \left\{
                \frac{1}{\mathcal{I}_{X_c}(n)} \right\} -
                \frac{1}{\mathbb{E}_{X_c\sim \psi} \left\{
                  \mathcal{I}_{X_c}(n) \right\} +
\mathcal{I}(\psi)} = \sigma^2_{MCR} - \sigma^2_{BCR}.
\end{align}
The first equality is from definition, the second equality is from the assumption that $\hat{\tau}^n_{Bayes}$ achieves the BCR bound, and the last inequality is from Eq.~(\ref{eq_sec_com_cr_2}). This concludes the result. 

   \bibliographystyle{aa} 
   \bibliography{bibliography} 

\begin{thebibliography}{41}
\expandafter\ifx\csname natexlab\endcsname\relax\def\natexlab#1{#1}\fi

\bibitem[{Adorf(1996)}]{adorf1996limits}
Adorf, H.-M. 1996, in ADASS, Vol. 101, 13

\bibitem[{Auer \& Van~Altena(1978)}]{auer1978digital}
Auer, L. \& Van~Altena, W. 1978, AJ, 83, 531

\bibitem[{Bastian(2004)}]{bastian2004maximum}
Bastian, U. 2004, 2004BASNOCODE, Gaia DPAC Public Documents
  (http://www.cosmos.esa.int/web/gaia/public-dpac-documents)

\bibitem[{Bendinelli {et~al.}(1987)Bendinelli, Parmeggiani, Piccioni, \&
  Zavatti}]{bendinelli1987newton}
Bendinelli, O., Parmeggiani, G., Piccioni, A., \& Zavatti, F. 1987, AJ, 94,
  1095

\bibitem[{Bouquillon {et~al.}(2016)Bouquillon, Mendez, Altmann, Carlucci,
  Barache, Taris, Andrei, \& Smart}]{Bouquillon2016}
Bouquillon, S., Mendez, R., Altmann, M., {et~al.} 2016, submitted to A\&A

\bibitem[{Chromey(2010)}]{chromey2010measure}
Chromey, F.~R. 2010, To measure the sky: an introduction to observational
  astronomy (Cambridge University Press)

\bibitem[{Cram{\'e}r(1946)}]{cramer1946contribution}
Cram{\'e}r, H. 1946, Scandinavian Actuarial Journal, 1946, 85

\bibitem[{Freyhammer {et~al.}(2001)Freyhammer, Andersen, Arentoft, Sterken, \&
  N{\o}rregaard}]{freyetal01}
Freyhammer, L.~M., Andersen, M.~I., Arentoft, T., Sterken, C., \&
  N{\o}rregaard, P. 2001, Experimental Astronomy, 12, 147

\bibitem[{Gawiser {et~al.}(2006)Gawiser, van Dokkum, Herrera, \&
  et~al.}]{gawiet06}
Gawiser, E., van Dokkum, P.~G., Herrera, D., \& et~al. 2006, ApJS, 162, 1

\bibitem[{Gray \& Davisson(2004)}]{gray2004introduction}
Gray, R.~M. \& Davisson, L.~D. 2004, An introduction to statistical signal
  processing (Cambridge University Press)

\bibitem[{H{\o}g(2011)}]{hog2011astrometry}
H{\o}g, E. 2011, Baltic Astronomy, 20, 221

\bibitem[{Howell(2006)}]{howell2006handbook}
Howell, S.~B. 2006, Handbook of CCD astronomy, Vol.~5 (Cambridge University
  Press)

\bibitem[{Jakobsen {et~al.}(1992)Jakobsen, Greenfield, \&
  Jedrzejewski}]{jakobsen1992cramer}
Jakobsen, P., Greenfield, P., \& Jedrzejewski, R. 1992, A\&A, 253, 329

\bibitem[{Kay(2010)}]{kay2010fundamentals}
Kay, S.~M. 2010, Fundamentals of statistical signal processing: estimation
  theory. (Prentice-Hall PTR)

\bibitem[{{King}(1971)}]{King1971}
{King}, I.~R. 1971, \pasp, 83, 199

\bibitem[{Lee \& van Altena(1983)}]{lee1983theoretical}
Lee, J.-F. \& van Altena, W. 1983, AJ, 88, 1683

\bibitem[{Lehmann \& Casella(1998)}]{lehmann1998theory}
Lehmann, E.~L. \& Casella, G. 1998, Theory of point estimation, Vol.~31
  (Springer Science \& Business Media)

\bibitem[{Lindegren(2008)}]{Lindegren2008}
Lindegren. 2008, GAIA-C3-TN-LU-LL-078, Gaia DPAC Public Documents
  (http://www.cosmos.esa.int/web/gaia/public-dpac-documents)

\bibitem[{Lindegren(1978)}]{lindegren1978photoelectric}
Lindegren, L. 1978, in IAU Colloq. 48: Modern Astrometry, Vol.~1, 197--217

\bibitem[{{Lindegren}(2010)}]{lindegren2010}
{Lindegren}, L. 2010, ISSI, 9, 279

\bibitem[{Lobos {et~al.}(2015)Lobos, Silva, Mendez, \&
  Orchard}]{lobos2015performance}
Lobos, R.~A., Silva, J.~F., Mendez, R.~A., \& Orchard, M. 2015, \pasp, 127,
  1166

\bibitem[{Mason(2008)}]{mason07}
Mason, E. 2008, in The 2007 ESO Instrument Calibration Workshop, ed. A.~Kaufer
  \& F.~Kerber, Vol. 107 (Springer-Verlag)

\bibitem[{M{\'e}ndez {et~al.}(2010)M{\'e}ndez, Costa, Pedreros, Moyano,
  Altmann, \& Gallart}]{mendez2010proper}
M{\'e}ndez, R.~A., Costa, E., Pedreros, M.~H., {et~al.} 2010, \pasp, 122, 853

\bibitem[{Mendez {et~al.}(2013)Mendez, Silva, \& Lobos}]{mendez2013analysis}
Mendez, R.~A., Silva, J.~F., \& Lobos, R. 2013, \pasp, 125, 580

\bibitem[{Mendez {et~al.}(2014)Mendez, Silva, Orostica, \&
  Lobos}]{mendez2014analysis}
Mendez, R.~A., Silva, J.~F., Orostica, R., \& Lobos, R. 2014, \pasp, 126, 798

\bibitem[{Michalik \& Lindegren(2016)}]{michalik2015quasars}
Michalik, D. \& Lindegren, L. 2016, A\&A, 586, A26

\bibitem[{Michalik {et~al.}(2015{\natexlab{a}})Michalik, Lindegren, \&
  Hobbs}]{michalik2015tycho}
Michalik, D., Lindegren, L., \& Hobbs, D. 2015{\natexlab{a}}, A\&A, 574, A115

\bibitem[{Michalik {et~al.}(2015{\natexlab{b}})Michalik, Lindegren, Hobbs, \&
  Butkevich}]{michalik2015gaia}
Michalik, D., Lindegren, L., Hobbs, D., \& Butkevich, A.~G. 2015{\natexlab{b}},
  A\&A, 583, A68

\bibitem[{Michalik {et~al.}(2014)Michalik, Lindegren, Hobbs, \&
  Lammers}]{michalik2014joint}
Michalik, D., Lindegren, L., Hobbs, D., \& Lammers, U. 2014, A\&A, 571, A85

\bibitem[{Mignard(2009)}]{mignard2009hundred}
Mignard, F. 2009, GAIA-C3-TN-OCA-FM-040, Gaia DPAC Public Documents
  (http://www.cosmos.esa.int/web/gaia/public-dpac-documents)

\bibitem[{Monet {et~al.}(2003)Monet, Levine, Canzian, Ables, Bird, Dahn,
  Guetter, Harris, Henden, Leggett, {et~al.}}]{monet2003usno}
Monet, D.~G., Levine, S.~E., Canzian, B., {et~al.} 2003, AJ, 125, 984

\bibitem[{Moon \& Stirling(2000)}]{moon2000mathematical}
Moon, T.~K. \& Stirling, W.~C. 2000, Mathematical methods and algorithms for
  signal processing, Vol.~1 (Prentice hall Upper Saddle River)

\bibitem[{Ochsenbein {et~al.}(2000)Ochsenbein, Bauer, \&
  Marcout}]{ochsenbein2000vizier}
Ochsenbein, F., Bauer, P., \& Marcout, J. 2000, A\&A, 143, 23

\bibitem[{Perlman(1974)}]{perlman1974jensen}
Perlman, M.~D. 1974, Journal of Multivariate Analysis, 4, 52

\bibitem[{Rao(1945)}]{radhakrishna1945information}
Rao, R.~C. 1945, Bulletin of the Calcutta Mathematical Society, 37, 81

\bibitem[{Tyson(1986)}]{tyson1986low}
Tyson, J.~A. 1986, JOSA A, 3, 2131

\bibitem[{van Altena \& Auer(1975)}]{van1975digital}
van Altena, W. \& Auer, L. 1975, in Image Processing Techniques in Astronomy
  (Springer), 411--418

\bibitem[{Van~Trees(2004)}]{van2004detection}
Van~Trees, H.~L. 2004, Detection, estimation, and modulation theory (John Wiley
  \& Sons)

\bibitem[{Walker(1987)}]{walker1987noao}
Walker, A. 1987, NOAO Newsletter, 10, 16

\bibitem[{Weinstein \& Weiss(1988)}]{weinstein1988general}
Weinstein, E. \& Weiss, A.~J. 1988, Information Theory, IEEE Transactions on,
  34, 338

\bibitem[{Winick(1986)}]{winick1986cramer}
Winick, K.~A. 1986, JOSA A, 3, 1809

\end{thebibliography}
\end{document}